%% file: arxiv.tex
\documentclass[conference]{IEEEtran}

\input{mydef}

\input{theoremstyle}

\begin{document}

\title{Fixpoint operators for $2$-categorical structures}

\author{\IEEEauthorblockN{Zeinab Galal}
\IEEEauthorblockA{LIP6, Sorbonne University}
}

  \IEEEoverridecommandlockouts
 \IEEEpubid{\makebox[\columnwidth]{979-8-3503-3587-3/23/\$31.00~
 \copyright2023 European Union \hfill} \hspace{\columnsep}\makebox[\columnwidth]{ }}
\maketitle
\thispagestyle{plain}
\pagestyle{plain}

\begin{abstract}

Fixpoint operators are tools to reason on recursive programs and data types obtained by induction (e.g. lists, trees) or coinduction (e.g. streams). They were given a categorical treatment with the notion of categories with fixpoints. A theorem by Plotkin and Simpson characterizes existence and uniqueness of fixpoint operators for categories satisfying some conditions on bifree algebras and recovers the standard examples of the category Cppo ($\omega$-complete pointed partial orders and continuous functions) in domain theory and the relational model in linear logic. 

We present a categorification of this result and develop the theory of $2$-categorical fixpoint operators where the $2$-dimensional framework allows to model the execution steps for languages with (co)inductive principles. We recover the standard categorical constructions of initial algebras and final coalgebras for endofunctors as well as fixpoints of generalized species and polynomial functors.
\end{abstract}

\IEEEpeerreviewmaketitle

\section*{Introduction}
\input{intro}

\section{Fixpoint operators for $1$-categories}\label{sec:1fix}
\input{1fixtheory}

\section{Bidimensional fixpoints operators}\label{sec:2fix}
\input{2fixdef}

\section{The Plotkin-Simpson theorem for $2$-categories}\label{sec:construction}
\input{construction}

\section{Examples}\label{sec:examples}
\input{examples}

\section*{Conclusion}
We have presented the theory of fixpoint operators for $2$-categories using a categorification of Plotkin-Simpson's theorem as a guideline to derive the equations on the structural $2$-cells in dimension $2$. The concrete $2$-categorical examples allow us to confirm that theses equations are verified.

In future work, we aim to extend our theory to \emph{parametrized} and \emph{guarded} fixpoint operators. Adding parameters allows to consider richer contexts for terms and guardedness restricts the possible infinite behavior of fixpoints to ensure properties such as solvability or productivity are satisfied.
Since a parametrized fixpoint operator verifying the Conway axioms is equivalent to a \emph{traced monoidal category} with the cartesian product as the chosen tensor, we aim to use our formalism to develop the theory of traced monoidal bicategories and establish new connections with cyclic $\lambda$-calculi \cite{hasegawa1997recursion,benton2003traced}. We also want to formulate the theory of $2$-dimensional fixpoints in a cartesian closed framework where the fixpoint operator is expressed internally as a family of $1$-cells of type $(A \Rightarrow A) \to A$ bringing us closer to the intuition of fixpoint combinators for $\lambda$-calculus. 

\section*{Acknowledgments}
I am grateful to the anonymous reviewers for their comments and to Marcelo Fiore, Nicola Gambino and Martin Hyland for valuable discussions on this work which was supported by EPSRC grant EP/V002309/1 and by the RealiSe PGRM Emergence project.

\bibliographystyle{IEEEtran}

\bibliography{biblio}

\onecolumn
\appendix
\input{appendix}

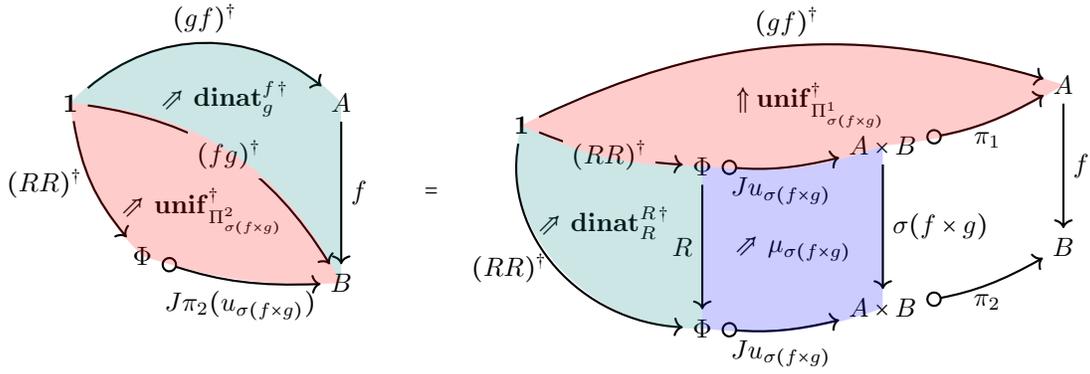
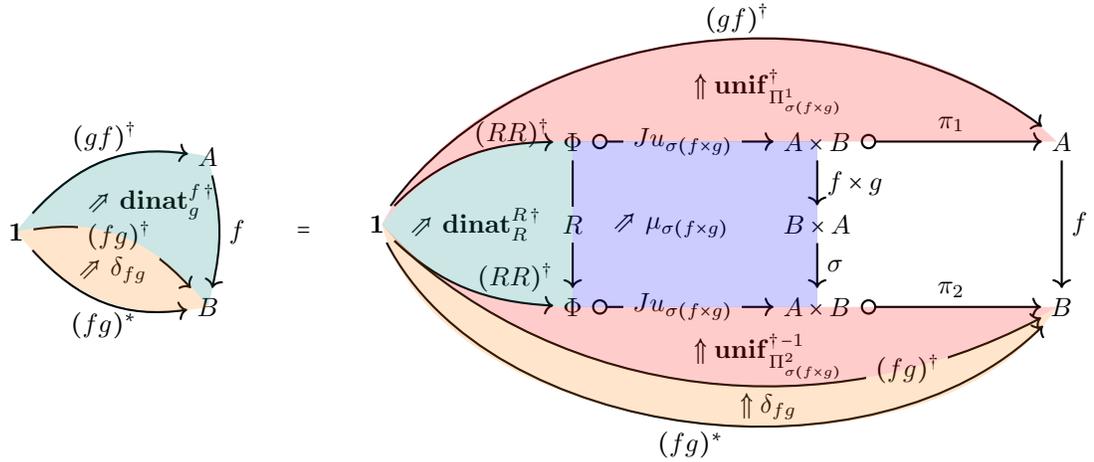
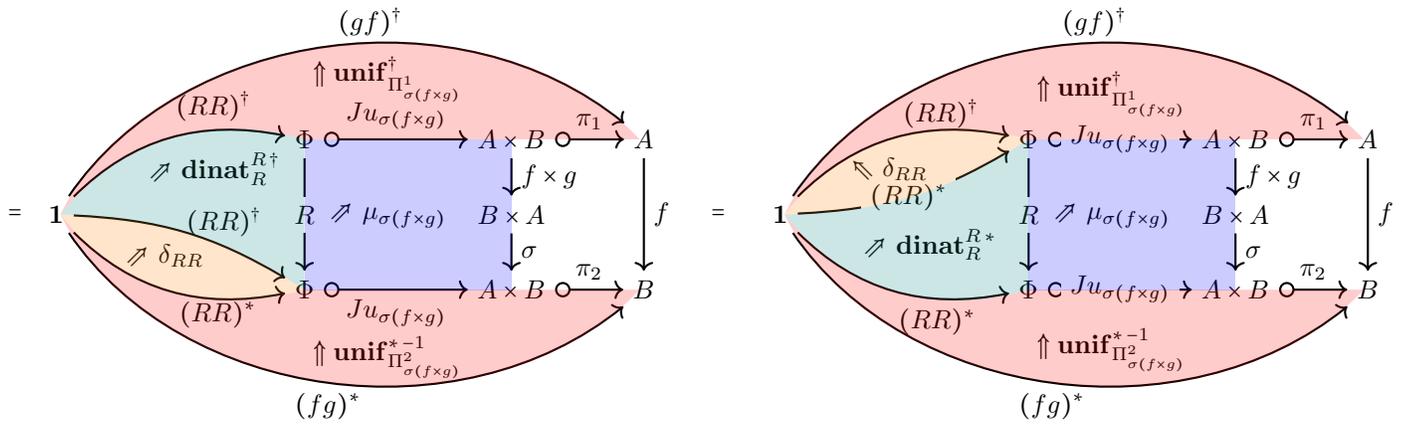
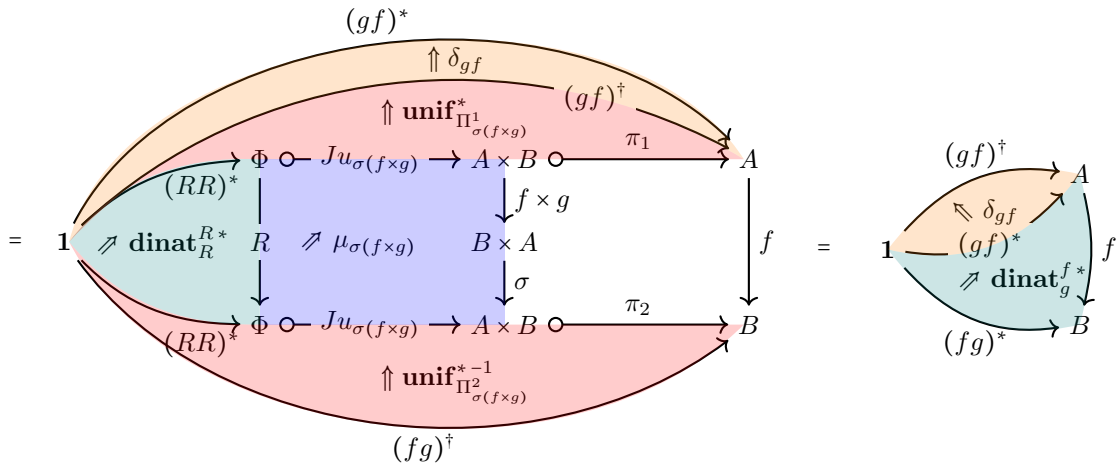
\end{document}

%% file: mydef.tex
\usepackage{mathrsfs}
\usepackage{cmll}
\usepackage{MnSymbol}
\usepackage{etex}
\usepackage{amsmath}
\usepackage{amsthm}
\usepackage{amsfonts}
\usepackage{mathtools} 
\usepackage{stmaryrd}

\usepackage{thmtools}
\usepackage{tikz,tikz-cd}
\usetikzlibrary{arrows,arrows.meta,positioning}

\usepackage{mathpartir}
\usepackage[utf8]{inputenc}
\usepackage[english]{babel}
\usepackage[draft]{optional}
\usepackage[colorinlistoftodos,prependcaption,textsize=tiny]{todonotes}
\usepackage{xcolor}
\usepackage{comment}
\usepackage[all]{xy}
\usepackage{braket}
\usepackage{thm-restate}
\usepackage{graphicx}
\graphicspath{  }

\newcommand{\lis}[1]{\langle #1 \rangle} 

\definecolor{bluem}{rgb}{0,0.3, 1.0}
\definecolor{brown}{rgb}{1, 0.5, 0}
\definecolor{purple}{rgb}{0.75, 0.0, 1.0}
\definecolor{green}{rgb}{0.2, 0.9, 0.5}
\definecolor{olive}{rgb}{0.5, 0.0, 0.5}
\definecolor{yellow}{rgb}{1, 0.9,0}
 
\newcommand{\id}{\mathrm{id}} 
\newcommand{\op}{\mathrm{op}}

\renewcommand{\circ}{}

\DeclareMathOperator{\One}{\mathbf{1}} 

\newcommand{\cat}[1]{\mathbb{#1}}
\newcommand{\bicat}[1]{\mathscr{#1}}

\newcommand{\Cat}{\mathbf{Cat}}
\newcommand{\CAT}{\mathbf{CAT}}
\renewcommand{\Set}{\mathbf{Set}}
\newcommand{\Rel}{\mathbf{Rel}}

\newcommand{\Cpo}{\mathbf{Cpo}}
\newcommand{\Cppo}{\mathbf{Cppo}}
\newcommand{\Lin}{\mathbf{Lin}}
\newcommand{\ScottL}{\mathbf{ScottL}}

\newcommand{\Prof}{\mathbf{Prof}}

\newcommand{\Cocont}{\mathbf{Cocont}}

\makeatletter
\def\slashedarrowfill@#1#2#3#4#5{%
	$\m@th\thickmuskip0mu\medmuskip\thickmuskip\thinmuskip\thickmuskip
	\relax#5#1\mkern-7mu%
	\cleaders\hbox{$#5\mkern-2mu#2\mkern-2mu$}\hfill
	\mathclap{#3}\mathclap{#2}%
	\cleaders\hbox{$#5\mkern-2mu#2\mkern-2mu$}\hfill
	\mkern-7mu#4$%
}
\def\rightslashedarrowfill@{%
	\slashedarrowfill@\relbar\relbar\mapstochar\rightarrow}
\newcommand\xslashedrightarrow[2][]{%
	\ext@arrow 0055{\rightslashedarrowfill@}{#1}{#2}}
\makeatother

\newcommand{\dinato}{
	\mathrel{\vbox{\offinterlineskip
			\ialign{\hfil##\hfil\cr
				\normalfont\scalebox{1.2}{.\;}\cr
				\noalign{\kern.05ex}
				$\Longrightarrow$\cr}
	}}%
}

\newcommand{\stdinato}{
	\mathrel{\vbox{\offinterlineskip
			\ialign{\hfil##\hfil\cr
				\normalfont\scalebox{1.2}{\,.\,.\;}\cr
				\noalign{\kern.05ex}
				$\Longrightarrow$\cr}
	}}%
}

\newcommand{\StDin}{\mathbf{StDin}} 

\DeclareMathOperator{\fix}{\mathbf{fix}} 
\DeclareMathOperator{\unif}{\mathbf{unif}} 
\DeclareMathOperator{\dinat}{\mathbf{dinat}} 
\DeclareMathOperator{\din}{\mathbf{din}} 
\newcommand{\Fixop}{\mathbf{Fix}}
\newcommand{\DinFixop}{\mathbf{DinFix}}

%% file: theoremstyle.tex
\theoremstyle{plain}
\newtheorem{theorem}{Theorem}[section]

\newtheorem{lemma}[theorem]{Lemma}
\newtheorem{proposition}[theorem]{Proposition}

\theoremstyle{definition}
\newtheorem{definition}[theorem]{Definition}

\theoremstyle{remark}
\newtheorem{remark}{Remark}

%% file: intro.tex
Fixpoints operators play an important role to model infinite computation in a wide range of computer science disciplines: design of programming languages, verification, model checking, databases, concurrency theory, type theory, etc. A fixpoint for a program $t$ is an input $x$ such that there is a calculation sequence between $x$ and $t (x)$. For example, a fundamental property of untyped $\lambda$-calculus is the existence of fixpoint combinators \emph{i.e.} terms $\mathbf{Y}$ such that for any $\lambda$-term $t$, there is a reduction path connecting the terms $\mathbf{Y} t $ and $t(\mathbf{Y} t )$. 

From a set-theoretic viewpoint, the standard notion of fixpoint for an endomap $f :A \to A$ is an element $x\in A$ such that $f (x) = x$ and it was axiomatized with the notion of categories with fixpoint operators \cite{simpson2000complete}. There is however no notion of categorical fixpoint operator taking into account the computational reduction steps and not collapsing them into strict equalities.  

The objective of this paper is to work in a $2$-dimensional framework to model explicitly the reductions of languages with fixpoints and study their coherences \emph{i.e.} the equations satisfied by the program computations steps. 
It fits into the line of research of \emph{categorifying} models of computation by replacing semantics where types are sets or preorders with richer categorical structures to establish stronger mathematical invariants.

Categories in dimension one have objects and morphisms that can be composed. We can consider an additional dimension with the notion of $2$-categories or bicategories which have objects, $1$-morphisms that can be composed and $2$-morphisms that can be composed in two different ways that verify compatibility conditions. These $2$-morphisms are thought of as morphisms between $1$-morphisms. When using bidimensional categorical structures to model computations, we can study program execution steps as primitive objects as they become explicit $2$-morphisms carrying information on program reductions. It has seen many applications in concurrency, game semantics, type theory, higher dimensional rewriting \cite{fiore2004axiomatic, cattani2005profunctors}, \cite{Fiorecartesian}, \cite{MazzaPellissierVial, TTCartClosedBicatFioreSaville2019,TemplateDiLL, CoherenceNormbyEvalFioreSaville2020, OlimpieriLics21, KerinecMO23}. 

When generalizing preorder semantics to richer categorical semantics, least fixpoints become initial algebras and greatest fixpoints become final coalgebras. In both cases, strict equalities for fixpoints $t(x)=x$ are now represented by explicit isomorphisms
\[
t(x) \xrightarrow{\text{algebra}} x \qquad\quad  x \xrightarrow{\text{coalgebra}} t(x)	\] capturing the dynamic aspect of fixpoint reductions.
Initial algebras correspond to recursive definition with induction as a logical reasoning principle. They are typically used to model finite data types (such as finite lists and trees) with a constructor operation.

Dually, final coalgebras are the counterpart of corecursion and coinduction where structures are described with a destructor or observer operation. They are now widely used in computer science to model state-based systems with circular or non-terminating behavior (automata, transitions systems, network dynamics etc.). The categorical coalgebraic framework is also a standard tool to study notions such as bisimilarity and trace equivalence. It has been used in functional programming to model lazy datatypes in languages such as Haskell or reactive programs in proof assistants such as Coq and Agda.

Settings where initial algebra and final coalgebras coincide because of limit-colimit coincidence arguments have also been studied to solve equations with mixed variance variables (such as reflexive objects $D \cong D\Rightarrow D$ for $\lambda$-calculus models \cite{scott1971toward}). These algebras are called \emph{bifree} and they provide a framework where inductive and coinductive arguments are equivalent. 
An important result by Plotkin and Simpson in this area states that provided some conditions on bifree algebras are satisfied, we obtain the existence of a unique uniform fixpoint operator  for 1-categories~\cite{simpson2000complete}. 

While using initial algebra or final coalgebra semantics to model infinite computations are now well-established traditions in computer science, there is no counterpart axiomatization of fixpoint operators for $2$-categorical structures and the goal of this paper is develop their theory. In order to axiomatize the notion of $2$-dimensional fixpoint operator, the main difficulty is to understand what axioms and coherences the rewriting $2$-morphisms should satisfy.  We both need to ensure that the equations are correct, \emph{i.e.} that they hold in concrete models, and we also need completeness properties, \emph{i.e.} we want to ensure that we have stated them all.

We proceed in two steps: we first categorify the restricted case of the Plotkin-Simpson theorem characterizing the existence of a unique uniform fixpoint operator for $1$-categories to extract an axiomatization of pseudo-fixpoint operator for $2$-categories. These operators form a category and uniqueness of the fixpoint operator is replaced by a contractibility property: there is a unique isomorphism between any two fixpoint operators. The second step is to generalize to models where fixpoints are not unique to verify that our axiomatization holds. Our motivation for choosing this approach is that the Plotkin-Simpson proof constructs explicitly a canonical fixpoint operator using the bifree algebras and then uses the fixpoint axioms to show that it is in fact unique. In the $2$-dimensional case, the pseudo-bifree algebras allow us to construct a canonical pseudo-fixpoint operator and in order to obtain a contractibility property, we need to impose certain coherence equations on the structural reduction $2$-morphisms which provides a guideline for the axiomatization of general pseudo-fixpoint operators.

\subsection*{Related works}

Categorification of recursive domain theory has an established history with the work of Lambek, Freyd, Lehmann, Adamek, Taylor, Fiore, Winskel and Cattani~\cite{Lambek1968, AlgebraicSpecDataTypesLehmann1981, AdamekCategoricalScott1997, TaylorLimColimCategories1988,Fiorethesis, RecDomainEqCattaniFioreWinskel98, BicategorySolRecursiveDomEqCattani2007}.
In this paper, we mostly use results by Cattani, Fiore and Winskel generalizing the notion of algebraically compact categories for enriched categories to enriched bicategories and proving limit-colimit coincidence theorems in this setting with applications to presheaf models of concurrency \cite{RecDomainEqCattaniFioreWinskel98, BicategorySolRecursiveDomEqCattani2007}.
From a syntactic viewpoint, Pitts presented a candidate $2$-dimensional type theory for fixpoints which can serve to prove coherence theorems for our notion of pseudo-fixpoint operators \cite{ElemCalculusApprox1987Pitts}.

Ponto and Shulman have also studied a categorification of the notion of fixpoint and trace for bicategories in a different direction \cite{ponto2007fixed, ponto2013shadows}. They consider the trace of endo-$2$-cells $\alpha : f \Rightarrow f$ for a $1$-cell $f :A \to B$; whereas in our case we still want to compute the trace or fixpoint of endo-$1$-cells but up to explicit rewriting $2$-cells. We aim to investigate whether the two approaches can be compared for the cartesian closed setting in future work.

\subsection*{Plan of the paper}
\begin{itemize}
	\item We start in Section \ref{sec:1fix} by recalling the standard theory of $1$-categorical fixpoint operators and give examples from domain theory and linear logic that can be recovered by the Plotkin-Simpson theorem.
	\item In Section \ref{sec:2fix}, we state the definitions of pseudo-fixpoint operators for $2$-categories with uniformity and dinaturality axioms.
	\item We prove in Section \ref{sec:construction} a generalization of the Plotkin-Simpson theorem for $2$-categories where pseudo-fixpoint operators now form a category and uniqueness of the fixpoint operator is replaced by a contractibility property.
	\item We show in Section \ref{sec:examples} how the notion of $2$-categorical fixpoint we developed in Section \ref{sec:2fix} is verified in well-known $2$-categorical models. 
\end{itemize}

%% file: 1fixtheory.tex
\begin{definition}
	Let $\cat{D}$ be a category with a terminal object~$\One$,~a \emph{fixpoint operator} on $\cat{D}$ is a family of functions $(-)^*: \cat{D}(A, A) \to \cat{D}(\One, A)$
	indexed by the objects $A$ of $\cat{D}$ verifying that for all morphisms $f :A \to A$, 
	\begin{equation}
		\tag{$\fix$}
		f^*= f \circ f^*.
	\end{equation}

\end{definition}
We can further require additional axioms such as uniformity and dinaturality which are usually satisfied in concrete models and can serve to characterize uniqueness properties of fixpoint operators~\cite{CharacterisationFixDinatSimpson1993}. We can also consider parametrized axioms giving the connection with traced monoidal categories~\cite{joyal1996traced} but we leave the parametrized case for an accompanying paper.

For a category with fixpoints, the uniformity principle (also called Plotkin's axiom) is relative to a subcategory of ``strict maps" and is used to characterize the fixpoint operator uniquely without relying on order-theoretic arguments~\cite{EilenbergUniformity, plotkindomains}.  For domain-like structures, strict maps are usually the ones which preserve bottom elements $\bot$ whereas general maps are just assumed to be Scott-continuous. Another possibility is to consider linear maps instead of strict maps, \emph{i.e.} maps which commute with all suprema not just directed ones and these are the typical examples in linear logic models with fixpoints. Freyd has also considered the case of a reflective subcategory for the subcategory of strict maps~\cite{freyd1991algebraically}.

\begin{definition}\label{def:1dunifdinat}
	Let $\cat{C}, \cat{D}$ be categories with terminal objects and $J : \cat{C} \to \cat{D}$ be an identity-on-objects functor preserving terminal objects.
	\begin{itemize}
		\item 
		A fixpoint operator $(-)^*$ on $\cat{D}$ is said to be \emph{uniform with respect to $J$} if for every $s: A \to B$ in $\cat{C}$ and $f :A \to A, g : B \to B$ in $\cat{D}$, we have:
		\begin{equation}
			\tag{$\unif$}
			J(s) \circ f = g \circ J(s) \quad \text{implies} \quad J(s) \circ f^* = g^*.
		\end{equation}
		\item A fixpoint operator $(-)^*$ on $\cat{D}$ is \emph{dinatural} if for every $f : A \to B$ and $g : B \to A$ in $\cat{D}$,
		\begin{equation}
			\tag{$\dinat$}
			(f \circ g)^*= f \circ (g \circ f)^*.
		\end{equation}
	\end{itemize}
\end{definition}

\begin{remark}
	We can in fact consider bijective-on-objects functors for $J : \cat{C} \to \cat{D}$ but we restrict to identity-on-objects functors to make the notation less cumbersome.
\end{remark}

 As mentioned in the introduction, our approach is to use a categorification of the Plotkin-Simpson theorem which characterizes existence and uniqueness of fixpoint operators for $1$-categories that are obtained as Kleisli categories of comonads satisfying some conditions on bifree algebras \cite{simpson2000complete}.  

Recall that for an endofunctor $T : \cat{D} \to \cat{D}$, a \emph{bifree $T$-algebra} (also called dialgebra, compact algebra or free bialgebra) is an initial $T$-algebra $(A, a: TA \to A)$ such that the inverse of $a$ is a final $T$-coalgebra $(A, a^{-1}: A \to TA)$.
\begin{theorem}[Plotkin-Simpson \cite{simpson2000complete}]\label{thm:PlotkinSimpson}
	Let $\cat{C}$ be a category equipped with a comonad $(T, \delta, \varepsilon)$ and a terminal object. We denote by $\cat{D}$ the co-Kleisli category $\cat{C}_{T}$ and by $J : \cat{C} \to \cat{D}$ the free functor induced by the comonadic adjunction. 
	\begin{enumerate}
		\item If the endofunctor $T$ has a bifree algebra, then $\cat{D}$ has a unique uniform (with respect to $J$) fixpoint operator.
		\item If $\cat{C}$ is cartesian and the endofunctor $T \circ T$ has a bifree algebra, then $\cat{D}$ has a unique uniform (with respect to $J$) dinatural fixpoint operator.
	\end{enumerate}
\end{theorem}
We proceed to give the general recipe to obtain bifree algebras for endofunctors in a suitable preorder-enriched setting in order to motivate the generalizations to dimension $2$ in Section~\ref{sec:examples}.

We denote by $\Cpo$ the category whose objects are $\omega$-complete partial orders and morphisms are Scott-continuous functions (monotone maps preserving colimits of $\omega$-chains). If we restrict the objects to \emph{pointed} cpo's (with a bottom element $\bot$), we denote this subcategory $\Cppo$. We can further restrict the morphisms to Scott-continuous functions that are \emph{strict} (preserving bottom elements) and we denote by $\Cppo_{\bot}$ the category of pointed cpo's and strict Scott-continuous functions. We are interested in $\Cppo_{\bot}$-enriched categories since they provide a well-behaved setting to compute bifree algebras as we will see below. Explicitly, a category $\cat{C}$ is $\Cppo_{\bot}$-enriched if each homset $\cat{C}(A,B)$ is a cpo with a bottom element $\bot$ and composition $\cat{C}(A,B) \times \cat{C}(B,C) \to \cat{C}(B,C)$
is Scott-continuous and strict in both components.
Examples of $\Cppo_{\bot}$-enriched categories include the category $\Cppo_{\bot}$ which is enriched over itself and the category $\Rel$ whose objects are sets and morphisms are binary relations. Another example is the category $\Lin$ of preorders and ideal relations (binary relations $R \subseteq A \times B$ between preorders that are up-closed in $A$ and down-closed in $B$). 
The category $\Rel$ can be viewed as the subcategory of $\Lin$ containing discrete preorders.

The following theorem is a consequence of the limit-colimit coincidence theorem by Smyth and Plotkin \cite{SmythPlotkin1982} which was generalized by Fiore \cite{Fiorethesis}:
\begin{theorem}\label{thm:bifree1d}
	Let $\cat{C}$ be a $\Cppo_{\bot}$-enriched category. If $\cat{C}$ has an initial object and $\omega$-colimits of chains of embeddings, then $\cat{C}$ is $\Cpo$-algebraically compact. It means that for every endofunctor $T : \cat{C} \to \cat{C}$, if $T$ is $\Cpo$-enriched, \emph{i.e.} for all $A,B$, the induced map
	\[
	\cat{C}(A,B) \longrightarrow 	\cat{C}(TA,TB) 
	\]
	preserves colimits of $\omega$-chains, then $T$ has a bifree algebra.
\end{theorem}

We proceed to give examples of well-known fixpoint operators that can be recovered from the Plotkin-Simpson theorem.
Consider the lifting comonad $(-)_{\bot}$ on $\Cppo_{\bot}$: this comonad freely adjoins a bottom element and its co-Kleisli category is isomorphic to $\Cppo$. Since the endofunctors $(-)_{\bot}$ and $(-)_{\bot} \circ (-)_{\bot}$ are $\Cppo$-enriched, the category $\Cppo$ has a unique dinatural fixpoint operator uniform with respect to the free functor $\Cppo_{\bot} \to \Cppo$ and it is given by the standard formula \[
f^* =  \bigvee\limits_{n\in \omega} f^n(\bot)
\] for $f : A \to A$ in $\Cppo$.
We can also consider the finite multiset comonad $\mathcal{M}_{\mathrm{fin}}$ on the category $\Rel$. It leads to a \emph{quantitative} model of linear logic where multisets allow to count multiplicities of the inputs for a term \cite{GIRARDLL}. As $\mathcal{M}_{\mathrm{fin}}$ and $\mathcal{M}_{\mathrm{fin}} \mathcal{M}_{\mathrm{fin}}$ have bifree algebras, we recover the fixpoint operator in the relational model \cite{grellois2015infinitary} and obtain that it is unique.

Lastly, the category $\Lin$ (also called the linear Scott category $\ScottL$) can be equipped with the $\vee$-semi-lattice comonad yielding a \emph{qualitative} model of linear logic where substitution allows for duplication and erasure. This comonad also verifies the necessary enrichment conditions to obtain a unique uniform dinatural fixpoint operator.

%% file: 2fixdef.tex
We state in this section the definition of fixpoint operators for $2$-categories. We will show in the remaining sections how the axioms we present arise from the generalization of the Plotkin-Simpson construction in dimension $2$ and how they are verified in concrete examples. 
When moving to a higher dimension, we have several possibilities depending on the degrees of strictness and direction of the $2$-morphisms (strict, pseudo, (op)lax).

In this paper, for space considerations, we will focus on one case: \emph{pseudo-fixpoint operators for $2$-categories} and leave the remaining cases and the bicategorical weakening for the long version. Even if we only consider the pseudo case where the $2$-cells are invertible, we will still provide a direction for the arrows to give a better understanding of where they come from and provide a guideline for the directed lax and oplax cases.

\begin{definition} \label{def:pseudoFixOp}
	Let $\bicat{D}$ be a $2$-category with a  terminal object~$\One$. A \emph{pseudo-fixpoint operator on $\bicat{D}$} consists of a family of functors indexed by the objects $A$ of $\bicat{D}$:
	\[
	(-)^*_A : \bicat{D}(A,A) \to \bicat{D}(\One, A)
	\]
	together with a family of natural isomorphisms $\fix_A$ (we will omit the subscripts for objects to simplify the notation) with components:
	\vspace{-0.3cm}
	\begin{center}
		\begin{tikzpicture}[thick, scale=0.7]
			\node (B) at (3,-1) {$A$};
			\node (C) at (0,0) {$\One$};
			\node (D) at (3,1) {$A$};
			
			\draw [->] (C)  to [bend left =-30] node [below] {$f^*$} (B);
			\draw [->] (C) to [bend left =30] node [above] {$f^*$} (D);
			\draw [->] (D)  to node [right] {$f$} (B);
			\node at (1.6,0)  {$\Nearrow\fix_f$};
			\fill[fill=bluem,fill opacity=0.2] (0,0) to [bend left =36] (3,1) to  (3,-1) to [bend left =36] (0,0);
		\end{tikzpicture}
	\end{center}
	\vspace{-0.3cm}
for a $1$-cell $f: A \to A$ in $\bicat{D}$.
			Naturality means that for an invertible $2$-cell $\alpha : f \Rightarrow g$ in $\bicat{D}$, we have:
		\vspace{-0.3cm}
\begin{center}
	\begin{tikzpicture}[thick, xscale=0.8,yscale=1]
		\node (B) at (3,0) {$A$};
		\node (C) at (0,1) {$\One$};
		\node (D) at (3,2) {$A$};
		
		\draw [->] (C)  to [bend left =-25] node [below] {$f^*$} (B);
		\draw [->] (C)  to [bend left =25] node [fill=white] {$g^*$} (B);
		\draw [->] (C) to [bend left =25] node [above] {$g^*$} (D);
		\draw [->] (D)  to [bend left =30] node [right] {$g$} (B);
		\node at (1.5,0.5)  {$\Nearrow\alpha^*$};
		\node at (2.2,1.4) {$\Nearrow\fix_g$};
		\fill[fill=bluem,fill opacity=0.2] (0.05,1) to [bend left =30] (3,2) to [bend left =38] (3,0) to [bend left =-29] (0.05,1);
		\fill[fill=red,fill opacity=0.2] (0.05,1) to [bend left =29] (3,0) to [bend left =30] (0.05,1);
		\node at (4.5,1) {$=$};
		
		\begin{scope}[xshift = 5.5 cm]
			\node (B) at (3,0) {$A$};
			\node (C) at (0,1) {$\One$};
			\node (D) at (3,2) {$A$};
			
			\draw [->] (C)  to [bend left =-25] node [below] {$f^*$} (B);
			\draw [->] (C)  to [bend left =-20] node [fill=white] {$f^*$} (D);
			\draw [->] (C) to [bend left =25] node [above] {$g^*$} (D);
			\draw [->] (D)  to [bend left =-30] node [left] {$f$} (B);
			\draw [->] (D)  to [bend left =30] node [right] {$g$} (B);
			\node at (1.6,0.5) {$\Nearrow\fix_f$};
			\node at (3,0.9) {$\Rightarrow $};
			\node at (3,1.1) {$\alpha $};
			\node at (1.6,1.6) {$\Nwarrow\alpha^*$};
			\fill[fill=bluem,fill opacity=0.1] (0.05,1) to [bend left =-23] (3,2) to [bend left =-38] (3,0) to [bend left =30] (0.05,1);
			\fill[fill=red,fill opacity=0.1] (3,2) to [bend left =38] (3,0) to [bend left =38] (3,2);
			\fill[fill=red,fill opacity=0.2] (0.05,1) to [bend left =30] (3,2) to [bend left =23] (0.05,1);

		\end{scope}
	\end{tikzpicture}
\end{center}
	\vspace{-0.3cm}
	\end{definition}

\begin{remark}
	Strictly speaking, our pseudo-fixpoint operators act on the sub-$2$-category of $\bicat{D}$ where we only consider invertible $2$-cells. Indeed, for a general $2$-cell $\alpha : f \Rightarrow g$, $\alpha^*$ is not defined unless $\alpha$ is invertible and we need to consider (op)lax-fixpoint operators to act on non-invertible $2$-cells. In order to lighten the notation, for a $2$-category $\bicat{D}$, we implicitly use the same notation for $\bicat{D}$ and its sub-$2$-category containing only invertible $2$-cells for the rest of the paper.
\end{remark}

In the preordered case, we can compare fixpoint operators pointwise and we are usually interested in the least and greatest fixpoints. In the categorified setting, fixpoint operators form a category so there can be more than one way of comparing two fixpoint operators. Morphisms of pseudo-fixpoint operators are transformations that commute with the structural $2$-cells $\fix$. Initial objects in this category correspond to least fixpoints while terminal objects correspond to greatest fixpoints.
The uniqueness property for fixpoint operators in the preorder setting now becomes a contractibility property for the category of fixpoint operators. A category is \emph{contractible} when it is not empty and for any two objects, there is a unique isomorphism between them (in particular, it is a groupoid).

\begin{definition}\label{def:morFixOp}
	Let $((-)^*, \fix^*)$ and $((-)^\dag, \fix^\dag)$ be two pseudo-fixpoint operators on a $2$-category $\bicat{D}$. A pseudo-morphism of pseudo-fixpoint operators $((-)^*, \fix^*)\to ((-)^\dag, \fix^\dag)$ consists of a family of natural isomorphisms
	\[
	\delta_A : (-)^*_A \Rightarrow (-)^\dag_A : \bicat{D}(A,A) \to \bicat{D}(\One, A)
	\]
	indexed by the objects $A $ of $\bicat{D}$ that commutes with the structural $2$-cells $\fix$, \emph{i.e.} it satisfies the following coherence for every $f: A \to A$:

		\begin{center}
		\begin{tikzpicture}[thick,yscale=0.9, xscale=0.8]

			\node (B) at (3,0) {$A$};
			\node (C) at (0,1) {$\One$};
			\node (D) at (3,2) {$A$};
			
			\draw [->] (C)  to [bend left =-25] node [below] {$f^*$} (B);
			\draw [->] (C) to [bend left =25] node [above] {$f^\dag$} (D);
			\draw [->] (C) to [bend left =25] node [fill=white] {$f^\dag$} (B);
			\draw [->] (D) to  [bend left =15] node [right] {$f$} (B);
			\node at (2.2,1.4) {$\Nearrow\fix^\dag_f$};
			\node at (1.5,0.5) {$\Nearrow\delta_f$};
			\node at (4.75,1) {$=$};
			\fill[fill=bluem,fill opacity=0.2] (0,1) to [bend left =30] (3,2) to [bend left =18]  (3,0) to [bend left =-30] (0,1);
			\fill[fill=brown,fill opacity=0.2] (0,1) to [bend left =30] (3,0) to [bend left =28] (0,1);
			\begin{scope}[xshift = 6 cm]
				\node (B) at (3,0) {$A$};
				\node (C) at (0,1) {$\One$};
				\node (D) at (3,2) {$A$};
				
				\draw [->] (C)  to [bend left =-25] node [below] {$f^*$} (B);
				\draw [->] (C) to [bend left =25] node [above] {$f^\dag$} (D);
				\draw [->] (C) to [bend left =-25] node [fill=white] {$f^*$} (D);
				\draw [->] (D) to [bend left =15] node [right] {$f$} (B);
				\node at (2.2,0.6) {$\Nearrow\fix_f^*$};
				\node at (1.5,1.5) {$\Nwarrow\delta_f$};
				\fill[fill=bluem,fill opacity=0.2] (0,1) to [bend left =-30] (3,2) to [bend left =18]  (3,0) to [bend left =30] (0,1);
				\fill[fill=brown,fill opacity=0.2] (0,1) to [bend left =30] (3,2) to [bend left =30] (0,1);
			\end{scope}
		\end{tikzpicture}
	\end{center}
	\vspace{-0.3cm}
	We denote by $\Fixop(\bicat{D})$ the category of pseudo-fixpoint operators on $\bicat{D}$.
\end{definition}

Dinatural transformations are used to model mixed variance operators using their argument both covariantly and contravariantly. The typical examples arising from semantics occur with (cartesian or monoidal) closed structure where the evaluation map is dinatural and with fixpoint operators.  In our setting, we use the notion of pseudo-dinatural transformation for $2$-functors.

\begin{definition}
	A pseudo-dinatural fixpoint operator on $\bicat{D}$ consists a pseudo-fixpoint operator $((-)^*, \fix)$ on $\bicat{D}$ together with a family of invertible $2$-cells
	
	\begin{center}
		\begin{tikzpicture}[thick, scale=0.8]
			\node (B) at (3,0) {$B$};
			\node (C) at (0,1) {$\One$};
			\node (D) at (3,2) {$A$};
			
			\draw [->] (C)  to [bend left =-30] node [below] {$(fg)^*$} (B);
			\draw [->] (C) to [bend left =30] node [above] {$(gf)^*$} (D);
			\draw [->] (D)  to node [right] {$f$} (B);
			\node at (1.6,1) {$\Nearrow\dinat_{g}^f$};
			
			\fill[fill=teal,fill opacity=0.2] (0,1) to [bend left =36] (3,2) to  (3,0) to [bend left =36] (0,1);
			
		\end{tikzpicture}
	\end{center}
	for $1$-cells $f: A \to B$ and $g: B\to A$ in $\bicat{D}$ satisfying the following axioms:
	\begin{enumerate}
		\item \textbf{pseudo-dinaturality axioms:}
		\begin{enumerate}
			\item Unity axiom: for a $1$-cell $f:A \to A$, we have
			\begin{center}
				\begin{tikzpicture}[thick, scale=0.7]
					\node (B) at (3,0) {$A$};
					\node (C) at (0,1) {$\One$};
					\node (D) at (3,2) {$A$};
					
					\draw [->] (C)  to [bend left =-30] node [below] {$f^*$} (B);
					\draw [->] (C) to [bend left =30] node [above] {$f^*$} (D);
					\draw [->] (D)  to node [right] {$1_A$} (B);
					\node at (1.6,1) {$\Nearrow \dinat^{1_A}_f$};
					\node at (5.5,1) {$= \qquad \id_{f^*}$};
					\fill[fill=teal,fill opacity=0.2] (0,1) to [bend left =36] (3,2) to  (3,0) to [bend left =36] (0,1);
				\end{tikzpicture}
			\end{center}
			\item $1$-naturality axiom: for all $1$-cells $f : A \to B$, $g:  B\to C$ and $h: C\to A$ in $\bicat{D}$:
			\begin{center}
				\begin{tikzpicture}[thick, yscale=0.8, xscale=0.8]
					\node (A) at (3,4) {$A$};
					\node (B) at (3,0) {$C$};
					\node (C) at (0,2) {$\One$};
					\node (D) at (3,2) {$B$};
					
					\draw [->] (C)  to [bend left =-30] node [below left] {$(gfh)^*$} (B);
					\draw [->] (C) to node [fill=white] {$(fhg)^*$} (D);
					\draw [->] (C) to [bend left =30] node [above left] {$(hgf)^*$} (A);
					\draw [->] (D)  to node [right] {$g$} (B);
					\draw [->] (A)  to node [right] {$f$} (D);
					\node at (1.8,1.1) {$\Nearrow\dinat^g_{fh}$};
					\node at (1.8,2.9) {$\Nearrow\dinat^f_{hg}$};
					\node at (4,2) {$=$};
					\fill[fill=teal,fill opacity=0.1] (0,2) to [bend left =36] (3,4) to  (3,2) to  (0,2);
					\fill[fill=teal,fill opacity=0.3] (0,2) to (3,2) to  (3,0) to [bend left =36] (0,2);
					\begin{scope}[xshift = 5 cm]
						\node (A) at (3,4) {$A$};
						\node (B) at (3,0) {$C$};
						\node (C) at (0,2) {$\One$};
						\node (D) at (3,2) {$B$};
						
						\draw [->] (C)  to [bend left =-30] node [below left] {$(gfh)^*$} (B);
						\draw [->] (C) to [bend left =30] node [above left] {$(hgf)^*$} (A);
						\draw [->] (D)  to node [right] {$g$} (B);
						\draw [->] (A)  to node [right] {$f$} (D);
						\node at (1.6,2) {$\Nearrow\dinat^{gf}_h$};
						\fill[fill=teal,fill opacity=0.2] (0,2) to [bend left =36] (3,4) to  (3,0) to [bend left =36] (0,2);
					\end{scope}
				\end{tikzpicture}
			\end{center}
			\item $2$-naturality axiom: for an invertible $2$-cell $\alpha : f \Rightarrow f': A \to B$ in $\bicat{D}$ and a $1$-cell $g : B \to A$ in $\bicat{D}$, we have
			\begin{center}
				\begin{tikzpicture}[thick, xscale=0.9,yscale=1.15]
					\node (B) at (3,0) {$B$};
					\node (C) at (0,1) {$\One$};
					\node (D) at (3,2) {$A$};
					
					\draw [->] (C)  to [bend left =-25] node [below] {$(fg)^*$} (B);
					\draw [->] (C)  to [bend left =25] node [fill=white] {$(f'g)^*$} (B);
					\draw [->] (C) to [bend left =25] node [above] {$(gf')^*$} (D);
					\draw [->] (D)  to [bend left =30] node [fill=white] {$f'$} (B);
					\node at (1.5,0.5)  {$\Nearrow(\alpha g)^*$};
					\node at (2,1.4) {$\Nearrow\dinat^{f'}_g$};
					\fill[fill=teal,fill opacity=0.2] (0.05,1) to [bend left =30] (3,2) to [bend left =38] (3,0) to [bend left =-29] (0.05,1);
					\fill[fill=red,fill opacity=0.2] (0.05,1) to [bend left =29] (3,0) to [bend left =30] (0.05,1); 
					\node at (3.75,1) {$=$};
					
					\begin{scope}[xshift = 4.2cm]
						\node (B) at (3,0) {$B$};
						\node (C) at (0,1) {$\One$};
						\node (D) at (3,2) {$A$};
						
						\draw [->] (C)  to [bend left =-25] node [below] {$(fg)^*$} (B);
						\draw [->] (C)  to [bend left =-20] node [fill=white] {$(gf)^*$} (D);
						\draw [->] (C) to [bend left =25] node [above] {$(gf')^*$} (D);
						\draw [->] (D)  to [bend left =-30] node [fill=white] {$f$} (B);
						\draw [->] (D)  to [bend left =30] node [fill=white] {$f'$} (B);
						\node at (1.6,0.6) {$\Nearrow\dinat^f_g$};
						\node at (3,0.9) {$\Rightarrow $};
						\node at (3,1.1) {$\alpha $};
						\node at (1.6,1.6) {$\Nwarrow(g\alpha)^*$};
						\fill[fill=teal,fill opacity=0.1] (0.05,1) to [bend left =-23] (3,2) to [bend left =-38] (3,0) to [bend left =30] (0.05,1);
						\fill[fill=red,fill opacity=0.1] (3,2) to [bend left =38] (3,0) to [bend left =38] (3,2);
						\fill[fill=red,fill opacity=0.2] (0.05,1) to [bend left =30] (3,2) to [bend left =23] (0.05,1);
						\node at (15, 0) {};
					\end{scope}

				\end{tikzpicture}
			\end{center}
		\end{enumerate}
		these three axioms induce a pseudo-dinatural transformation:
		\[
		\dinat : \bicat{D}(-,-) \dinato \bicat{D}(\One, -) : \bicat{D}^{\op} \times \bicat{D} \to \CAT
		\]
		\item \textbf{coherence between $\dinat$ and $\fix$}: for all $1$-cells $f : A \to B$ and $g: B\to A$ in $\bicat{D}$, we have:
		\begin{center}
			\begin{tikzpicture}[thick, yscale=0.8, xscale=0.8]
				\node (A) at (3,4) {$B$};
				\node (B) at (3,0) {$B$};
				\node (C) at (0,2) {$\One$};
				\node (D) at (3,2) {$A$};
				
				\draw [->] (C)  to [bend left =-30] node [below left] {$(f\circ g)^*$} (B);
				\draw [->] (C) to node [fill=white] {$(g\circ f)^*$} (D);
				\draw [->] (C) to [bend left =30] node [above left] {$(f\circ g)^*$} (A);
				\draw [->] (D)  to node [right] {$f$} (B);
				\draw [->] (A)  to node [right] {$g$} (D);
				\node at (1.8,1.1) {$\Nearrow\dinat^f_g$};
				\node at (1.8,2.9) {$\Nearrow\dinat^g_f$};
				\node at (4.25,2) {$=$};
				\fill[fill=teal,fill opacity=0.1] (0,2) to [bend left =36] (3,4) to  (3,2) to  (0,2);
				\fill[fill=teal,fill opacity=0.3] (0,2) to (3,2) to  (3,0) to [bend left =36] (0,2);
				\begin{scope}[xshift = 5.5 cm]
					\node (A) at (3,4) {$B$};
					\node (B) at (3,0) {$B$};
					\node (C) at (0,2) {$\One$};
					\node (D) at (3,2) {$A$};
					
					\draw [->] (C)  to [bend left =-30] node [below left] {$(f\circ g)^*$} (B);
					\draw [->] (C) to [bend left =30] node [above left] {$(f\circ g)^*$} (A);
					\draw [->] (D)  to node [right] {$f$} (B);
					\draw [->] (A)  to node [right] {$g$} (D);
					\node at (1.8,2) {$\Nearrow\fix_{f g}$};
					\fill[fill=bluem,fill opacity=0.2] (0,2) to [bend left =36] (3,4) to  (3,0) to [bend left =36] (0,2);
				\end{scope}
			\end{tikzpicture}
		\end{center}
	\end{enumerate}

\end{definition}

\begin{remark}
	Note that the coherence axiom between $\dinat$ and $\fix$ above implies that 
	\begin{center}
		\begin{tikzpicture}[thick, scale=0.8]
			\node (B) at (3,0) {$A$};
			\node (C) at (0,1) {$\One$};
			\node (D) at (3,2) {$A$};
			
			\draw [->] (C)  to [bend left =-30] node [below] {$f^*$} (B);
			\draw [->] (C) to [bend left =30] node [above] {$f^*$} (D);
			\draw [->] (D)  to node [right] {$f$} (B);
			\node at (1.6,1) {$\Nearrow\dinat_{1_A}^f$};
			\node at (5,1) {$=$};
			\fill[fill=teal,fill opacity=0.2] (0,1) to [bend left =36] (3,2) to  (3,0) to [bend left =36] (0,1);
			\begin{scope}[xshift = 6.5 cm]
				\node (B) at (3,0) {$A$};
				\node (C) at (0,1) {$\One$};
				\node (D) at (3,2) {$A$};
				
				\draw [->] (C)  to [bend left =-30] node [below] {$f^*$} (B);
				\draw [->] (C) to [bend left =30] node [above] {$f^*$} (D);
				\draw [->] (D)  to node [right] {$f$} (B);
				\node at (1.6,1) {$\Nearrow\fix_f$};
				\fill[fill=bluem,fill opacity=0.2] (0,1) to [bend left =36] (3,2) to  (3,0) to [bend left =36] (0,1);
			\end{scope}
		\end{tikzpicture}
	\end{center}
	since $\dinat^{1_A}_f=\id_{f^*}$.                                                                                                                                                                                                                                          In the $1$-categorical setting, if the axiom for dinaturality $f \circ (g\circ f)^* = (f\circ g)^*$  holds, the fixpoint axiom $f \circ f^* = f^*$ becomes redundant (it suffices to take $g=\id_A$). In the $2$-dimensional case, the $2$-cells $\fix$ are entirely determined by the $2$-cells $\dinat$ and we therefore just write $((-)^*, \dinat)$ for a pseudo-dinatural fixpoint operator instead of  $((-)^*, \fix, \dinat)$.
\end{remark}

	\begin{definition}\label{def:morDinFixOp}
		Let $((-)^*, \dinat^*)$ and $((-)^\dag, \dinat^\dag)$ be two pseudo-dinatural fixpoint operators on a $2$-category $\bicat{D}$. A pseudo-morphism of pseudo-dinatural fixpoint operators $((-)^*, \dinat^*)\to ((-)^\dag, \dinat^\dag)$ consists of a family of natural isomorphisms

		\[
		\delta_A : (-)^*_A \Rightarrow (-)^\dag_A : \bicat{D}(A,A) \to \bicat{D}(\One, A)
		\]
		indexed by the objects $A $ of $\bicat{D}$ that commutes with the structural $2$-cells $\dinat$, \emph{i.e.} it satisfies the following coherence for every $f: A \to B$ and $g: B \to A$:
		
		\begin{center}
			\begin{tikzpicture}[thick,yscale=1.05, xscale=0.9]
				
				\node (B) at (3,0) {$B$};
				\node (C) at (0,1) {$\One$};
				\node (D) at (3,2) {$A$};
				
				\draw [->] (C)  to [bend left =-25] node [below] {$(fg)^*$} (B);
				\draw [->] (C) to [bend left =25] node [above] {$(gf)^\dag$} (D);
				\draw [->] (C) to [bend left =25] node [fill=white] {$(fg)^\dag$} (B);
				\draw [->] (D) to  [bend left =15] node [right] {$f$} (B);
				\node at (2.1,1.4) {$\Nearrow\dinat_g^{f\,\dag}$};
				\node at (1.5,0.5) {$\Nearrow\delta_{fg}$};
				\node at (4.25,1) {$=$};
				\fill[fill=teal,fill opacity=0.2] (0,1) to [bend left =30] (3,2) to [bend left =18]  (3,0) to [bend left =-30] (0,1);
				\fill[fill=brown,fill opacity=0.2] (0,1) to [bend left =30] (3,0) to [bend left =28] (0,1);
				\begin{scope}[xshift = 5 cm]
					\node (B) at (3,0) {$B$};
					\node (C) at (0,1) {$\One$};
					\node (D) at (3,2) {$A$};
					
					\draw [->] (C)  to [bend left =-25] node [below] {$(fg)^*$} (B);
					\draw [->] (C) to [bend left =25] node [above] {$(gf)^\dag$} (D);
					\draw [->] (C) to [bend left =-25] node [fill=white] {$(gf)^*$} (D);
					\draw [->] (D) to [bend left =15] node [right] {$f$} (B);
					\node at (2.1,0.6) {$\Nearrow \dinat_g^{f\,*}$};
					\node at (1.5,1.5) {$\Nwarrow\delta_{gf}$};
					\fill[fill=teal,fill opacity=0.2] (0,1) to [bend left =-30] (3,2) to [bend left =18]  (3,0) to [bend left =30] (0,1);
					\fill[fill=brown,fill opacity=0.2] (0,1) to [bend left =30] (3,2) to [bend left =30] (0,1);
				\end{scope}
			\end{tikzpicture}
		\end{center}
		\vspace{-0.3cm}
		We denote by $\DinFixop(\bicat{D})$ the category of pseudo-dinatural fixpoint operators on $\bicat{D}$.
	\end{definition}	
Dinatural transformations do not compose in general and therefore they do not form a category. In order to make the notion compositional, \emph{strong dinatural transformations} were introduced. It was noted by Mulry that the uniformity axiom for fixpoints (Definition \ref{def:1dunifdinat}) can be reformulated by requiring $(-)^*$ to be part of a strong dinatural transformation with respect to strict maps~\cite{mulry1992strong} and we use this characterization when moving to dimension~$2$. While generalizations of dinatural and extranatural transformations for $2$-categorical structures have been considered before \cite{bozapalides1976theorie,bozapalides1977fins, climent20102, corner2017universal, hirata2022notes}, to our knowledge, there is no existing notion of strong dinatural transformations in dimension $2$.

\begin{definition}\label{def:2fixunif}
	Let $J : \bicat{C} \to \bicat{D}$ be an identity-on-objects $2$-functor (strictly) preserving terminal objects. A pseudo-fixpoint operator on $\bicat{D}$ that is \emph{uniform with respect to $J$} consists of a pseudo-fixpoint operator $((-)^*, \fix)$ as in Definition~\ref{def:pseudoFixOp} together with a family of $2$-cells
		\vspace{-0.3cm}
		\begin{center}
		\begin{tikzpicture}[thick, scale=0.7]

				\node (B) at (3,0) {$B$};
				\node (C) at (0,1) {$\One$};
				\node (D) at (3,2) {$A$};
				
				\draw [->] (C)  to [bend left =-30] node [below] {$g^*$} (B);
				\draw [->] (C) to [bend left =30] node [above] {$f^*$} (D);
				\draw [o->] (D)  to node [right] {$Js$} (B);
				\node at (1.6,1) {$\Swarrow \unif_{\gamma}$};
				\fill[fill=red,fill opacity=0.2] (0,1) to [bend left =36] (3,2) to  (3,0) to [bend left =36] (0,1);
		\end{tikzpicture}
	\end{center}
	\vspace{-0.3cm}
	for every $1$-cells $s : A \to B$ in $\bicat{C}$, $f:A \to A$ and $g : B \to B$ in $\bicat{D}$ and invertible $2$-cell $\gamma$ as below:
		\vspace{-0.3cm}
		\begin{center}
		\begin{tikzpicture}[thick,scale =0.4]

				\node (A) at (0,0) {$B$};
				\node (B) at (3,0) {$B$};
				\node (C) at (0,3) {$A$};
				\node (D) at (3,3) {$A$};
				
				\draw [o->] (C) -- node [left] {$Js$} (A);
				\draw [->] (A) -- node [below] {$g$} (B);
				\draw [->] (C) -- node [above] {$f$} (D);
				\draw [o->] (D) -- node [right] {$Js$} (B);
				\node at (1.5,1.5) {$\Swarrow \gamma$};
				\fill[fill=yellow,fill opacity=0.2] (0,0) to (3,0) to  (3,3) to (0,3);
			
		\end{tikzpicture}
	\end{center}
	\vspace{-0.3cm}
	satisfying the following axioms:
	\begin{enumerate}
		\item \textbf{strong pseudo-dinaturality:}
			\begin{enumerate}
			\item Unity axiom: we have
			\begin{center}
				\begin{tikzpicture}[thick, scale=0.7]
					\node (B) at (3,0) {$A$};
					\node (C) at (0,1) {$\One$};
					\node (D) at (3,2) {$A$};
					
					\draw [->] (C)  to [bend left =-30] node [below] {$f^*$} (B);
					\draw [->] (C) to [bend left =30] node [above] {$f^*$} (D);
					\draw [o->] (D)  to node [right] {$J1_A$} (B);
					\node at (1.6,1) {$\Swarrow \unif_{\id_f}$};
					\node at (6,1) {$= \qquad \id_{f^*}$};
					\fill[fill=red,fill opacity=0.2] (0,1) to [bend left =36] (3,2) to  (3,0) to [bend left =36] (0,1);
				\end{tikzpicture}
			\end{center}
		
			\item $1$-naturality: for two squares in $\bicat{D}$
			\begin{center}
				\begin{tikzpicture}[thick,scale =0.4]
					\node (A) at (0,0) {$B$};
					\node (B) at (3,0) {$B$};
					\node (C) at (0,3) {$A$};
					\node (D) at (3,3) {$A$};
					
					\draw [o->] (C) -- node [left] {$Js$} (A);
					\draw [->] (A) -- node [below] {$g$} (B);
					\draw [->] (C) -- node [above] {$f$} (D);
					\draw [o->] (D) -- node [right] {$Js$} (B);
					\node at (1.6,1.5) {$\Swarrow \gamma$};
					\fill[fill=yellow,fill opacity=0.1] (0,0) to (3,0) to  (3,3) to (0,3);
					\begin{scope}[xshift=7cm]
						\node (A) at (0,0) {$C$};
						\node (B) at (3,0) {$C$};
						\node (C) at (0,3) {$B$};
						\node (D) at (3,3) {$B$};
						
						\draw [o->] (C) -- node [left] {$Jr$} (A);
						\draw [->] (A) -- node [below] {$h$} (B);
						\draw [->] (C) -- node [above] {$g$} (D);
						\draw [o->] (D) -- node [right] {$Jr$} (B);
						\node at (1.6,1.5) {$\Swarrow \rho$};
						\fill[fill=yellow,fill opacity=0.3] (0,0) to (3,0) to  (3,3) to (0,3);
					\end{scope}
				\end{tikzpicture}
			\end{center}
			we have:
			
			\begin{center}
				\begin{tikzpicture}[thick, scale=0.8]
					\node (A) at (3,4) {$A$};
					\node (B) at (3,0) {$C$};
					\node (C) at (0,2) {$\One$};
					\node (D) at (3,2) {$B$};
					
					\draw [->] (C)  to [bend left =-30] node [below left] {$h^*$} (B);
					\draw [->] (C) to [bend left =30] node [above left] {$f^*$} (A);
					\draw [o->] (D)  to node [right] {$Jr$} (B);
					\draw [o->] (A)  to node [right] {$Js$} (D);
					\node at (1.5,2) {$\Swarrow \unif_{\gamma*_v \rho}$};
					\node at (4,2) {$=$};
					\fill[fill=red,fill opacity=0.2] (0,2) to [bend left =36] (3,4) to  (3,0) to [bend left =36] (0,2);
					\begin{scope}[xshift = 5cm]
						
						\node (A) at (3,4) {$A$};
						\node (B) at (3,0) {$C$};
						\node (C) at (0,2) {$\One$};
						\node (D) at (3,2) {$B$};
						
						\draw [->] (C)  to [bend left =-30] node [below left] {$h^*$} (B);
						\draw [->] (C) to node [fill=white] {$g^*$} (D);
						\draw [->] (C) to [bend left =30] node [above left] {$f^*$} (A);
						\draw [o->] (D)  to node [right] {$Jr$} (B);
						\draw [o->] (A)  to node [right] {$Js$} (D);
						\node at (1.8,1.1) {$\Swarrow \unif_\rho$};
						\node at (1.8,2.9) {$\Swarrow \unif_\gamma$};
						\fill[fill=red,fill opacity=0.1] (0,2) to [bend left =36] (3,4) to  (3,2) to  (0,2);
						\fill[fill=red,fill opacity=0.3] (0,2) to (3,2) to  (3,0) to [bend left =36] (0,2);
					\end{scope}
				\end{tikzpicture}
			\end{center}
			where $\gamma*_v \rho$ denotes the $2$-cell corresponding to stacking the two squares vertically as follows:
			\begin{center}
				\begin{tikzpicture}[thick, xscale=0.5, yscale=0.5]
					\node (A) at (0,0) {$C$};
					\node (B) at (3,0) {$C$};
					\node (C) at (0,3) {$A$};
					\node (D) at (3,3) {$A$};
					\node at (5.5,1.5) {$:=$};
					
					\draw [o->] (C) to node [left] {$Jrs$} (A);
					\draw [->] (A) -- node [below] {$h$} (B);
					\draw [->] (C) -- node [above] {$f$} (D);
					\draw [o->] (D) to node [right] {$Jrs$} (B);
					\node at (1.5,1.5) {$\Swarrow \gamma*_v \rho$};
					\fill[fill=yellow,fill opacity=0.2] (0,0) to (3,0) to  (3,3) to (0,3);
					
					\begin{scope}[xshift = 6.5 cm, yshift=1.5cm]
						\node (A) at (1,0) {$B$};
						\node (B) at (4,0) {$B$};
						\node (A1) at (1,-3) {$C$};
						\node (B1) at (4,-3) {$C$};
						\node (C) at (1,3) {$A$};
						\node (D) at (4,3) {$A$};
						
						\draw [o->] (C)  to node [left] {$Js$} (A);
						\draw [o->] (A)  to node [left] {$Jr$} (A1);
						\draw [->] (A) -- node [fill=white] {$g$} (B);
						\draw [->] (A1) -- node [below] {$h$} (B1);
						\draw [->] (C) -- node [above] {$f$} (D);
						\draw [o->] (D)  to node [right] {$Js$} (B);
						\draw [o->] (B)  to node [right] {$Jr$} (B1);
						\fill[fill=yellow,fill opacity=0.1] (1,0) to (4,0) to  (4,3) to (1,3);
						\fill[fill=yellow,fill opacity=0.3] (1,-3) to (4,-3) to  (4,0) to (1,0);
						\node at (2.6,1.5) {$\Swarrow\gamma$};
						\node at (2.6,-1.5) {$\Swarrow\rho$};
						
					\end{scope}
				\end{tikzpicture}

			\end{center}
				\item $2$-naturality: for every invertible $2$-cell $\theta: s \Rightarrow r$ in $\bicat{C}$ such that 
			\vspace{-0.3cm}
			\begin{center}
				\begin{tikzpicture}[thick, scale=0.6]
					\node (A) at (1,0) {$B$};
					\node (B) at (4,0) {$B$};
					\node (C) at (1,3) {$A$};
					\node (D) at (4,3) {$A$};
					
					\draw [o->] (C)  to [bend left =-35] node [left] {$Jr$} (A);
					\draw [o->] (C)  to [bend left =35] node [right] {$Js$} (A);
					\draw [->] (A) -- node [below] {$g$} (B);
					\draw [->] (C) -- node [above] {$f$} (D);
					\draw [o->] (D)  to [bend left =35] node [fill=white] {$Js$} (B);
					\node at (1,1.7)  {$\Leftarrow$};
					\node at (1,1.2) {$J\theta$};
					\node at (3.2,1.5) {$\Swarrow \gamma$};
					\fill[fill=yellow,fill opacity=0.3] (1,0) to (4,0) to [bend left =-50] (4,3) to (1,3) to [bend left =50] (1,0);
					\fill[fill=green,fill opacity=0.2] (1,0) to [bend left =50] (1,3) to [bend left =50] (1,0);
					\node (X) at (5.75,1.5) {$=$};
					
					\begin{scope}[xshift = 7.5 cm]
						\node (A) at (0,0) {$B$};
						\node (B) at (3,0) {$B$};
						\node (C) at (0,3) {$A$};
						\node (D) at (3,3) {$A$};

						\draw [o->] (C) to[bend left =-35] node [fill=white] {$Jr$} (A);
						\draw [->] (A) -- node [below] {$g$} (B);
						\draw [->] (C) -- node [above] {$f$} (D);
						\draw [o->] (D) to [bend left =35] node [fill=white] {$Js$} (B);
						\draw [o->] (D) to [bend left =-35] node [left] {$Jr$} (B);
						\node at (3,1.7)  {$\Leftarrow$};
						\node at (3,1.2) {$J\theta$};
						\node at (0.6,1.5) {$\Swarrow \rho$};
						\fill[fill=yellow,fill opacity=0.1] (0,0) to (3,0) to [bend left =50] (3,3) to (0,3) to [bend left =-50] (0,0);
						\fill[fill=green,fill opacity=0.2] (3,0) to [bend left =50] (3,3) to [bend left =50] (3,0);
					\end{scope}
				\end{tikzpicture}
			\end{center}
			\vspace{-0.3cm}
			we have 
			\vspace{-0.3cm}
			\begin{center}
				\begin{tikzpicture}[thick,xscale=0.8, yscale=0.9]

					\node (B) at (3,0) {$B$};
					\node (C) at (0,1) {$\One$};
					\node (D) at (3,2) {$A$};
					\node at (4.25,1) {$=$};
					\draw [->] (C)  to [bend left =-30] node [below] {$g^*$} (B);
					\draw [->] (C) to [bend left =30] node [above] {$f^*$} (D);
					\draw [o->] (D) to [bend left =40] node [fill=white] {$Js$} (B);
					\node at (1.9,1) {$\Swarrow\unif_{\gamma}$};
					\fill[fill=red,fill opacity=0.3] (0,1) to [bend left =36] (3,2) to [bend left =60]  (3,0) to [bend left =36] (0,1);
					\begin{scope}[xshift = 5 cm]
								\node (B) at (3,0) {$B$};
					\node (C) at (0,1) {$\One$};
					\node (D) at (3,2) {$A$};
					
					\draw [->] (C)  to [bend left =-30] node [below] {$g^*$} (B);
					\draw [->] (C) to [bend left =30] node [above] {$f^*$} (D);
					\draw [o->] (D)  to [bend left =-40] node [fill=white] {$Jr$} (B);
					\draw [o->] (D) to [bend left =40] node [fill=white] {$Js$} (B);
					\node at (1.2,1) {$\Swarrow\unif_{\rho}$};
					\node at (3,1.1)  {$\Leftarrow$};
					\node at (3,0.8) {$J\theta$};

					\fill[fill=red,fill opacity=0.1] (0,1) to [bend left =36] (3,2) to [bend left =-60]  (3,0) to [bend left =36] (0,1);
					\fill[fill=green,fill opacity=0.2] (3,0) to [bend left =60] (3,2) to [bend left =60] (3,0);
					\end{scope}
				\end{tikzpicture}
			\end{center}
			\vspace{-0.2cm}
			\item If the following equality holds,
			\begin{center}
				\begin{tikzpicture}[thick, scale=0.6]
				\node (A) at (1,0) {$B$};
				\node (B) at (4,0) {$B$};
				\node (C) at (1,3) {$A$};
				\node (D) at (4,3) {$A$};
				\node at (5.5,1.5) {$=$};
				\draw [o->] (C)  to node [left] {$J s$} (A);
				\draw [->] (A) to  [bend left =-35] node [below] {$k$} (B);
				\draw [->] (C) to [bend left =35]  node [above] {$f$} (D);
				\draw [->] (C) to [bend left =-35]  node [below] {$h$} (D);
				\draw [o->] (D)  to node [right] {$Js$} (B);
				\fill[fill=olive,fill opacity=0.3] (1.05,3) to [bend left =50] (4,3) to [bend left =50] (1.05,3);
				\fill[fill=yellow,fill opacity=0.3] (1,0) to  [bend left =-50]  (4,0) to  (4,3) to [bend left =50] (1,3) to (1,0);
				\node at (2.5,1) {$\Swarrow\rho$};
				\node at (2.5,3) {$\Downarrow \alpha$};
				\begin{scope}[xshift = 6cm]
				\node (A) at (1,0) {$B$};
				\node (B) at (4,0) {$B$};
				\node (C) at (1,3) {$A$};
				\node (D) at (4,3) {$A$};
				
				\draw [o->] (C)  to node [left] {$J s$} (A);
				\draw [->] (A) to  [bend left =-35] node [below] {$k$} (B);
				\draw [->] (C) to [bend left =35]  node [above] {$f$} (D);
				\draw [->] (A) to [bend left =35]  node [above] {$g$} (B);
				\draw [o->] (D)  to node [right] {$Js$} (B);
				
				\node at (2.5,2) {$\Swarrow\gamma$};
				\node at (2.5,0) {$\Downarrow \beta$};
				\fill[fill=yellow,fill opacity=0.1] (1,0) to  [bend left =50]  (4,0) to  (4,3) to [bend left =-50] (1,3) to (1,0);

				\fill[fill=olive,fill opacity=0.1] (1.05,0) to [bend left =50] (4,0) to [bend left =50] (1.05,0);

					\end{scope}
				\end{tikzpicture}
			\end{center}
			then we have:
			\begin{center}
				\begin{tikzpicture}[thick,yscale=0.9, xscale=0.8]
						\node (B) at (3,0) {$B$};
					\node (C) at (0,1) {$\One$};
					\node (D) at (3,2) {$A$};
					
					\draw [->] (C)  to [bend left =-25] node [below] {$k^*$} (B);
					\draw [->] (C) to [bend left =25] node [above] {$f^*$} (D);
					\draw [->] (C) to [bend left =-25] node [fill=white] {$h^*$} (D);
					\draw [o->] (D) to [bend left =25] node [right] {$Js$} (B);
					\node at (2.2,0.6) {$\Swarrow \unif_{\rho}$};
					\node at (1.5,1.5) {$\Searrow\alpha^*$};
					\fill[fill=red,fill opacity=0.3] (0.05,1) to [bend left =-30] (3,2) to [bend left =36]  (3,0) to [bend left =30] (0,1);
					\fill[fill=olive,fill opacity=0.3] (0.05,1) to [bend left =30] (3,2) to [bend left =30] (0,1);
					\node at (4.25,1) {$=$};
					\begin{scope}[xshift = 5 cm]
						\node (B) at (3,0) {$B$};
						\node (C) at (0,1) {$\One$};
						\node (D) at (3,2) {$A$};
						
						\draw [->] (C)  to [bend left =-25] node [below] {$k^*$} (B);
						\draw [->] (C) to [bend left =25] node [above] {$f^*$} (D);
						\draw [->] (C) to [bend left =25] node [fill=white] {$g^*$} (B);
						\draw [o->] (D) to [bend left =25] node [right] {$Js$} (B);
						\node at (2.2,1.3) {$\Swarrow\unif_{\gamma}$};
						\node at (1.5,0.5) {$\Swarrow\beta^*$};

						\fill[fill=red,fill opacity=0.1] (0.05,1) to [bend left =30] (3,2) to [bend left =36] (3,0) to [bend left =-29] (0.05,1);
						\fill[fill=olive,fill opacity=0.1] (0.05,1) to [bend left =29] (3,0) to [bend left =30] (0.05,1);

					\end{scope}
				\end{tikzpicture}
			\end{center}
		\end{enumerate}
			These four axioms induce a strong pseudo-dinatural transformation:
		\[
		\unif :  \bicat{D}(J(-),J(-)) \stdinato \bicat{D}(\One, J(-)) : \bicat{C}^{\op} \times \bicat{C} \to \CAT
		\]
		\item \textbf{coherence between $\fix$ and $\unif$:}
		 for an invertible $2$-cell as below
		 \vspace{-0.4cm}
		\begin{center}
			\begin{tikzpicture}[thick,scale =0.4]
				\node (A) at (0,0) {$B$};
				\node (B) at (3,0) {$B$};
				\node (C) at (0,3) {$A$};
				\node (D) at (3,3) {$A$};
				
				\draw [o->] (C) -- node [left] {$J(s)$} (A);
				\draw [->] (A) -- node [below] {$g$} (B);
				\draw [->] (C) -- node [above] {$f$} (D);
				\draw [o->] (D) -- node [right] {$J(s)$} (B);
				\node at (1.5,1.5) {$\Swarrow \gamma$};
				\fill[fill=yellow,fill opacity=0.2] (0,0) to (3,0) to  (3,3) to (0,3);
			\end{tikzpicture}
		\end{center}
			 \vspace{-0.5cm}
		we have:
		\begin{center}
			\begin{tikzpicture}[thick, scale=0.9]
				\node (A) at (0.8,0.8) {$A$};
				\node (B) at (3,0.5) {$B$};
				\node (C) at (0,2.5) {$\One$};
				\node (D) at (3,2.5) {$B$};

				\draw [->] (C)  to [bend left =25] node [fill=white] {$g^*$} (B);
				\draw [->] (C) to [bend left =30] node [above] {$g^*$} (D);
				\draw [->] (C)  to [bend left =-15] node [below left] {$f^*$} (A);
				\draw [o->] (A)  to [bend left =-10] node [below] {$Js$} (B);
				\draw [->] (D) to node [right] {$g$} (B);
				\node at (1.9,2.4) {$\Nearrow\fix_g$};
				\node at (1.35,1.4) {$\Nearrow\unif_{\gamma}$};
				
				\fill[fill=bluem,fill opacity=0.2] (0,2.5) to [bend left =36] (3,2.5) to  (3,0.5) to [bend left =-30] (0,2.5);
				\fill[fill=red,fill opacity=0.2] (0,2.5) to [bend left =30] (3,0.5) to  [bend left =10]  (0.8,0.8) to [bend left =20] (0,2.5);
				\node at (3.75,1.5) {$=$};
				
				\begin{scope}[xshift = 3.5 cm, scale=0.9]
					\node (B) at (3,0) {$A$};
					\node (B1) at (5,0.5) {$B$};
					\node (C) at (1,2.5) {$\One$};
					\node (D) at (3,2) {$A$};
					\node (D1) at (5,2.5) {$B$};
					
					\draw [->] (C)  to [bend left =-40] node [below left] {$f^*$} (B);
					\draw [o->] (B)  to [bend left =-10] node [below] {$Js$} (B1);
					\draw [->] (C)  to [bend left =-10] node [fill=white] {$f^*$} (D);
					\draw [o->] (D)  to [bend left =-10] node [fill=white] {$Js$} (D1);
					\draw [->] (C) to [bend left =40] node [above] {$g^*$} (D1);
					\draw [->] (D)  to node [right] {$f$} (B);
					\draw [->] (D1)  to node [right] {$g$} (B1);
					\node at (2.1,1.2) {$\Nearrow\fix_f$};
					\node at (4,1.2) {$\Nearrow\gamma$};
					\node at (3.2,2.7) {$\Uparrow\unif_{\gamma} $};
					\fill[fill=bluem,fill opacity=0.1] (1,2.5) to [bend left =-15] (3,2) to  (3,0) to [bend left =48] (1,2.5);
					\fill[fill=red,fill opacity=0.2] (1,2.5) to [bend left =-15] (3,2) to [bend left =-15] (5,2.5) to [bend left =-48] (1,2.5);
					\fill[fill=yellow,fill opacity=0.2] (3,2) to [bend left =-15] (5,2.5) to  (5,0.5) to [bend left =15] (3,0);
				\end{scope}
			\end{tikzpicture}
		\end{center}
	\end{enumerate}
\end{definition}

\begin{definition}\label{def:morUnifFixOp}
	Let $((-)^*, \fix^*, \unif^*)$ and $((-)^\dag, \fix^\dag, \unif^\dag)$ be two pseudo-fixpoint operators uniform with respect to $J : \bicat{C} \to \bicat{D}$. A pseudo-morphism of uniform fixpoint operators from $((-)^*, \fix^*, \unif^*)$ to $((-)^\dag, \fix^\dag, \unif^\dag)$ is a pseudo-morphism of fixpoint operators $\delta : ((-)^*, \fix^*)\to ((-)^\dag, \fix^\dag)$ as in Definition \ref{def:morFixOp} satisfying the additional coherence for every square in $\bicat{D}$:
	\vspace{-0.3cm}
	 \begin{center}
	\begin{tikzpicture}[thick,scale =0.4]
		\node (A) at (0,0) {$B$};
		\node (B) at (3,0) {$B$};
		\node (C) at (0,3) {$A$};
		\node (D) at (3,3) {$A$};
		
		\draw [o->] (C) -- node [left] {$J(s)$} (A);
		\draw [->] (A) -- node [below] {$g$} (B);
		\draw [->] (C) -- node [above] {$f$} (D);
		\draw [o->] (D) -- node [right] {$J(s)$} (B);
		\node at (1.5,1.5) {$\Swarrow \gamma$};
		\fill[fill=yellow,fill opacity=0.2] (0,0) to (3,0) to  (3,3) to (0,3);
	\end{tikzpicture}
	\end{center}
\vspace{-0.3cm}
		we have:
		\vspace{-0.2cm}
		\begin{center}
			\begin{tikzpicture}[thick,yscale=0.95, xscale=0.8]
			\node (B) at (3,0) {$B$};
			\node (C) at (0,1) {$\One$};
			\node (D) at (3,2) {$A$};
			
			\draw [->] (C)  to [bend left =-25] node [below] {$g^\dag$} (B);
			\draw [->] (C) to [bend left =25] node [above] {$f^*$} (D);
			\draw [->] (C) to [bend left =25] node [fill=white] {$g^*$} (B);
			\draw [o->] (D) to [bend left =25] node [right] {$Js$} (B);
			\node at (2.2,1.3) {$\Swarrow\unif^*_{\gamma}$};
			\node at (1.5,0.5) {$\Swarrow\delta_g$};
			\node at (4.75,1) {$=$};
			\fill[fill=red,fill opacity=0.2] (0.05,1) to [bend left =30] (3,2) to [bend left =36] (3,0) to [bend left =-29] (0.05,1);
			\fill[fill=brown,fill opacity=0.2] (0.05,1) to [bend left =29] (3,0) to [bend left =30] (0.05,1);
			\begin{scope}[xshift = 6 cm]
			\node (B) at (3,0) {$B$};
			\node (C) at (0,1) {$\One$};
			\node (D) at (3,2) {$A$};
			
			\draw [->] (C)  to [bend left =-25] node [below] {$g^\dag$} (B);
			\draw [->] (C) to [bend left =25] node [above] {$f^*$} (D);
			\draw [->] (C) to [bend left =-25] node [fill=white] {$f^\dag$} (D);
			\draw [o->] (D) to [bend left =25] node [right] {$Js$} (B);
			\node at (2.2,0.6) {$\Swarrow \unif^\dag_{\gamma}$};
			\node at (1.5,1.5) {$\Searrow\delta_f$};
				\fill[fill=red,fill opacity=0.2] (0.05,1) to [bend left =-30] (3,2) to [bend left =36]  (3,0) to [bend left =30] (0,1);
			\fill[fill=brown,fill opacity=0.2] (0.05,1) to [bend left =30] (3,2) to [bend left =30] (0,1);
			\end{scope}
			\end{tikzpicture}
		\end{center}
		
		We denote by $\Fixop(\bicat{D}, J)$ the category of pseudo-fixpoint operators on $\bicat{D}$ uniform with respect to $J$.
\end{definition}

\begin{definition}
	Let $J : \bicat{C} \to \bicat{D}$ be a identity-on-objects $2$-functor (strictly) preserving terminal objects. 
	A \emph{pseudo-dinatural fixpoint operator uniform with respect to $J$} consists of a pseudo-dinatural fixpoint operator $((-)^*, \dinat)$ on $\bicat{D}$ together with a strong dinatural transformation 
	\[
	\unif :  \bicat{D}(J(-),J(-)) \stdinato \bicat{D}(\One, J(-),) : \bicat{C}^{\op} \times \bicat{C} \to \CAT
	\]
	satisfying the following additional coherence between $\dinat$ and $\unif$: for two squares in $\bicat{D}$ of the form
	\vspace{-0.2cm}
		\begin{center}
		\begin{tikzpicture}[thick,scale =0.5]
		\node (A) at (0,0) {$C$};
		\node (B) at (3,0) {$D$};
		\node (C) at (0,3) {$A$};
		\node (D) at (3,3) {$B$};
		\node (E) at (6,0) {$C$};
		\node (F) at (6,3) {$A$};
		
		\draw [o->] (C) -- node [left] {$Js$} (A);
		\draw [->] (A) -- node [below] {$h$} (B);
		\draw [->] (C) -- node [above] {$f$} (D);
		\draw [o->] (D) -- node [fill=white] {$Jr$} (B);
		\node at (1.6,1.5) {$\Swarrow \gamma$};
		\fill[fill=yellow,fill opacity=0.1] (0,0) to (3,0) to  (3,3) to (0,3);
		\fill[fill=yellow,fill opacity=0.3] (3,0) to (6,0) to  (6,3) to (3,3);
		\node at (-2.3,1.5) {$\rho\star\gamma=$};

		\draw [->] (D) -- node [above] {$g$} (F);
		\draw [->] (B) -- node [below] {$k$} (E);
		\draw [o->] (F) -- node [right] {$Js$} (E);
		\node at (4.6,1.5) {$\Swarrow \rho$};

		\end{tikzpicture}
	\end{center} 
	\vspace{-0.2cm}
we have:
 		\begin{center}
 			\begin{tikzpicture}[thick, yscale=1,xscale=1]
 				\node (A) at (0.8,0.8) {$B$};
 				\node (B) at (3,0.5) {$D$};
 				\node (C) at (0,2.5) {$\One$};
 				\node (D) at (3,2.5) {$C$};

 				\draw [->] (C)  to [bend left =25] node [fill=white] {$(hk)^*$} (B);
 				\draw [->] (C) to [bend left =40] node [above] {$(kh)^*$} (D);
 				\draw [->] (C)  to [bend left =-15] node [fill=white] {$(fg)^*$} (A);
 				\draw [o->] (A)  to [bend left =-10] node [below] {$Jr$} (B);
 				\draw [->] (D) to node [fill=white] {$h$} (B);
	\node at (1.7,2.55) {$\Nearrow\dinat_{k}^{h}$};
 				\node at (1.45,1.3) {$\Nearrow\unif_{ \gamma\star\rho}  $};
 				
 				\fill[fill=teal,fill opacity=0.2] (0,2.5) to [bend left =48] (3,2.5) to  (3,0.5) to [bend left =-30] (0,2.5);
 				\fill[fill=red,fill opacity=0.2] (0,2.5) to [bend left =30] (3,0.5) to  [bend left =10]  (0.8,0.8) to [bend left =20] (0,2.5);
 				\node at (3.5,1.5) {$=$};
 				
 				\begin{scope}[xshift = 3 cm, yscale=0.9]
 					\node (B) at (3,0) {$B$};
 					\node (B1) at (5,0.5) {$D$};
 					\node (C) at (1,2.5) {$\One$};
 					\node (D) at (3,2) {$A$};
 					\node (D1) at (5,2.5) {$C$};
 					
 					\draw [->] (C)  to [bend left =-50] node [fill=white] {$(fg)^*$} (B);
 					\draw [o->] (B)  to [bend left =-10] node [below] {$Jr$} (B1);
 					\draw [->] (C)  to [bend left =-10] node [fill=white] {$(gf)^*$} (D);
 					\draw [o->] (D)  to [bend left =-10] node [below] {$Js$} (D1);
 					\draw [->] (C) to [bend left =43] node [above] {$(kh)^*$} (D1);
 					\draw [->] (D)  to node [right] {$f$} (B);
 					\draw [->] (D1)  to node [fill=white] {$h$} (B1);
				\node at (2.1,1.3) {$\Nearrow\dinat_{g}^{f}$};
 					\node at (4,1.2) {$\Nearrow\gamma$};
 					\node at (3.2,2.75) {$\Uparrow\unif_{\rho\star\gamma}  $};
 					\fill[fill=teal,fill opacity=0.2] (1,2.5) to [bend left =-15] (3,2) to  (3,0) to [bend left =65 ] (1,2.5);
 					\fill[fill=red,fill opacity=0.2] (1,2.5) to [bend left =-15] (3,2) to [bend left =-15] (5,2.5) to [bend left =-55] (1,2.5);
 					\fill[fill=yellow,fill opacity=0.1] (3,2) to [bend left =-15] (5,2.5) to  (5,0.5) to [bend left =15] (3,0);
 				\end{scope}
 			\end{tikzpicture}
 		\end{center}
 	where $\gamma \star \rho$ corresponds to the following $2$-cell:
 		\vspace{-0.2cm}
 	\begin{center}
 		\begin{tikzpicture}[thick,scale =0.5]
 			\node (A) at (0,0) {$D$};
 			\node (B) at (3,0) {$C$};
 			\node (C) at (0,3) {$B$};
 			\node (D) at (3,3) {$A$};
 			\node (E) at (6,0) {$D$};
 			\node (F) at (6,3) {$B$};
 			
 			\draw [o->] (C) -- node [left] {$Jr$} (A);
 			\draw [->] (A) -- node [below] {$k$} (B);
 			\draw [->] (C) -- node [above] {$g$} (D);
 			\draw [o->] (D) -- node [fill=white] {$Js$} (B);
 			\node at (1.6,1.5) {$\Swarrow \rho$};
 			\fill[fill=yellow,fill opacity=0.3] (0,0) to (3,0) to  (3,3) to (0,3);
 			\fill[fill=yellow,fill opacity=0.1] (3,0) to (6,0) to  (6,3) to (3,3);
 			\node at (-2.3,1.5) {$\gamma\star \rho=$};

 			\draw [->] (D) -- node [above] {$f$} (F);
 			\draw [->] (B) -- node [below] {$h$} (E);
 			\draw [o->] (F) -- node [right] {$Jr$} (E);
 			\node at (4.6,1.5) {$\Swarrow \gamma$};
 			
 		\end{tikzpicture}
 	\end{center} 
 	\vspace{-0.2cm}
 	Note that if we restrict to the case where $A=B$, $C=D$, $g=1_A$, $k=1_C$, $r=s$ and $\rho=\id$, we obtain the coherence axiom between $\fix$ and $\unif$ in Definition \ref{def:2fixunif}. Pseudo-fixpoint operators uniform with respect to $J$ are therefore a special case of pseudo-dinatural fixpoint operators uniform with respect to $J$ as expected.
\end{definition}

	\begin{definition}
		Let $((-)^*, \dinat^*, \unif^*)$ and $((-)^\dag, \dinat^\dag, \unif^\dag)$ be two pseudo-dinatural fixpoint operators uniform with respect to $J : \bicat{C} \to \bicat{D}$. A pseudo-morphism of uniform dinatural fixpoint operators from $((-)^*, \dinat^*, \unif^*)$ to $((-)^\dag, \dinat^\dag, \unif^\dag)$ is a pseudo-morphism of dinatural fixpoint operators $\delta : ((-)^*, \dinat^*)\to ((-)^\dag, \dinat^\dag)$ as in Definition \ref{def:morDinFixOp} that is also a pseudo-morphism of uniform fixpoint operators as in Definition \ref{def:morUnifFixOp}, \emph{i.e.} $\delta$ commutes with both the dinaturality and uniformity structural $2$-cells $\dinat$ and $\unif$.

		We denote by $\DinFixop(\bicat{D}, J)$ the category of pseudo-dinatural fixpoint operators on $\bicat{D}$ uniform with respect to $J$.
	\end{definition}	
	
Before stating the main theorem of the paper, we recall the notion of pseudo-bifree algebras for $2$-functors.

	\begin{definition}\label{def:pseudoInitAlg}
		For a $2$-functor $\oc : \bicat{C} \to \bicat{C}$, a \emph{pseudo-initial algebra} is a $1$-cell $R:\oc \Phi \to \Phi$ such that for every $1$-cell $f : \oc A \to A$, there exists a pseudo-morphism of algebras $(u_f, \mu_f) : R \to f$, \emph{i.e.} a $1$-cell $u_f : \Phi \to A$ and a $2$-cell 
		\vspace{-0.2cm}
		\begin{center}
			\begin{tikzpicture}[thick,scale =0.5]
			\node (A) at (0,0) {$\oc A$};
			\node (B) at (3,0) {$A$};
			\node (C) at (0,3) {$\oc \Phi$};
			\node (D) at (3,3) {$\Phi$};
			
			\draw [->] (C) -- node [left] {$\oc u_f$} (A);
			\draw [->] (A) -- node [below] {$f$} (B);
			\draw [->] (C) -- node [above] {$R$} (D);
			\draw [->] (D) -- node [right] {$u_f $} (B);
			\node at (1.5,1.5) {$\Swarrow \mu_f$};
		\fill[fill=blue,fill opacity=0.2] (0,0) to (3,0) to  (3,3) to (0,3);
			\end{tikzpicture}
		\end{center}
		\vspace{-0.2cm}
	verifying the following universal property: for any pseudo-algebra $1$-cells $(v,\nu), (w, \omega) : R \to f$, there is a unique invertible algebra $2$-cell $\phi : (v,\nu)\Rightarrow (w, \omega)$, \emph{i.e.} a unique invertible $2$-cell $\phi : v \Rightarrow w$ in $\bicat{C}$ such that:
	\vspace{-0.2cm}
	\begin{center}
		\begin{tikzpicture}[thick, scale=0.55]
		\node (A) at (1,0) {$\oc A$};
		\node (B) at (4,0) {$A$};
		\node (C) at (1,3) {$\oc \Phi$};
		\node (D) at (4,3) {$\Phi$};
		
		\draw [->] (C)  to [bend left =-35] node [left] {$\oc w$} (A);
		\draw [->] (C)  to [bend left =35] node [right] {$\oc v$} (A);
		\draw [->] (A) -- node [below] {$f$} (B);
		\draw [->] (C) -- node [above] {$R$} (D);
		\draw [->] (D)  to [bend left =35] node [right] {$v$} (B);
		\node at (1,1.6)  {$\Leftarrow$};
		\node at (1,1.2) {$\oc\phi$};
		\node at (3.1,1.5) {$\Swarrow\nu$};
		\fill[fill=blue,fill opacity=0.1] (1,0) to (4,0) to [bend left =-47] (4,3) to (1,3) to [bend left =47] (1,0);
		\fill[fill=red,fill opacity=0.2] (1,0) to [bend left =47] (1,3) to [bend left =47] (1,0);
		\node (X) at (6,1.5) {$=$};

		\begin{scope}[xshift = 8 cm]
		\node (A) at (0,0) {$\oc A$};
		\node (B) at (3,0) {$A$};
		\node (C) at (0,3) {$\oc \Phi$};
		\node (D) at (3,3) {$\Phi$};

		\draw [->] (C) to[bend left =-35] node [left] {$\oc w$} (A);
		\draw [->] (A) -- node [below] {$f$} (B);
		\draw [->] (C) -- node [above] {$R$} (D);
		\draw [->] (D) to [bend left =35] node [right] {$v$} (B);
		\draw [->] (D) to [bend left =-35] node [left] {$w$} (B);
		\node at (3,1.6)  {$\Leftarrow$};
		\node at (3,1.2) {$\phi$};
		\node at (0.75,1.5) {$\Swarrow \omega$};
		\fill[fill=blue,fill opacity=0.3] (0,0) to (3,0) to [bend left =47] (3,3) to (0,3) to [bend left =-47] (0,0);
		\fill[fill=red,fill opacity=0.2] (3,0) to [bend left =47] (3,3) to [bend left =47] (3,0);
		\end{scope}
		\end{tikzpicture}
	\end{center}
		\vspace{-0.2cm}
	\end{definition}
We can similarly define a dual notion of pseudo-final $\oc$-coalgebra. Lambek's theorem stating that an initial algebra or a final coalgebra is an invertible morphism is generalized to an adjoint equivalence:
\begin{lemma}[\cite{BicategorySolRecursiveDomEqCattani2007}] \label{lem:PseudoInitAdjEquiv}
	If $R: \oc \Phi \to \Phi$ is a pseudo-initial $\oc$-algebra, then it is part of an adjoint equivalence $(R : \oc \Phi \to \Phi, L : \Phi \to \oc \Phi, \eta: \id \xRightarrow{\cong} RL, \varepsilon: L R \xRightarrow{\cong} \id)$.
\end{lemma}

\begin{definition}
	We say that $R : \oc \Phi \to \Phi$ is a \emph{pseudo-bifree algebra} if $R$ is a pseudo-initial algebra and its (uniquely determined) left adjoint $L$ is a pseudo-final coalgebra.
\end{definition}

We can now state the main theorem of this paper which is categorification of Theorem \ref{thm:PlotkinSimpson}:
\begin{theorem}\label{thm:main}
	Let $\bicat{C}$ be a $2$-category equipped with a (strict) $2$-comonad $(\oc, \delta, \varepsilon)$ and a (strict) terminal object $\One$. We denote by $\bicat{D}$ the co-Kleisli $2$-category $\bicat{C}_{\oc}$ and by $J : \bicat{C} \to \bicat{D}$ the free functor induced by the comonadic adjunction. 
	\begin{enumerate}
		\item If the endofunctor $\oc$ has a pseudo-bifree algebra, then the category $\Fixop(\bicat{D}, J)$ of pseudo-fixpoint operators on $\bicat{D}$ uniform with respect to $J$ is contractible.
		\item If $\bicat{C}$ is cartesian and the endofunctor $\oc \oc$ has a pseudo-bifree algebra, then the category $\DinFixop(\bicat{D}, J)$ of pseudo-dinatural fixpoint operators on $\bicat{D}$ uniform with respect to $J$ is contractible.
	\end{enumerate}
\end{theorem}
We proceed to prove this theorem in the next section by first constructing an explicit pseudo-fixpoint operator from the bifree algebras and showing that it verifies the required axioms and then proving the contractibility property \emph{i.e.} for any other pseudo-fixpoint operator, there is a unique isomorphism between them.

%% file: construction.tex
In this section, we fix a $2$-category $\bicat{C}$ equipped with a (strict) $2$-comonad $(\oc, \delta, \varepsilon)$ and a (strict) terminal object $\One$. We denote by $\bicat{D}$ the co-Kleisli $2$-category $\bicat{C}_{\oc}$ and by $J : \bicat{C} \to \bicat{D}$ the free functor induced by the comonadic adjunction. We assume further that the endofunctor $\oc$ has a pseudo-bifree algebra $R : \oc \Phi \to \Phi$. 
	
The following lemma is simply a reformulation of Definition \ref{def:pseudoInitAlg} from $\bicat{C}$ to the co-Kleisli $\bicat{D}$:
	\begin{lemma}\label{lem:fixInitial}
		For any $1$-cell $f : A \to A$ in $\bicat{D}$, there exists a $1$-cell $u_f : \Phi \to A$ in $\bicat{C}$ and a $2$-cell $\mu_f$ 
		\begin{center}
			\begin{tikzpicture}[thick,scale =0.5]
			\node (A) at (0,0) {$A$};
			\node (B) at (3,0) {$A$};
			\node (C) at (0,3) {$\Phi$};
			\node (D) at (3,3) {$\Phi$};
			
			\draw [o->] (C) -- node [left] {$J(u_f)$} (A);
			\draw [->] (A) -- node [below] {$f$} (B);
			\draw [->] (C) -- node [above] {$R$} (D);
			\draw [o->] (D) -- node [right] {$J(u_f)$} (B);
			\node at (1.6,1.5) {$\Swarrow \mu_f$};
			\fill[fill=blue,fill opacity=0.2] (0,0) to (3,0) to  (3,3) to (0,3);
			\end{tikzpicture}
		\end{center}
	 in $\bicat{D}$ verifying the following property: for any $1$-cells $v,w: \Phi \to A$ in $\bicat{C}$ and $2$-cells $\nu: J(v) R \Rightarrow f J(v)$ and $\omega: J(w) R \Rightarrow f J(w)$ in $\bicat{D}$, there exists a unique invertible $\phi: v \Rightarrow w$ in $\bicat{C}$ such that 
	 \begin{center}
	 	\begin{tikzpicture}[thick, scale=0.6]
	 	\node (A) at (1,0) {$A$};
	 	\node (B) at (4,0) {$A$};
	 	\node (C) at (1,3) {$\Phi$};
	 	\node (D) at (4,3) {$\Phi$};
	 	
	 	\draw [o->] (C)  to [bend left =-35] node [left] {$Jw$} (A);
	 	\draw [o->] (C)  to [bend left =35] node [right] {$Jv$} (A);
	 	\draw [->] (A) -- node [below] {$f$} (B);
	 	\draw [->] (C) -- node [above] {$R$} (D);
	 	\draw [o->] (D)  to [bend left =35] node [right] {$Jv$} (B);
	 	\node at (1,1.6)  {$\Leftarrow$};
	 	\node at (1,1.2) {$J\phi$};
	 	\node at (3.2,1.5) {$\Swarrow \nu$};
	 	\fill[fill=blue,fill opacity=0.1] (1,0) to (4,0) to [bend left =-49] (4,3) to (1,3) to [bend left =48] (1,0);
	 	\fill[fill=red,fill opacity=0.2] (1,0) to [bend left =48] (1,3) to [bend left =48] (1,0);
	 	\node (X) at (6,1.5) {$=$};
	 	\node (X) at (6,1.5) {$=$};
	 	
	 	\begin{scope}[xshift = 8 cm]
	 	\node (A) at (0,0) {$A$};
	 	\node (B) at (3,0) {$A$};
	 	\node (C) at (0,3) {$\Phi$};
	 	\node (D) at (3,3) {$\Phi$};

	 	\draw [o->] (C) to[bend left =-35] node [left] {$Jw$} (A);
	 	\draw [->] (A) -- node [below] {$f$} (B);
	 	\draw [->] (C) -- node [above] {$R$} (D);
	 	\draw [o->] (D) to [bend left =35] node [right] {$Jv$} (B);
	 	\draw [o->] (D) to [bend left =-35] node [left] {$Jw$} (B);
	 	\node at (3,1.6)  {$\Leftarrow$};
	 	\node at (3,1.2) {$J\phi$};
	 	\node at (0.55,1.5) {$\Swarrow \omega$};
		\fill[fill=blue,fill opacity=0.3] (0,0) to (3,0) to [bend left =50] (3,3) to (0,3) to [bend left =-50] (0,0);
	\fill[fill=red,fill opacity=0.2] (3,0) to [bend left =50] (3,3) to [bend left =50] (3,0);
	 	\end{scope}
	 	\end{tikzpicture}
	 \end{center}
	\end{lemma}

The next lemma uses the fact that $R$ is pseudo-bifree and therefore part of an adjoint equivalence where the left adjoint $L$ is pseudo-final.
\begin{restatable}{lemma}{fixFinal}\label{lem:fixFinal}
There exists a $1$-cell $t: \One \to \Phi$ and an invertible $2$-cell $\tau: t \Rightarrow Rt$ in $\bicat{D}$
\vspace{-0.2cm}
\begin{center}
	\begin{tikzpicture}[thick, scale=0.7]
		\node (B) at (3,0) {$\Phi$};
		\node (C) at (0,1) {$\One$};
		\node (D) at (3,2) {$\Phi$};
		
		\draw [->] (C)  to [bend left =-30] node [below] {$t$} (B);
		\draw [->] (C) to [bend left =30] node [above] {$t$} (D);
		\draw [->] (D)  to node [right] {$R$} (B);
		\node at (1.6,1)  {$\Nearrow\tau$};
		\fill[fill=cyan,fill opacity=0.2] (0,1) to [bend left =36] (3,2) to  (3,0) to [bend left =36] (0,1);
	\end{tikzpicture}
\end{center} 
\vspace{-0.2cm}
satisfying the following property: for any $1$-cells $v, w: \One \to \Phi$ in $\bicat{D}$ and invertible $2$-cells $\nu: v \Rightarrow R v$ and $\omega: w \Rightarrow R w$, there exists a unique invertible $2$-cell $\psi : v \Rightarrow w$ in $\bicat{D}$ such that
\begin{center}
	\begin{tikzpicture}[thick,yscale=0.95, xscale=0.8]
		\node (B) at (3,0) {$\Phi$};
		\node (C) at (0,1) {$\One$};
		\node (D) at (3,2) {$\Phi$};
		
		\draw [->] (C)  to [bend left =-25] node [below] {$v$} (B);
		\draw [->] (C) to [bend left =25] node [above] {$w$} (D);
		\draw [->] (C) to [bend left =25] node [fill=white] {$w$} (B);
		\draw [->] (D) to [bend left =25] node [right] {$R$} (B);
		\node at (2.2,1.3) {$\Nearrow\omega$};
		\node at (1.5,0.5) {$\Nearrow\psi$};
		\node at (4.75,1) {$=$};
		\fill[fill=cyan,fill opacity=0.1] (0.05,1) to [bend left =30] (3,2) to [bend left =36] (3,0) to [bend left =-29] (0.05,1);
		\fill[fill=magenta,fill opacity=0.2] (0.05,1) to [bend left =29] (3,0) to [bend left =30] (0.05,1);
		\begin{scope}[xshift = 6 cm]
			\node (B) at (3,0) {$\Phi$};
			\node (C) at (0,1) {$\One$};
			\node (D) at (3,2) {$\Phi$};
			
			\draw [->] (C)  to [bend left =-25] node [below] {$v$} (B);
			\draw [->] (C) to [bend left =25] node [above] {$w$} (D);
			\draw [->] (C) to [bend left =-25] node [fill=white] {$v$} (D);
			\draw [->] (D) to [bend left =25] node [right] {$R$} (B);
			\node at (2.2,0.6) {$\Nearrow\nu$};
			\node at (1.5,1.5) {$\Nwarrow\psi$};
			\fill[fill=cyan,fill opacity=0.3] (0.05,1) to [bend left =-30] (3,2) to [bend left =36]  (3,0) to [bend left =30] (0,1);
			\fill[fill=magenta,fill opacity=0.2] (0.05,1) to [bend left =30] (3,2) to [bend left =30] (0,1);
		\end{scope}
	\end{tikzpicture}
\end{center}    
\end{restatable}

Using these two lemmas, we can now construct the pseudo-fixpoint operator on $\bicat{D}$. We define a family of functors indexed by the objects $A \in \bicat{D}$
	\[
		(-)^*_A : \bicat{D}(A,A) \to \bicat{D}(\One, A)
	\]
	mapping a $1$-cell $f : A \to A$ to 
	\begin{center}
	\begin{tikzpicture}[thick]
		\node (A) at (0,0) {$f^* := \One$};
		\node (B) at (2,0) {$\Phi$};
		\node (C) at (4,0) {$A$};
		\draw [o->] (B) to node [above] {$J(u_f)$} (C);
		\draw [->] (A)  to node [above] {$t$} (B);
	\end{tikzpicture}
\end{center} 
	 where $u_f$ and $t$ are obtained from Lemmas \ref{lem:fixInitial} and \ref{lem:fixFinal}. For a $2$-cell $\alpha: f \Rightarrow g$ in $\bicat{D}(A,A)$, define $\alpha^* : f^* \Rightarrow g^*$ as
	 \vspace{-0.3cm}
	\begin{center}
		\begin{tikzpicture}[thick,scale=0.8]
		\node (A) at (-0.5,0) {$\One$};
		\node (B) at (1.5,0) {$\Phi$};
		\node (C) at (4,0) {$A$};
		
		\draw [o->] (B)  to [bend left =-37] node [below] {$J(u_g)$} (C);
		\draw [o->] (B) to [bend left =37] node [above] {$J(u_f)$} (C);
		\draw [->] (A)  to node [above] {$t$} (B);
		\node at (2.7,0) {$\Downarrow J\phi$};
		\fill[fill=red,fill opacity=0.2] (1.59,0) to [bend left =60] (3.9,0) to [bend left =60] (1.59,0);
		\end{tikzpicture}
	\end{center} 
	 \vspace{-0.3cm}
	where $\phi$ is the unique $2$-cell $u_f \Rightarrow u_g$ in $\bicat{C}$ such that
	 \vspace{-0.4cm}
	\begin{center}
		\begin{tikzpicture}[thick, yscale=0.7, xscale=0.7]
		\node (A) at (1,0) {$A$};
		\node (B) at (4,0) {$A$};
		\node (C) at (1,3) {$\Phi$};
		\node (D) at (4,3) {$\Phi$};
		
		\draw [o->] (C)  to [bend left =-35] node [left] {$J u_g$} (A);
		\draw [o->] (C)  to [bend left =35] node [right] {$J u_f $} (A);
		\draw [->] (A) to  [bend left =30] node [fill=white] {$f$} (B);
		\draw [->] (A) to  [bend left =-30] node [below] {$g$} (B);
		\draw [->] (C) -- node [above] {$R$} (D);
		\draw [o->] (D)  to [bend left =35] node [fill=white] {$J u_f$} (B);
		\node at (1,1.5)  {$\Leftarrow$};
		\node at (1,1.1) {$J\phi$};

		\node at (3.4,1.5) {$\Swarrow\mu_f$};
		\node at (2.5,0) {$\Downarrow \alpha$};
		
		\node (X) at (5.7,1.5) {$=$};
		\fill[fill=blue,fill opacity=0.2] (1,0) to [bend left =38] (4,0) to [bend left =-47] (4,3) to  (1,3) to [bend left =47] (1,0);
		\fill[fill=red,fill opacity=0.2] (1,0) to [bend left =47] (1,3) to [bend left =47] (1,0);
		\fill[fill=red,fill opacity=0.1] (1,0) to [bend left =38] (4,0) to [bend left =38] (1,0);
		\begin{scope}[xshift = 7.4 cm]
		\node (A) at (0,0) {$A$};
		\node (B) at (3,0) {$A$};
		\node (C) at (0,3) {$\Phi$};
		\node (D) at (3,3) {$\Phi$};

		\draw [o->] (C) to[bend left =-35] node [fill=white] {$Ju_g$} (A);
		\draw [->] (A) to  node [below] {$g$} (B);
		\draw [->] (C) -- node [above] {$R$} (D);
		\draw [o->] (D) to [bend left =35] node [right] {$Ju_f$} (B);
		\draw [o->] (D) to [bend left =-35] node [left] {$Ju_g$} (B);
		\node at (3,1.5)  {$\Leftarrow$};
		\node at (3,1.1) {$J\phi$};

		\node at (0.5,1.5) {$\Swarrow\mu_g$};
		\fill[fill=blue,fill opacity=0.2] (0,0) to (3,0) to [bend left =47] (3,3) to (0,3) to [bend left =-47] (0,0);
		\fill[fill=red,fill opacity=0.2] (3,0) to [bend left =47] (3,3) to [bend left =47] (3,0);
		\end{scope}
		\end{tikzpicture}
	\end{center}
		 \vspace{-0.3cm}
	which exists by Lemma \ref{lem:fixInitial}.
	
	We can now define the fixpoint $2$-cell $\fix_f : f \circ f^* \Rightarrow f^*$ for a $1$-cell $f: A \to A$ in $\bicat{D}$ as:
	\begin{center}
		\begin{tikzpicture}[thick, scale=0.7]
		\node (B) at (3,-1) {$A$};
		\node (C) at (0,0) {$\One$};
		\node (D) at (3,1) {$A$};
		
		\draw [->] (C)  to [bend left =-30] node [below] {$f^*$} (B);
		\draw [->] (C) to [bend left =30] node [above] {$f^*$} (D);
		\draw [->] (D)  to node [right] {$f$} (B);
		\node at (1.6,0)  {$\Nearrow\fix_f$};
		\fill[fill=bluem,fill opacity=0.2] (0,0) to [bend left =36] (3,1) to  (3,-1) to [bend left =36] (0,0);
		\node at (4,0) {$:=$};
		\begin{scope}[xshift=5cm]
		\node (B) at (2.5,-1.1) {$\Phi$};
	\node (A) at (0,0) {$\One$};
	\node (C) at (2.5,1.1) {$\Phi$};
	\node (D) at (5,1.1) {$A$};
	\node (E) at (5,-1.1) {$A$};

	\draw [->] (A)  to [bend left =-20] node [below] {$t$} (B);
	\draw [->] (A) to  [bend left =20] node [above] {$t$} (C);
	\draw [->] (C)  to node [right] {$R$} (B);
	\node at (1.5,0) {$\Nearrow\tau$};
	
	\draw [o->] (C) -- node [above] {$J u_f$} (D);
	\draw [->] (D) -- node [right] {$f$} (E);
	\draw [o->] (B) -- node [below] {$J u_f $} (E);
	\node at (4,0) {$\Nearrow\mu_f$};
	\fill[fill=cyan,fill opacity=0.2] (0,0) to [bend left =26] (2.5,1.1) to  (2.5,-1.1) to [bend left =26] (0,0);
			\fill[fill=blue,fill opacity=0.2] (2.5,-1.1) to (5,-1.1) to  (5,1.1) to (2.5,1.1);
		\end{scope}
	
		\end{tikzpicture}
	\end{center} 
where $\mu_f$ and $\tau$ are obtained from Lemmas \ref{lem:fixInitial} and \ref{lem:fixFinal}.
 To obtain that $((-)^*, \fix)$ is a pseudo-fixpoint operator on $\bicat{D}$, it only remains to show that $\fix$ is natural:
 
 \begin{restatable}[Naturality of $\fix$]{lemma}{fixNat}\label{lem:Nat}

		For a $2$-cell $\alpha: f \Rightarrow g$ in $\bicat{D}$, we have $\fix_g \circ \: \alpha^* = (\alpha \cdot \alpha^*)\circ \fix_f$.
\end{restatable}

	Now that we have a constructed a pseudo-fixpoint operator on $\bicat{D}$, we want to show that it is uniform with respect to the free functor $J : \bicat{C} \to \bicat{D}$ from the base $2$-category $\bicat{C}$ to the co-Kleisli $\bicat{D} = \bicat{C}_{\oc}$.
	
	Assume that there exist a $1$-cell $s :A  \to B$ in $\bicat{C}$ and $f:A \to A$, $g: B\to B$ in $\bicat{D}$ and a $2$-cell 
	\vspace{-0.2cm}
	 \begin{center}
		\begin{tikzpicture}[thick,scale =0.4]
			\node (A) at (0,0) {$B$};
			\node (B) at (3,0) {$B$};
			\node (C) at (0,3) {$A$};
			\node (D) at (3,3) {$A$};
			
			\draw [o->] (C) -- node [left] {$J(s)$} (A);
			\draw [->] (A) -- node [below] {$g$} (B);
			\draw [->] (C) -- node [above] {$f$} (D);
			\draw [o->] (D) -- node [right] {$J(s)$} (B);
			\node at (1.5,1.5) {$\Swarrow \gamma$};
			\fill[fill=yellow,fill opacity=0.2] (0,0) to (3,0) to  (3,3) to (0,3);
		\end{tikzpicture}
	\end{center}
	 \vspace{-0.3cm}
	in $\bicat{D}$. By Lemma \ref{lem:fixInitial}, there exists a unique $2$-cell $\phi : s \circ u_f  \Rightarrow u_g$ in $\bicat{C}$ such that:
	 \vspace{-0.2cm}
	\begin{center}
		\begin{tikzpicture}[thick, xscale=0.65,yscale=0.5]
			\node (A) at (1.9,0) {$A$};
			\node (B) at (4.9,0) {$A$};
			\node (A1) at (1,-3) {$B$};
			\node (B1) at (4,-3) {$B$};
			\node (C) at (1,3) {$\Phi$};
			\node (D) at (4,3) {$\Phi$};
			
			\draw [o->] (C)  to [bend left =-25] node [left] {$Ju_g$} (A1);
			\draw [o->] (C)  to [bend left =15] node [fill=white] {$Ju_f$} (A);
			\draw [o->] (A)  to [bend left =15] node [fill=white] {$Js$} (A1);
			\draw [->] (A) -- node [fill=white] {$f$} (B);
			\draw [->] (A1) -- node [below] {$g$} (B1);
			\draw [->] (C) -- node [above] {$R$} (D);
			\draw [o->] (D)  to [bend left =15] node [right] {$Ju_f$} (B);
			\draw [o->] (B)  to [bend left =15] node [right] {$Js$} (B1);
			\node at (1,-0.25)  {$\Leftarrow$};
			\node at (1,0.25) {$J\phi$};
			\fill[fill=red,fill opacity=0.2] (1,3) to [bend left =18] (1.9,0) to [bend left =18] (1,-3) to [bend left =30] (1,3);
			\fill[fill=blue,fill opacity=0.2] (1,3) to [bend left =18] (1.9,0) to  (4.9,0) to [bend left =-20] (4,3) to (1,3);
			\fill[fill=yellow,fill opacity=0.2] (1.9,0) to [bend left =18] (1,-3) to (4,-3) to [bend left =-20] (4.9,0);
			\node at (3.4,1.5) {$\Swarrow\mu_f$};
			
			\node at (3.4,-1.5) {$\Swarrow\gamma$};
			\node at (5.75,0) {$=$};
			
			\begin{scope}[xshift = 7.5cm, yshift=-2cm, yscale=1.3]
				\node (A) at (0,-0.5) {$B$};
				\node (B) at (3,-0.5) {$B$};
				\node (C) at (0,3.5) {$\Phi$};
				\node (D) at (3,3.5) {$\Phi$};
				\node (E) at (3.8,1.5) {$A$};

				\draw [o->] (C) to[bend left =-35] node [fill=white] {$Ju_g$} (A);
				\draw [->] (A) -- node [below] {$g$} (B);
				\draw [->] (C) -- node [above] {$R$} (D);
				\draw [o->] (D) to [bend left =13] node [right] {$Ju_f$} (E);
				\draw [o->] (E) to [bend left =13] node [right] {$Js$} (B);
				\draw [o->] (D) to [bend left =-33] node [fill=white] {$Ju_g$} (B);
				\node at (3,1.15)  {$\Leftarrow$};
				\node at (3,1.65) {$J\phi$};
				\fill[fill=blue,fill opacity=0.2] (0,-0.5) to (3,-0.5) to [bend left =44] (3,3.5) to (0,3.5) to [bend left =-44] (0,-0.5);
				\fill[fill=red,fill opacity=0.2] (3,3.5) to [bend left =44] (3,-0.5) to [bend left =44] (3,3.5);
				\node at (0.6,1.5) {$\Swarrow \mu_g$};
			\end{scope}
		\end{tikzpicture}
	\end{center}
	\vspace{-0.2cm}
	
	Define $\unif_\gamma$ as:
	\vspace{-0.2cm}
	\begin{center}
	\begin{tikzpicture}[thick, scale=0.7]
	\node (B) at (3,0) {$B$};
	\node (C) at (0,1) {$\One$};
	\node (D) at (3,2) {$A$};
	
	\draw [->] (C)  to [bend left =-30] node [below] {$g^*$} (B);
	\draw [->] (C) to [bend left =30] node [above] {$f^*$} (D);
	\draw [o->] (D)  to node [right] {$Js$} (B);
	\node at (1.6,1) {$\Swarrow \unif_\gamma$};
	\node at (4.5,1) {$:=$};
	\fill[fill=red,fill opacity=0.2] (0,1) to [bend left =36] (3,2) to  (3,0) to [bend left =36] (0,1);
	\begin{scope}[xshift = 5.5 cm]
	\node (A) at (0,1) {$\One$};
	\node (B) at (2,1) {$\Phi$};
	\node (C) at (5,2) {$A$};
	\node (D) at (5,0) {$B$};
	
	\draw [o->] (B)  to [bend left =-30] node [below] {$Ju_g$} (D);
	\draw [o->] (B) to [bend left =30] node [above] {$Ju_f$} (C);
	\draw [o->] (C) to  node [right] {$Js$} (D);
	\draw [->] (A)  to node [above] {$t$} (B);
	\node at (3.75,1) {$\Swarrow J\phi$};
		\fill[fill=red,fill opacity=0.2] (2,1) to [bend left =38] (5,2) to  (5,0) to [bend left =38] (2,1);
	\end{scope}
	\end{tikzpicture}
\end{center}
\vspace{-0.2cm}
\begin{restatable}{proposition}{unifstrongpsdinat}\label{prop:unifstrongpsdinat}
	The $2$-cells $\unif_\gamma$ we constructed yield a strong pseudo-dinatural transformation and they verify the coherence axiom between $\fix$ and $\unif$.
\end{restatable}

We have constructed a uniform pseudo-fixpoint operator showing that the category $\Fixop(\bicat{D}, J)$ is inhabited, it remains to show that it is contractible.

Assume that there exists another pseudo-fixpoint operator $((-)^\dag, \fix^\dag, \unif^\dag)$ on $\bicat{D}$ that is uniform with respect to $J : \bicat{C} \to \bicat{D}$. We want to show that there is a unique isomorphism of uniform pseudo-fixpoint operators $\delta : ((-)^* \fix^*, \unif^*)\to ((-)^\dag, \fix^\dag, \unif^\dag)$.

By Lemma \ref{lem:fixFinal}, there exists a unique invertible $2$-cell $\delta_0 : t \Rightarrow R^\dag$ such that 
\begin{center}
	\begin{tikzpicture}[thick,scale=0.9]
	\node (B) at (3,0) {$\Phi$};
	\node (C) at (0,1) {$\One$};
	\node (D) at (3,2) {$\Phi$};
	
	\draw [->] (C)  to [bend left =-30] node [below] {$t$} (B);
	\draw [->] (C)  to [bend left =30] node [fill=white] {$R^\dag$} (B);
	\draw [->] (C) to [bend left =30] node [above] {$R^\dag$} (D);
	\draw [->] (D)  to node [right] {$R$} (B);
	\node at (1.3,0.5)  {$\Nearrow\delta_0$};
	\node at (2.2,1.5) {$\Nearrow\fix_R^\dag$};
	\fill[fill=bluem,fill opacity=0.2] (0.05,1) to [bend left =38] (3,2) to (3,0) to [bend left =-35] (0.05,1);
	\fill[fill=magenta,fill opacity=0.2] (0.05,1) to [bend left =35] (3,0) to [bend left =36] (0.05,1);
	\node at (4,1) {$=$};
	
	\begin{scope}[xshift = 5 cm]
	\node (B) at (3,0) {$\Phi$};
	\node (C) at (0,1) {$\One$};
	\node (D) at (3,2) {$\Phi$};
	
	\draw [->] (C)  to [bend left =-30] node [below] {$t$} (B);
	\draw [->] (C)  to [bend left =-30] node [fill=white] {$t$} (D);
	\draw [->] (C) to [bend left =30] node [above] {$R^\dag$} (D);
	\draw [->] (D)  to node [right] {$R$} (B);
	\node at (2,0.5) {$\Nearrow\tau$};
	\node at (1.5,1.5) {$\Nwarrow\delta_0$};
	\fill[fill=cyan,fill opacity=0.2] (0.05,1) to [bend left =-35] (3,2) to (3,0) to [bend left =38] (0,1);
\fill[fill=magenta,fill opacity=0.2] (0.05,1) to [bend left =36] (3,2) to [bend left =35] (0,1);
	\end{scope}
	\end{tikzpicture}
\end{center}
Now, for every $1$-cell $f: A \to A$, there exists by Lemma \ref{lem:fixInitial} a $1$-cell $u_f: \Phi \to A$ in $\bicat{C}$ and a $2$-cell 
\vspace{-0.2cm}
	\begin{center}
	\begin{tikzpicture}[thick,scale =0.5]
	\node (A) at (0,0) {$A$};
	\node (B) at (3,0) {$A$};
	\node (C) at (0,3) {$\Phi$};
	\node (D) at (3,3) {$\Phi$};
	
	\draw [o->] (C) -- node [left] {$J(u_f)$} (A);
	\draw [->] (A) -- node [below] {$f$} (B);
	\draw [->] (C) -- node [above] {$R$} (D);
	\draw [o->] (D) -- node [right] {$J(u_f)$} (B);
	\node at (1.6,1.5) {$\Swarrow \mu_f$};
	\fill[fill=blue,fill opacity=0.2] (0,0) to (3,0) to  (3,3) to (0,3);
	\end{tikzpicture}
\end{center}
\vspace{-0.2cm}
in $\bicat{D}$. Since $(-)^\dag$ is uniform with respect to $J$, we have a $2$-cell $\unif^\dag_{\mu_f} : Ju_f \circ R^\dag \Rightarrow f^\dag$ and we define $\delta_f : f^* \Rightarrow f^\dag$ as:
\vspace{-0.2cm}
\begin{center}
	\begin{tikzpicture}[thick,yscale=1, xscale=0.8]

	\node at (0,1) {$\delta_f\quad :=$};
	
	\begin{scope}[xshift = 1.5 cm]
	\node (B) at (3,0) {$B$};
	\node (C) at (0,1) {$\One$};
	\node (D) at (3,2) {$A$};
	
	\draw [->] (C)  to [bend left =-25] node [below] {$f^\dag$} (B);
	\draw [->] (C) to [bend left =25] node [above] {$t$} (D);
	\draw [->] (C) to [bend left =-25] node [fill=white] {$R^\dag$} (D);
	\draw [o->] (D) to [bend left =25] node [right] {$Ju_f$} (B);
	\node at (2.1,0.6) {$\Swarrow\unif^\dag_{\mu_f}$};
	\node at (1.5,1.5) {$\Searrow\delta_0$};
		\fill[fill=red,fill opacity=0.1] (0.05,1) to [bend left =-30] (3,2) to [bend left =36]  (3,0) to [bend left =30] (0,1);
	\fill[fill=magenta,fill opacity=0.2] (0.05,1) to [bend left =30] (3,2) to [bend left =30] (0,1);
	\end{scope}
	\end{tikzpicture}
\end{center}
\vspace{-0.3cm}
\begin{restatable}{proposition}{contractiblefix}\label{prop:contractiblefix}
		The category $\Fixop(\bicat{D}, J)$ is contractible \emph{i.e.} $\delta$ is an isomorphism of uniform pseudo-fixpoint operators and it is unique.
\end{restatable}

\section{Dinaturality}

To obtain that the pseudo-fixpoint operator is pseudo-dinatural, we want to construct an invertible dinaturality $2$-cell for every $1$-cell $f : A \to B$ in $\bicat{D}$, 
\begin{center}
	\begin{tikzpicture}[thick, xscale=1.1]
		\node (A) at (-0.25,1) {$\bicat{D}(B,A)$};
		\node (B) at (1.5,0) {$\bicat{D}(B,B)$};
		\node (B1) at (4,0) {$\bicat{D}(\One,B)$};
		\node (D) at (1.5,2) {$\bicat{D}(A,A)$};
		\node (D1) at (4,2) {$\bicat{D}(\One,A)$};
		
		\draw [->] (A)  to [bend left =-10] node [below left] {$\bicat{D}(B,f)$} (B);
		\draw [->] (B)  to node [below] {$(-)^*_B$} (B1);
		\draw [->] (D)  to node [above] {$(-)^*_A$} (D1);
		\draw [->] (A) to [bend left =10] node [above left] {$\bicat{D}(f,A)$} (D);
		\draw [->] (D1)  to  node [right] {$\bicat{D}(\One,f)$} (B1);
		\node at (2.2,1) {$\Nearrow\dinat^f$};
		\fill[fill=teal,fill opacity=0.2] (-0.15,1) to [bend left =20] (1.5,2) to  (4,2) to (4,0) to  (1.5,0) to [bend left =20] (-.15,1);
	\end{tikzpicture}
\end{center}
with components
\vspace{-0.2cm}
	\begin{center}
	\begin{tikzpicture}[thick, scale=0.7]
		\node (B) at (3,0) {$B$};
		\node (C) at (0,1) {$\One$};
		\node (D) at (3,2) {$A$};
		
		\draw [->] (C)  to [bend left =-30] node [below] {$(fg)^*$} (B);
		\draw [->] (C) to [bend left =30] node [above] {$(gf)^*$} (D);
		\draw [->] (D)  to node [right] {$f$} (B);
		\node at (1.6,1) {$\Nearrow\dinat_{g}^f$};
		
		\fill[fill=teal,fill opacity=0.2] (0,1) to [bend left =36] (3,2) to  (3,0) to [bend left =36] (0,1);
	\end{tikzpicture}
\end{center}
\vspace{-0.2cm}
for $g$ in $\bicat{D}(B,A)$. To do so, we need to assume further that the endofunctor $\oc \oc : \bicat{C} \to \bicat{C}$ has a pseudo-bifree algebra and that $2$-category $\bicat{C}$ is cartesian. Note that it implies that co-Kleisli $\bicat{D}$ is cartesian as well and that the free functor $J : \bicat{C} \to \bicat{D}$ preserves the cartesian structure.

The following lemma uses the same argument as the strict case by Freyd \cite{freyd1991algebraically}.
\begin{lemma}
	Assume that $\oc\oc : \bicat{C} \to \bicat{C}$ has a pseudo bifree algebra. Then, if $R : \oc \Phi \to \Phi$ is a pseudo bifree $\oc$-algebra, 
		\[
		\oc \oc \Phi \xrightarrow{\oc R} \oc \Phi \xrightarrow{R} \Phi
		\]
	is a pseudo bifree $\oc\oc$-algebra.
\end{lemma}

	\begin{restatable}{lemma}{dinatbifree}\label{lem:dinatbifree}
		For $1$-cells $v,w: \One \to \Phi$ in $\bicat{D}$ and $2$-cells 
		\begin{center}
			\begin{tikzpicture}[thick, scale=0.7]
				\node (B) at (3,0) {$\Phi$};
				\node (C) at (0,1) {$\One$};
				\node (D) at (3,2) {$\Phi$};
				
				\draw [->] (C)  to [bend left =-30] node [below] {$v$} (B);
				\draw [->] (C) to [bend left =30] node [above] {$v$} (D);
				\draw [->] (D)  to node [right] {$R \circ R$} (B);
				\node at (1.6,1) {$\Nearrow\nu$};
				\fill[fill=purple,fill opacity=0.1] (0.05,1) to [bend left =38] (3,2) to  (3,0) to [bend left =38] (0,1);
				\node at (5.2,1) {and};
				
				\begin{scope}[xshift = 6.5 cm]
					\node (B) at (3,0) {$\Phi$};
					\node (C) at (0,1) {$\One$};
					\node (D) at (3,2) {$\Phi$};
					
					\draw [->] (C)  to [bend left =-30] node [below] {$w$} (B);
					\draw [->] (C) to [bend left =30] node [above] {$w$} (D);
					\draw [->] (D)  to node [right] {$R\circ R$} (B);
					\node at (1.6,1) {$\Nearrow\omega$};
				\fill[fill=purple,fill opacity=0.2] (0.05,1) to [bend left =38] (3,2) to  (3,0) to [bend left =38] (0,1);
				\end{scope}
			\end{tikzpicture}
		\end{center}
		in $\bicat{D}$, there exists a unique invertible $2$-cell $\lambda: w \Rightarrow v$ in $\bicat{D}$ such that 
			\begin{center}
			\begin{tikzpicture}[thick,yscale=0.8, xscale=0.7]
				\node (B) at (3,0) {$\Phi$};
				\node (C) at (0,1) {$\One$};
				\node (D) at (3,2) {$\Phi$};
				
				\draw [->] (C)  to [bend left =-25] node [below] {$w$} (B);
				\draw [->] (C) to [bend left =25] node [above] {$v$} (D);
				\draw [->] (C) to [bend left =25] node [fill=white] {$v$} (B);
				\draw [->] (D) to [bend left =25] node [right] {$R\circ R$} (B);
				\node at (2.2,1.3) {$\Nearrow\nu$};
				\node at (1.5,0.5) {$\Nearrow\lambda$};
				\node at (5.5,1) {$=$};
				\fill[fill=purple,fill opacity=0.1] (0.05,1) to [bend left =30] (3,2) to [bend left =36] (3,0) to [bend left =-29] (0.05,1);
				\fill[fill=teal,fill opacity=0.2] (0.05,1) to [bend left =29] (3,0) to [bend left =30] (0.05,1);
				\begin{scope}[xshift = 6.8 cm]
					\node (B) at (3,0) {$\Phi$};
					\node (C) at (0,1) {$\One$};
					\node (D) at (3,2) {$\Phi$};
					
					\draw [->] (C)  to [bend left =-25] node [below] {$w$} (B);
					\draw [->] (C) to [bend left =25] node [above] {$v$} (D);
					\draw [->] (C) to [bend left =-25] node [fill=white] {$w$} (D);
					\draw [->] (D) to [bend left =25] node [right] {$R\circ R$} (B);
					\node at (2.2,0.6) {$\Nearrow\omega$};
					\node at (1.5,1.5) {$\Nwarrow\lambda$};
					\fill[fill=purple,fill opacity=0.2] (0.05,1) to [bend left =-30] (3,2) to [bend left =36]  (3,0) to [bend left =30] (0,1);
					\fill[fill=teal,fill opacity=0.2] (0.05,1) to [bend left =30] (3,2) to [bend left =30] (0,1);
				\end{scope}
			\end{tikzpicture}
		\end{center}   
	\end{restatable}

	To construct the dinaturality $2$-cells, we first start by constructing a $2$-cell $\din : t \Rightarrow (R\circ R)^*$. Let $\din$ be the unique invertible $2$-cell from $t$ to $J(u_{R \circ R}) \circ t$ (obtained from Lemma \ref{lem:dinatbifree}) such that:
	\begin{center}
		\begin{tikzpicture}[thick, yscale=0.9, xscale=0.75]
			\node (B) at (3,1) {$\Phi$};
			\node (B1) at (5.5,0.25) {$\Phi$};
			\node (C) at (0.5,1.25) {$\One$};
			\node (D) at (3,3) {$\Phi$};
			\node (D1) at (5.5,2.25) {$\Phi$};
			
			\draw [->] (C)  to [bend left =5] node [fill=white] {$t$} (B);

			\draw [->] (C)  to [bend left =20] node [above] {$t$} (D);
			\draw [o->] (D)  to [bend left =5] node [above, sloped] {$Ju_{R\circ R}$} (D1);
			\draw [->] (C) to [bend left =-30] node [below] {$t$} (B1);
			\draw [->] (D)  to node [fill=white] {$R$} (B);
			\draw [->] (D1)  to node [fill=white] {$R\circ R$} (B1);
			\node at (2.2,1.8) {$\Nearrow\tau$};
			\node at (4.25,1.9) {$\Nearrow\mu_{R\circ R}$};
			\node at (3.2,0.4) {$\Nearrow\din$};
			\draw [o->] (B)  to [bend left =5] node [above, sloped] {$Ju_{R\circ R}$} (B1);
			\fill[fill=cyan,fill opacity=0.2] (0.55,1.25) to [bend left =25] (3,3) to  (3,1) to [bend left =-6] (0.55,1.25);
			\fill[fill=blue,fill opacity=0.2] (3,3) to  (3,1) to [bend left =6] (5.5,0.25) to (5.5,2.25) to [bend left =-6] (3,3);
			\fill[fill=teal,fill opacity=0.2] (0.55,1.25) to [bend left =6] (3,1) to [bend left =6] (5.5,0.25) to [bend left =35]  (0.55,1.25);
			\node at (6.5,1.25) {$=$};
			
			\begin{scope}[xshift = 6 cm]
				
				\node (A) at (5.5,2.25){$\Phi$};
				\node (B) at (5.5,0.25) {$\Phi$};
				\node (C) at (1,1.25) {$\One$};
				\node (D) at (5.5,1.25) {$\Phi$};
				\node (E) at (3,3) {$\Phi$};
				
				\draw [->] (C)  to [bend left =-30] node [below left] {$t$} (B);
				\draw [->] (C) to [bend left =-10] node [fill=white] {$t$} (D);
				\draw [->] (C) to [bend left =5] node [fill=white] {$t$} (A);
				\draw [->] (D)  to node [right] {$R$} (B);
				\draw [->] (A)  to node [right] {$R$} (D);
				\node at (3.5,0.5) {$\Nearrow \tau$};
				\node at (3.5,1.5) {$\Nearrow \tau$};
				\draw [o->] (E)  to [bend left =10] node [above, sloped] {$Ju_{R\circ R}$} (A);
				\draw [->] (C)  to [bend left =20] node [above] {$t$} (E);
				\node at (3.2,2.4) {$\Nearrow\din$};
				\fill[fill=cyan,fill opacity=0.2] (1.05,1.25) to [bend left =6] (5.5,2.25) to (5.5,0.25) to [bend left =35]  (1.05,1.25);
				\fill[fill=teal,fill opacity=0.2] (1.05,1.25) to [bend left =25] (3,3) to  [bend left =13] (5.5,2.25)  to [bend left =-6] (1.05,1.25);
			\end{scope}
		\end{tikzpicture}
	\end{center}

\begin{figure*}[h]
  \caption{Construction of the dinaturality $2$-cells}
	\begin{center}
		\begin{tikzpicture}[thick]	
			\begin{scope}[scale=0.8]
				\node (B) at (3,0) {$B$};
				\node (C) at (0,1) {$\One$};
				\node (D) at (3,2) {$A$};
				
				\draw [->] (C)  to [bend left =-30] node [below] {$(fg)^*$} (B);
				\draw [->] (C) to [bend left =30] node [above] {$(gf)^*$} (D);
				\draw [->] (D)  to node [right] {$f$} (B);
				\node at (1.6,1) {$\Nearrow\dinat_{g}^f$};
				\node at (4.5,1) {$:=$};
				\fill[fill=teal,fill opacity=0.2] (0,1) to [bend left =36] (3,2) to  (3,0) to [bend left =36] (0,1);
			\end{scope}
			\begin{scope}[xshift = 6.5 cm, xscale=1.2]
				\node (A) at (-1.5,1) {$\One$};
				\node (A1) at (4,1) {$B\times A$};
				\node (B) at (4,0) {$A\times B$};
				\node (B1) at (5.6,0) {$B$};
				\node (B0) at (1.5,0) {$\Phi$};
				\node (C) at (1.5,2) {$\Phi$};
				\node (D) at (4,2) {$A\times B$};
				\node (D1) at (5.6,2) {$A$};
				
				\draw [o->] (B0)  to node [below] {$Ju_{\sigma(f\times g)}$} (B);
				\draw [o->] (B)  to  node [below] {$\pi_2$} (B1);
				\draw [o->] (C)  to node [above] {$Ju_{\sigma(f\times g)}$} (D);
				\draw [o->] (D)  to  node [above] {$\pi_1$} (D1);
				
				\draw [->] (D)  to node [right] {$f\times g$} (A1);
				\draw [->] (A1)  to node [right] {$\sigma$} (B);
				\draw [->] (D1)  to node [right] {$f$} (B1);
				\draw [->] (C)  to node [fill=white] {$R$} (B0);
				\node at (2.5,1) {$\Nearrow\mu_{\sigma(f\times g)}$};
				\node at (4.8,1) {$=$};
				\node at (2.5,2.8) {$\Uparrow\unif_{\Pi^1_{\sigma(f\times g)} }$};
				\node at (2.5,-0.9) {$\Uparrow\unif^{-1}_{\Pi^2_{\sigma(f\times g)} }$};
				\draw [->] (A)  to [bend left =-30] node [below right] {$(RR)^*$} (B0);
				\draw [->] (A)  to [bend left =15] node [fill=white] {$t$} (B0);
				\draw [->] (A) to [bend left =-15] node [fill=white] {$t$} (C);
				\draw [->] (A) to [bend left =30] node [above right] {$(RR)^*$} (C);
				\node at (0.7,1) {$\Nearrow\tau$};
				\node at (-0.2,0.45) {$\Nearrow\din^{-1}$};
				\node at (-0.2,1.6) {$\Nwarrow\din$};
				\draw [->] (A) to [bend left =50] node [above] {$(gf)^*$} (D1);
				\draw [->] (A) to [bend left =-50] node [below] {$(fg)^*$} (B1);
				\fill[fill=cyan,fill opacity=0.2] (-1.45,1) to [bend left =-17] (1.5,2) to (1.5,0) to [bend left =-17] (-1.45,1);
				\fill[fill=teal,fill opacity=0.2] (-1.45,1) to [bend left =17] (1.5,0) to [bend left =36] (-1.45,1);
				\fill[fill=red,fill opacity=0.2] (-1.45,1) to [bend left =-36] (1.5,0) to (5.55,0) to [bend left =58] (-1.45,1);
				\fill[fill=red,fill opacity=0.2] (-1.45,1) to [bend left =36] (1.5,2) to (5.55,2) to [bend left =-58] (-1.45,1);
				\fill[fill=teal,fill opacity=0.2] (-1.45,1) to [bend left =36] (1.5,2) to [bend left =17] (-1.45,1);
				\fill[fill=blue,fill opacity=0.2] (1.5,2) to (4,2) to  (4,0) to (1.5,0);
			\end{scope}
		\end{tikzpicture}
	\end{center}
  \label{fig:dinat}
\end{figure*}
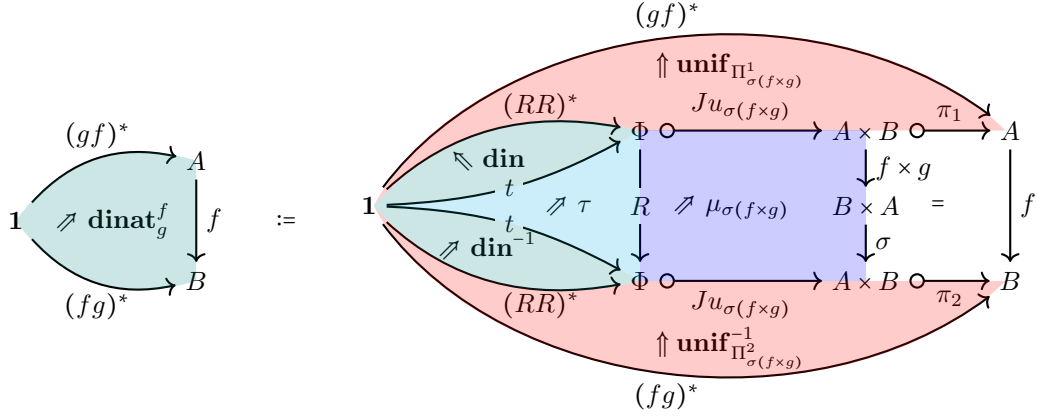

	 Once this isomorphism is obtained, we construct the dinaturality $2$-cells $\dinat_g^f$ using uniformity. Consider the following endo-$1$-cell:
	\begin{center}
	\begin{tikzpicture}[thick]
		\node (A) at (0,0) {$A \times B$};
		\node (B) at (2,0) {$B \times A$};
		\node (C) at (4,0) {$A \times B$};
		\draw [o->] (B) to node [above] {$\sigma$} (C);
		\draw [->] (A)  to node [above] {$f \times g$} (B);
	\end{tikzpicture}
\end{center} 
	where $\sigma$ is the symmetry obtained in the standard way as the pairing $\lis{\pi_1, \pi_2}$. Using Lemma \ref{lem:fixInitial}, there is a $1$-cell $u_{\sigma(f\times g)} : \Phi \to A \times B$ in $\bicat{C}$ and a $2$-cell $\mu_{\sigma(f\times g)}$:
		\begin{center}
		\begin{tikzpicture}[thick,yscale =0.6, xscale=0.8]
			\node (A) at (0,0) {$A$};
			\node (B) at (3,0) {$A$};
			\node (C) at (0,3) {$\Phi$};
			\node (D) at (3,3) {$\Phi$};
			
			\draw [o->] (C) -- node [left] {$J(u_{\sigma(f\times g)})$} (A);
			\draw [->] (A) -- node [below] {$f$} (B);
			\draw [->] (C) -- node [above] {$R$} (D);
			\draw [o->] (D) -- node [right] {$J(u_{\sigma(f\times g)})$} (B);
			\node at (1.6,1.5) {$\Swarrow \mu_{\sigma(f\times g)}$};
			\fill[fill=blue,fill opacity=0.2] (0,0) to (3,0) to  (3,3) to (0,3);
		\end{tikzpicture}
	\end{center}
	From the two squares
		\begin{center}
		\begin{tikzpicture}[thick, xscale=0.9, yscale=0.6]
			
			\node (B) at (3,0) {$ \Phi$};
			\node (E) at (6,0) {$\Phi$};
			\node (F) at (6,-3) {$A \times B$};
			\node (F1) at (6,-6) {$A$};
			\node (F2) at (0,-3) {$A \times B$};
			\node (G) at (0,0) {$\Phi$};
			\node (H) at (3,-3) {$A \times B$};
			\node (I) at (0,-6) {$A$};
			\node (J) at (3,-6) {$B$};
			
			\draw [o->] (B)  to node [fill=white] {$Ju_{\sigma (f \times g)}$} (H);
			\draw [->] (G)  to node [above] {$R$} (B);
			\draw [->] (I)  to node [below] {$f$} (J);
			\draw [o->] (G)  to node [left] {$Ju_{\sigma (f \times g)}$} (F2);
			\node at (1.3,-1.5) {$\Swarrow \mu_{\sigma (f \times g)}$};
			\draw [o->] (E) -- node [right] {$Ju_{\sigma (f \times g)}$} (F);
			\draw [o->] (F) -- node [right] {$\pi_1$} (F1);
			\draw [o->] (F2) -- node [left] {$\pi_1$} (I);
			\draw [o->] (H) -- node [right] {$\pi_2$} (J);
			\draw [->] (J) -- node [below] {$g$} (F1);
			\draw [->] (H) -- node [below] {$\sigma (f \times g)$} (F);
			\draw [->] (F2) -- node [below] {$\sigma (f \times g)$} (H);
			\draw [->] (B) -- node [above] {$R$} (E);
			\node at (4.7,-1.5) {$\Swarrow \mu_{\sigma (f \times g)}$};
			\node at (-1.5, -3) {$\Pi^1_{\sigma(f\times g)} :=$};
			\fill[fill=blue,fill opacity=0.2] (0,0) to (6,0) to  (6,-3) to (0,-3);
		\end{tikzpicture}

	\begin{tikzpicture}[thick, xscale=0.9, yscale=0.6]
		
		\node (B) at (3,0) {$ \Phi$};
		\node (E) at (6,0) {$\Phi$};
		\node (F) at (6,-3) {$A \times B$};
		\node (F1) at (6,-6) {$B$};
		\node (F2) at (0,-3) {$A \times B$};
		\node (G) at (0,0) {$\Phi$};
		\node (H) at (3,-3) {$A \times B$};
		\node (I) at (0,-6) {$B$};
		\node (J) at (3,-6) {$A$};

		\draw [o->] (B)  to node [fill=white] {$Ju_{\sigma (f \times g)}$} (H);
		\draw [->] (G)  to node [above] {$R$} (B);
		\draw [->] (I)  to node [below] {$g$} (J);
		\draw [o->] (G)  to node [left] {$Ju_{\sigma (f \times g)}$} (F2);
		\node at (1.3,-1.5) {$\Swarrow \mu_{\sigma (f \times g)}$};
		\draw [o->] (E) -- node [right] {$Ju_{\sigma (f \times g)}$} (F);
		\draw [o->] (F) -- node [right] {$\pi_2$} (F1);
		\draw [o->] (F2) -- node [left] {$\pi_2$} (I);
		\draw [o->] (H) -- node [right] {$\pi_1$} (J);
		\draw [->] (J) -- node [below] {$f$} (F1);
		
		\draw [->] (H) -- node [below] {$\sigma (f \times g)$} (F);
		\draw [->] (F2) -- node [below] {$\sigma (f \times g)$} (H);
		\draw [->] (B) -- node [above] {$R$} (E);
		\node at (4.7,-1.5) {$\Swarrow \mu_{\sigma (f \times g)}$};
		\fill[fill=blue,fill opacity=0.2] (0,0) to (6,0) to  (6,-3) to (0,-3);
		\node at (-1.5, -3) {$\Pi^2_{\sigma(f\times g)} :=$};
	\end{tikzpicture}
\end{center}

we obtain uniformity $2$-cells 
	\begin{center}
	\begin{tikzpicture}[thick, xscale=0.95, yscale=0.95]
		\node (B) at (3,-0.5) {$A$};
		\node (C) at (0,1) {$\One$};
		\node (D) at (3,2.5) {$\Phi$};
		\node (A) at (3, 1) {$A\times B$};
		
		\draw [->] (C)  to [bend left =-30] node [below] {$(gf)^*$} (B);
		\draw [->] (C) to [bend left =30] node [above] {$(RR)^*$} (D);
		\draw [o->] (D)  to node [right] {$Ju_{\sigma(f\times g)}$} (A);
		\draw [o->] (A)  to node [right] {$J\pi_1$} (B);
		\node at (1.3,1) {$\Swarrow\unif_{\Pi^1_{\sigma(f\times g)} }$};
		\fill[fill=red,fill opacity=0.2] (0.05,1) to [bend left =38] (3,2.5) to  (3,-0.5) to [bend left =38] (0,1);
		\node at (4.25,1) {and};
		
		\begin{scope}[xshift = 5cm]
					\node (B) at (3,-0.5) {$B$};
			\node (C) at (0,1) {$\One$};
			\node (D) at (3,2.5) {$\Phi$};
			\node (A) at (3, 1) {$A\times B$};
			
			\draw [->] (C)  to [bend left =-30] node [below] {$(fg)^*$} (B);
			\draw [->] (C) to [bend left =30] node [above] {$(RR)^*$} (D);
			\draw [o->] (D)  to node [right] {$Ju_{\sigma(f\times g)}$} (A);
			\draw [o->] (A)  to node [right] {$J\pi_2$} (B);
			\node at (1.3,1) {$\Swarrow\unif_{\Pi^2_{\sigma(f\times g)} }$};
			\fill[fill=red,fill opacity=0.2] (0.05,1) to [bend left =38] (3,2.5) to  (3,-0.5) to [bend left =38] (0,1);
		\end{scope}
	\end{tikzpicture}
\end{center}
We can now construct the general dinaturality $2$-cell $\dinat_g^f$ as in Figure \ref{fig:dinat}.

Note that for $\fix$ and $\unif$, we wrote the direction of $2$-cells without needing to take inverses whereas for dinaturality, a back-and-forth is needed to construct the $2$-cells. It means that for the general directed case, we need both the lax and oplax directions in order to obtain dinaturality.

\begin{restatable}{proposition}{unifpsdinat}\label{prop:unifpsdinat}
The $2$-cells $\dinat^f_g$ we constructed yield a pseudo-dinatural transformation and they verify the coherence axiom between $\dinat$ and $\unif$.
\end{restatable}

We have now shown that the category of uniform pseudo-dinatural fixpoint operators $\DinFixop(\bicat{D}, J)$ is inhabited, it remains to show that it is contractible to complete the proof of Theorem \ref{thm:main}.

Assume that there is another uniform pseudo-dinatural fixpoint operators $((-)^\dag, \dinat^\dag, \unif^\dag)$. By Proposition \ref{prop:contractiblefix}, we already have a constructed morphism of uniform pseudo-fixpoint operators $\delta : ((-)^*, \fix^*, \unif^*) \to ((-)^\dag, \fix^\dag, \unif^\dag)$ where the $2$-cells $\fix$ are induced from the $2$-cells $\dinat$. Therefore, it only remains to show:

\begin{restatable}{proposition}{contractibledinat}\label{prop:contractibledinat}
	The category $\DinFixop(\bicat{D}, J)$ is contractible \emph{i.e.} $\delta$ commutes with the dinaturality $2$-cells and it is unique.
\end{restatable}

%% file: examples.tex
We start by presenting examples where the pseudo-fixpoint operators are obtained as instances of Theorem \ref{thm:main} and therefore for which the category of pseudo-fixpoint operators is contractible. In particular, the least and greatest fixpoint (\emph{i.e.} the initial and final object of the category of pseudo-fixpoints) are isomorphic.  We consider afterwards the example of polynomial functors which are not an instance of our theorem but where the axioms for pseudo-fixpoints presented in Section \ref{sec:2fix} are verified.

\subsection{Limit-colimit coincidence theorem for $2$-categorical structures}
We give a brief reminder of the general recipe to obtain pseudo-bifree algebras for endofunctors on $2$-categories using the machinery developed by Cattani, Fiore and Winskel \cite{Fiorethesis, RecDomainEqCattaniFioreWinskel98, BicategorySolRecursiveDomEqCattani2007}.
Instead of considering preorder-enriched categories, we move to $2$-categories or bicategories whose hom-categories have colimits of $\omega$-chains and initial objects. The colimits of $\omega$-chains of embedding-projection pairs in the preorder-enriched setting are generalized to pseudo-colimits of $\omega$-chains of co-reflections (adjunctions with invertible unit) in the categorified setting.

We say that a $2$-category or bicategory $\bicat{C}$ is $\Cat_{\omega}$-enriched ($\Cat_{\omega,\bot}$-enriched) if for all objects $A$ and $B$, its hom-category $\bicat{C}(A,B)$ have colimits of $\omega$-chains (and initial objects) and whose composition functors
\[
\bicat{C}(A,B) \times \bicat{C}(B,C) \to \bicat{C}(B,C)
\]
preserve of colimits $\omega$-chains (and initial objects) in both variables.
The analogue of the category $\Cppo$ is the $2$-category $\Cat_{\omega}$ given by
\begin{itemize}
	\item $0$-cells: categories with colimits of $\omega$-chains and initial objects;
	\item $1$-cells: functors preserving colimits of $\omega$-chains;
	\item $2$-cells: natural transformations.
\end{itemize}
and its sub-$2$-category $\Cat_{\omega,\bot}$ which restricts the $1$-morphisms to those preserving initial objects (corresponding to the category $\Cppo_{\bot}$ in the preorder setting).

In order to obtain pseudo-bifree algebras, we make use of the following theorem:
\begin{theorem}[\cite{BicategorySolRecursiveDomEqCattani2007}]\label{thm:bifree2d}
	Let $\bicat{C}$ be a $\Cat_{\omega,\bot}$-enriched $2$-category. If $\bicat{C}$ has a pseudo-initial object and pseudo-$\omega$-colimits of chains of coreflections (adjunctions with invertible units), then $\bicat{C}$ is $\Cat_{\omega}$-pseudo-algebraically compact. It means that for every pseudo-functor $T : \bicat{C} \to \bicat{C}$, if $T$ is $\Cat_{\omega}$-enriched, \emph{i.e.} for all $A,B$, the induced functor
	\[
	\bicat{C}(A,B) \longrightarrow 	\bicat{C}(TA,TB) 
	\]
	preserves colimits of $\omega$-chains, then $T$ has a pseudo-bifree algebra.
\end{theorem}

\subsection{Categorical domain theory}
The canonical example is the $2$-category $\Cat_{\omega,\bot}$ with the lifting $2$-comonad $(-)_{\bot}:\Cat_{\omega,\bot} \to \Cat_{\omega,\bot}$ which freely adjoins initial objects. Both $(-)_{\bot}$ and $(-)_{\bot}\circ (-)_{\bot}$ are $\Cat_{\omega}$-enriched and therefore the coKleisli $2$-category (which is equivalent to $\Cat_{\omega}$) verifies that its category of pseudo-dinatural fixpoint operators uniform (with respect to the free functor $\Cat_{\omega,\bot} \to \Cat_{\omega}$) is contractible. The standard Lambek construction for initial algebras for a finitary endofunctor $f : \cat{A} \to \cat{A}$ on a category $\cat{A}$ with an initial object $\bot$ and $\omega$-colimits by calculating the colimit of the diagram
\[
\bot \to f(\bot) \to f^2(\bot) \to \dots
\]
provides both the initial and final pseudo-fixpoint operators~\cite{lambek1968fixpoint}. As a consequence of Theorem \ref{thm:main}, Lambek's construction verifies the dinaturality axioms and is uniform with respect to sub-$2$-category of functors preserving initial objects up to isomorphism.

We can also recover Adamek's result for \emph{Scott-complete categories} which can be viewed as a categorification of Scott domains \cite{AdamekCategoricalScott1997}. He considers the $2$-category $\mathbf{SCC}$ given by
	\begin{itemize}
		\item $0$-cells: Scott-complete categories (i.e finitely accessible category such that every diagram with a cocone has a colimit);
		\item $1$-cells: functors preserving directed colimits;
		\item $2$-cells: natural transformations.
	\end{itemize}
	and its sub-$2$-category $\mathbf{SCC}_{\bot}$ whose $1$-cells are restricted to functors preserving directed colimits and initial objects. The lifting $2$-comonad on $\mathbf{SCC}_{\bot}$ has the required bifree algebras so that the co-Kleisli $\mathbf{SCC}$ has a contractible category of pseudo-dinatural fixpoint operators uniform with respect to the free $2$-functor $\mathbf{SCC}_{\bot} \to \mathbf{SCC}$.

\subsection{Profunctors and linear logic models}

Another example of $\Cat_{\omega,\bot}$-enriched bicategory is the bicategory of profunctors denoted by $\Prof$ \cite{BenabouDist}. For small categories $\cat{A}$ and $\cat{B}$, a profunctor $P$ from $\cat{A}$ to $\cat{B}$ is a functor $P : \cat{A} \times \cat{B}^{\op} \to \Set$ or equivalently a functor from $\cat{A}$ to the presheaf category $\hat{\cat{B}}$. Profunctors can be seen as a categorification of $\Rel$ as a relation $R \subseteq A\times  B$ corresponds to a profunctor between discrete categories such that each component is either the empty set or a singleton.

Many pseudo-comonad structures were considered on $\Prof$ leading to models of linear logic with various notion of substitution \cite{cattani2005profunctors, Fiorecartesian, FSCD2020}. In particular, the free symmetric monoidal completion comonad on $\Prof$ yields the model of \emph{generalized species of structures} which encompasses Joyal’s combinatorial species and is also a categorification of the relational model of linear logic \cite{Fiorecartesian}. The morphisms in the co-Kleisli bicategory correspond to the notion of \emph{analytic functors} which are generalized power series with quotients \cite{fiore2014analytic}. We can also consider the pseudo-comonads freely adjoining finite coproducts or finite colimits to generalize the category $\Lin$ with the $\vee$-semi-lattice comonad in Section \ref{sec:1fix} or simply the comonad freely adjoining an initial object. We refer the reader to \cite{fiore2018relative} for a general treatment of pseudo-(co)-monads on $\Prof$ and to \cite{OlimpieriLics21} for other examples with applications to intersection typing systems.

The pseudo-comonads we consider verify the necessary conditions on bifree algebras as they are all $\Cat_{\omega}$-enriched and Theorem \ref{thm:bifree2d} has been extended to bicategories and pseudo-functors \cite{BicategorySolRecursiveDomEqCattani2007}. For the colimit-completion cases, it is a straightforward consequence of the commutation of colimits and it only needs to be checked by hand for the free symmetric strict monoidal case.

Strictly speaking, we have only provided the notion of pseudo-fixpoint operators for $2$-categories and not for bicategories. Even if we strictify the bicategory $\Prof$ to its biequivalent $2$-category $\Cocont$ (concontinuous functors between presheaf categories), the comonads we consider are pseudo and their corresponding co-Kleisli are therefore bicategories and not $2$-categories.

For space considerations, we do not give the proof of Theorem \ref{thm:main} for bicategories in this paper but only state that we obtain as a corollary that species with the free symmetric strict monoidal completion, initial object, finite coproduct, finite colimit pseudo-comonads are all instances of this construction and therefore all have a contractible category of pseudo-dinatural fixpoint operators uniform with respect to the inclusion of profunctors into these generalized species.

\subsection{Polynomial functors}

Initial algebras (well-founded trees) and final coalgebras (non-well-founded trees) for polynomial functors have been extensively studied \cite{moerdijk2000wellfounded, CoalgebraicInfTreesAczel2001, gambino2004wellfounded, van2007non} and are standard tools to model (co)inductive types in dependent type theories. If we fix a locally cartesian closed category $\cat{C}$, for objects $I$ and $J$ in $\cat{C}$ a polynomial from $I$ to $J$ is a diagram of shape 
\vspace{-0.1cm}
\begin{center}
	\begin{tikzpicture}[thick,xscale=0.6, yscale=0.7]
		\node (A) at (0.5,0) {$I$};
		\node (B) at (2,1) {$E$};
		\node (C) at (4,1) {$B$};
				\node (D) at (5.5,0) {$J$};
		\draw [->] (B) to node [above] {$p$} (C);
		\draw [->] (B)  to node [above] {$s$} (A);
			\draw [->] (C)  to node [above] {$t$} (D);
	\end{tikzpicture}
\end{center} 
\vspace{-0.1cm}
in the category $\cat{C}$. It induces a polynomial functor $\cat{C}/I \to \cat{C}/J$ between the slice categories which form a $2$-category with $2$-cells given by cartesian transformations. We say that $\cat{C}$ has $\mathbf{W}$-types if all polynomial endofunctors have initial algebras and that it has $\mathbf{M}$-types if they have final coalgebras.

It is well-known that $\mathbf{W}$-types and $\mathbf{M}$-types do not coincide and they are therefore not an instance of the contractibility property of Theorem \ref{thm:main}. This example is therefore an adequate test to verify that the axioms we stated in Section \ref{sec:2fix} are not just valid in the restricted contractible case but provide a general notion of pseudo-fixpoint operators.

The $2$-naturality axiom of the operator computing initial algebras (or final coalgebras) is well known and is sometimes called the ``functoriality property". Dinaturality for the initial algebra or final coalgebra operators is also known in the literature \cite{freyd1991algebraically}. The statement is that for functors $F : \cat{A} \to \cat{B}$, $G: \cat{B} \to \cat{A}$ such that $GF$ and $FG$ have initial algebras, if $GF A \xrightarrow{a} A$ is $GF$-initial, then $FGF A \xrightarrow{ Fa} FA$ is $FG$-initial. To our knowledge, the axioms of a pseudo-dinatural transformation, while straightforward to check, have not been stated explicitly.

Uniformity on the other hand depends on how the initial algebras (or final coalgebras) are computed in the ambiant $2$-category. In general, if initial algebras are obtained as colimit constructions, then natural candidates for the $2$-category of strict maps are functors preserving initial objects or cocontinuous functors whenever these notions are well-defined. Dually, if the pseudo-fixpoint operator is obtained by computing final coalgebras as certain limits then we can consider terminal object preserving functors or continuous functors as strict maps.
In the case of polynomial functors over $\Set$, $\mathbf{W}$-types are uniform with respect to spans \emph{i.e.} polynomial functors of shape
\vspace{-0.5cm}
\begin{center}
	\begin{tikzpicture}[thick,xscale=0.6, yscale=0.7]
		\node (A) at (0.5,0) {$I$};
		\node (B) at (2,1) {$E$};
		\node (C) at (4,1) {$B$};
		\node (D) at (5.5,0) {$J$};
		\draw [->] (B) to node [above] {$\cong$} (C);
		\draw [->] (B)  to node [above] {$s$} (A);
		\draw [->] (C)  to node [above] {$t$} (D);
	\end{tikzpicture}
\end{center} 
\vspace{-0.3cm}
and $\mathbf{M}$-types are uniform with respect to monomials which are polynomial functors of shape
\vspace{-0.1cm}
\begin{center}
	\begin{tikzpicture}[thick,xscale=0.6, yscale=0.7]
		\node (A) at (0.5,0) {$I$};
		\node (B) at (2,1) {$E$};
		\node (C) at (4,1) {$1$};
		\node (D) at (5.5,0) {$J$};
		\draw [->] (B) to node [above] {$\oc$} (C);
		\draw [->] (B)  to node [above] {$s$} (A);
		\draw [->] (C)  to node [above] {$t$} (D);
	\end{tikzpicture}
\end{center} 
\vspace{-0.2cm}
where $1$ is a singleton set.

%% file: appendix.tex
\subsection{Dinatural  and strong dinatural transformations for $2$-categories}

We start by recalling the $1$-categorical notions of dinatural and strong dinatural transformations and proceed with the $2$-categorical generalizations.
\subsubsection{Dinatural transformations}\label{sec:2Dinatdef}

\begin{definition}
	For categories $\cat{C}$, $\cat{D}$ and functors $F,G: \cat{C}^{\op} \times \cat{C} \to \cat{D}$, a \emph{dinatural transformation} $\theta: F \dinato G$ consists of a family of $1$-cells $\{ \theta_c : F(c,c)\to G(c,c) \}_{c \in \bicat{C}}$ indexed by the objects $c$ in $\cat{C}$ such that for every morphism $f: c \to d$ in $\cat{C}$, the following hexagon commutes:
		\begin{center}
			\begin{tikzpicture}[thick, xscale=0.9, yscale=1.1]
				\node (A) at (0.25,1) {$F(d,c)$};
				\node (B) at (2,0) {$F(d,d)$};
				\node (B1) at (4,0) {$G(d,d)$};
				\node (C) at (5.75,1) {$G(c,d)$};
				\node (D) at (2,2) {$F(c,c)$};
				\node (D1) at (4,2) {$G(c,c)$};
				
				\draw [->] (A)  to [bend left =-30] node [below, sloped] {\footnotesize$F(d,f)$} (B);
				\draw [->] (B)  to [bend left =-5] node [below] {$\theta_d$} (B1);
				\draw [->] (D)  to [bend left =5] node [above] {$\theta_c$} (D1);
				\draw [->] (B1)  to [bend left =-30] node [below, sloped] {\footnotesize$G(f,d)$} (C);
				\draw [->] (A) to [bend left =30] node [above, sloped] {\footnotesize$F(f,c)$} (D);
				\draw [->] (D1)  to [bend left =30] node [above, sloped] {\footnotesize$G(c,f)$} (C);
			\end{tikzpicture}
		\end{center}
\end{definition}

\begin{definition}
	For $2$-categories $\bicat{C}$, $\bicat{D}$ and $2$-functors $F,G: \bicat{C}^{\op} \times \bicat{C} \to \bicat{D}$, a \emph{lax dinatural transformation} $\theta: F \dinato G$ consists of:
	\begin{itemize}
		\item a family of $1$-cells $\{ \theta_c : F(c,c)\to G(c,c) \}_{c \in \bicat{C}}$ indexed by the objects $c$ in $\bicat{C}$;
		\item for every $1$-cell $f: c \to d$ in $\bicat{C}$, a $2$-cell:
		\begin{center}
			\begin{tikzpicture}[thick, yscale=1.1]
				\node (A) at (0.25,1) {$F(d,c)$};
				\node (B) at (2,0) {$F(d,d)$};
				\node (B1) at (4,0) {$G(d,d)$};
				\node (C) at (5.75,1) {$G(c,d)$};
				\node (D) at (2,2) {$F(c,c)$};
				\node (D1) at (4,2) {$G(c,c)$};
				
				\draw [->] (A)  to [bend left =-30] node [below, sloped] {\footnotesize$F(d,f)$} (B);
				\draw [->] (B)  to [bend left =-5] node [below] {$\theta_d$} (B1);
				\draw [->] (D)  to [bend left =5] node [above] {$\theta_c$} (D1);
				\draw [->] (B1)  to [bend left =-30] node [below, sloped] {\footnotesize$G(f,d)$} (C);
				\draw [->] (A) to [bend left =30] node [above, sloped] {\footnotesize$F(f,c)$} (D);
				\draw [->] (D1)  to [bend left =30] node [above, sloped] {\footnotesize$G(c,f)$} (C);
				\node at (3,1) {$\Downarrow \theta_{f}$};
			\end{tikzpicture}
		\end{center}
	\end{itemize}
	satisfying the following axioms:
		\begin{enumerate}
			\item \emph{unity:} for every object $c \in \bicat{C}$, $\theta_{1_c}=\id_{\theta_c}$,
			\item \emph{$1$-naturality:} for every $1$-cells $f: c\to d$ and $g: d \to e$ in $\bicat{C}$,
			\begin{center}
				\begin{tikzpicture}[thick, xscale=1.1, yscale=1.2]
					\node at (-3,0) {$\theta_{gf}\quad =$};
					\node (A0) at (-1.5,0) {$F(e,c)$};
					\node (A) at (0.25,1) {$F(d,c)$};
					\node (A1) at (0.25,-1) {$F(e,d)$};
					\node (B) at (2,0) {$F(d,d)$};
					\node (B1) at (4,0) {$G(d,d)$};
					\node (B2) at (4,-2) {$G(e,e)$};
					\node (C) at (5.75,1) {$G(c,d)$};
					\node (C1) at (5.75,-1) {$G(d,e)$};
					\node (C2) at (2,-2) {$F(e,e)$};
					\node (E) at (7.5,0) {$G(c,e)$};
					\node (D) at (2,2) {$F(c,c)$};
					\node (D1) at (4,2) {$G(c,c)$};
					
					\draw [->] (A)  to [bend left =20] node [above, sloped] {\footnotesize$F(d,f)$} (B);
					\draw [->] (A1)  to [bend left =-30] node [below, sloped] {\footnotesize$F(e,g)$} (C2);
					\draw [->] (B)  to [bend left =-5] node [below] {$\theta_d$} (B1);
					\draw [->] (C2)  to [bend left =-5] node [below] {$\theta_e$} (B2);
					\draw [->] (B2)  to [bend left =-30] node [below, sloped] {\footnotesize$G(g,e)$} (C1);
					\draw [->] (D)  to [bend left =5] node [above] {$\theta_c$} (D1);
					\draw [->] (B1)  to [bend left =20] node [above, sloped] {\footnotesize$G(f,d)$} (C);
					\draw [->] (A0)  to [bend left =20] node [above, sloped] {\footnotesize$F(g,c)$} (A);
					\draw [->] (C)  to [bend left =20] node [above, sloped] {\footnotesize$G(c,g)$} (E);
					\draw [->] (B1)  to [bend left =-20] node [below, sloped] {\footnotesize$G(d,g)$} (C1);
					\draw [->] (A0)  to [bend left =-20] node [below, sloped] {\footnotesize$F(e,f)$} (A1);
					\draw [->] (C1)  to [bend left =-20] node [below, sloped] {\footnotesize$G(f,e)$} (E);
					\draw [->] (A1)  to [bend left =-20] node [below, sloped] {\footnotesize$F(g,d)$} (B);
					\draw [->] (A) to [bend left =30] node [above, sloped] {\footnotesize$F(f,c)$} (D);					
					\draw [->] (D1)  to [bend left =30] node [above, sloped] {\footnotesize$G(c,f)$} (C);
					\node at (3,1.2) {$\Downarrow \theta_{f}$};		
					\node at (3,-1.2) {$\Downarrow \theta_{g}$};
					\node at (5.75,0) {$=$};
					\node at (0.25,0) {$=$};
				\end{tikzpicture}
			\end{center}
			\item \emph{$2$-naturality:} for every $2$-cell $\alpha: f\Rightarrow f'$ in $\bicat{C}$:
			\begin{center}
				\begin{tikzpicture}[thick, yscale=1.3,xscale=1.1]
					\node (A) at (0.25,1) {$F(d,c)$};
					\node (B) at (2,0) {$F(d,d)$};
					\node (B1) at (4,0) {$G(d,d)$};
					\node (C) at (5.75,1) {$G(c,d)$};
					\node (D) at (2,2) {$F(c,c)$};
					\node (D1) at (4,2) {$G(c,c)$};
					
					\draw [->] (A)  to [bend left =-30] node [below, sloped] {\footnotesize$F(d,f')$} (B);
					\draw [->] (B)  to [bend left =-5] node [below] {$\theta_d$} (B1);
					\draw [->] (D)  to [bend left =5] node [above] {$\theta_c$} (D1);
					\draw [->] (B1)  to [bend left =-30] node [below, sloped] {\footnotesize$G(f',d)$} (C);
					\draw [->] (A) to [bend left =35] node [above, sloped] {\footnotesize$F(f,c)$} (D);
					\draw [->] (A) to [bend left =-35] node [below, sloped] {\footnotesize$F(f',c)$} (D);
					\draw [->] (D1)  to [bend left =35] node [above, sloped] {\footnotesize$G(c,f)$} (C);
					\draw [->] (D1)  to [bend left =-35] node [below, sloped] {\footnotesize$G(c,f')$} (C);
					\node at (3,1) {$\Downarrow \theta_{f'}$};
					\node[rotate=-45] at (1.2,1.6) {\footnotesize$ F(\alpha,c)$};
					\node at (0.95,1.45) {\footnotesize$\Searrow $};
					\node at (4.95,1.45) {\footnotesize$\Swarrow $};
					\node[rotate=45] at (4.75,1.6) {\footnotesize$G(c, \alpha)$};
					\node at (6.75,1) {$=$};
					\begin{scope}[xshift=7.5cm]
						\node (A) at (0.25,1) {$F(d,c)$};
						\node (B) at (2,0) {$F(d,d)$};
						\node (B1) at (4,0) {$G(d,d)$};
						\node (C) at (5.75,1) {$G(c,d)$};
						\node (D) at (2,2) {$F(c,c)$};
						\node (D1) at (4,2) {$G(c,c)$};
						
						\draw [->] (A)  to [bend left =-35] node [below, sloped] {\footnotesize$F(d,f')$} (B);
						\draw [->] (B)  to [bend left =-5] node [below] {$\theta_d$} (B1);
						\draw [->] (D)  to [bend left =5] node [above] {$\theta_c$} (D1);
						\draw [->] (B1)  to [bend left =-35] node [below, sloped] {\footnotesize$G(f',d)$} (C);
						\draw [->] (A) to [bend left =30] node [above, sloped] {\footnotesize$F(f,c)$} (D);
						\draw [->] (A) to [bend left =35] node [above, sloped] {\footnotesize$F(d,f)$} (B);
						\draw [->] (D1)  to [bend left =30] node [above, sloped] {\footnotesize$G(c,f)$} (C);
						\draw [->] (B1)  to [bend left =35] node [above, sloped] {\footnotesize$G(f,d)$} (C);
						\node at (3,1) {$\Downarrow \theta_{f}$};
						\node[rotate=-45] at (4.95,0.6) {\footnotesize$ G(\alpha,d)$};
						\node at (4.75,0.45) {\footnotesize$\Searrow $};
						\node at (1.25,0.45) {\footnotesize$\Swarrow $};
						\node[rotate=45] at (1,0.6) {\footnotesize$F(d, \alpha)$};
						
					\end{scope}
				\end{tikzpicture}
			\end{center}
		\end{enumerate}
	
\end{definition}

For an \emph{oplax dinatural transformation}, the $2$-cells $\theta_f$ go in the opposite direction. When the $2$-cells $\theta_f$ are invertible, we obtain the notion of \emph{pseudo dinatural transformation} and when they are strict identities, we obtain \emph{strict dinatural transformations}.

\begin{definition}
	A \emph{modification between lax dinatural transformation} $\Phi : \theta \Rrightarrow
	\delta : F \dinato G: \bicat{C}^{\op} \times \bicat{C} \to \bicat{D}$ consists of a family of $2$-cells $\{\Phi_c : \theta_c \Rightarrow \delta_c \}_{c \in \bicat{C}}$ such that for every $1$-cell $f : c\to d$ in $\bicat{C}$ the following equality holds:
		\begin{center}
			\begin{tikzpicture}[thick, yscale=1.1, xscale=1]
				\node (A) at (0.25,1) {$F(d,c)$};
				\node (B) at (2,0) {$F(d,d)$};
				\node (B1) at (4,0) {$G(d,d)$};
				\node (C) at (5.75,1) {$G(c,d)$};
				\node (D) at (2,2) {$F(c,c)$};
				\node (D1) at (4,2) {$G(c,c)$};
				
				\draw [->] (A)  to [bend left =-30] node [below, sloped] {\footnotesize$F(d,f)$} (B);
				\draw [->] (B)  to [bend left =-35] node [below] {$\delta_d$} (B1);
				\draw [->] (B)  to [bend left =35] node [above] {$\theta_d$} (B1);
				\draw [->] (D)  to [bend left =5] node [above] {$\theta_c$} (D1);
				\draw [->] (B1)  to [bend left =-30] node [below, sloped] {\footnotesize$G(f,d)$} (C);
				\draw [->] (A) to [bend left =30] node [above, sloped] {\footnotesize$F(f,c)$} (D);
				\draw [->] (D1)  to [bend left =30] node [above, sloped] {\footnotesize$G(c,f)$} (C);
				\node at (3,1.3) {$\Downarrow \theta_{f}$};
				\node at (3,0) {$\Downarrow \Phi_d$};
				\node at (7,1) {$=$};
				
				\begin{scope}[xshift=8cm]
					\node (A) at (0.25,1) {$F(d,c)$};
					\node (B) at (2,0) {$F(d,d)$};
					\node (B1) at (4,0) {$G(d,d)$};
					\node (C) at (5.75,1) {$G(c,d)$};
					\node (D) at (2,2) {$F(c,c)$};
					\node (D1) at (4,2) {$G(c,c)$};
					
					\draw [->] (A)  to [bend left =-30] node [below, sloped] {\footnotesize$F(d,f)$} (B);
					\draw [->] (B)  to [bend left =-5] node [below] {$\delta_d$} (B1);
					\draw [->] (D)  to [bend left =35] node [above] {$\theta_c$} (D1);
					\draw [->] (D)  to [bend left =-35] node [below] {$\delta _c$} (D1);
					\draw [->] (B1)  to [bend left =-30] node [below, sloped] {\footnotesize$G(f,d)$} (C);
					\draw [->] (A) to [bend left =30] node [above, sloped] {\footnotesize$F(f,c)$} (D);
					\draw [->] (D1)  to [bend left =30] node [above, sloped] {\footnotesize$G(c,f)$} (C);
					\node at (3,0.7) {$\Downarrow \delta _{f}$};
					\node at (3,2) {$\Downarrow \Phi_c$};
				\end{scope}
			\end{tikzpicture}
		\end{center}
\end{definition}
Similarly to the $1$-categorical case, we cannot compose lax dinatural transformations horizontally and therefore the following data
\begin{itemize}
	\item $0$-cells: mixed variance $2$-functors $F,G :  \bicat{C}^{\op} \times \bicat{C} \to \bicat{D}$;
	\item $1$-cells: lax dinatural transformations $\theta : F \dinato G$;
	\item $2$-cells: modifications $\Phi: \theta \Rrightarrow \delta$ between them.
\end{itemize}
does not constitute a $2$-category and we need the notion of strong lax dinatural transformation to make it compositional.\\

\subsubsection{Strong dinatural transformations}\label{sec:Strong2Dinatdef}

\begin{definition}
		For categories $\cat{C}$, $\cat{D}$ and functors $F,G: \cat{C}^{\op} \times \cat{C} \to \cat{D}$, a \emph{strong dinatural transformation} $\gamma: F \stdinato G$ consists of a family of $1$-cells $\{ \gamma_c : F(c,c)\to G(c,c) \}_{c \in \cat{C}}$ indexed by the objects $c$ in $\cat{C}$ such that for every morphism $f: c \to d$ in $\cat{C}$ and for every span $F(c,c) \xleftarrow{q_c} Q \xrightarrow{q_d} F(d,d)$ in $\cat{D}$, if the square
		\begin{center}
			\begin{tikzpicture}[thick, xscale=0.9]
				\node (A) at (0,1) {$Q$};
				\node (B) at (2,0) {$F(d,d)$};
				\node (C) at (4,1) {$F(c,d)$};
				\node (D) at (2,2) {$F(c,c)$};
				
				\draw [->] (A)  to [bend left =-30] node [below] {$q_d$} (B);
				\draw [->] (B)  to [bend left =-30] node [below, sloped] {\footnotesize$F(f,d)$} (C);
				\draw [->] (A) to [bend left =30] node [above] {$q_c$} (D);
				\draw [->] (D)  to [bend left =30] node [above, sloped] {\footnotesize$F(c,f)$} (C);
			\end{tikzpicture}
		\end{center}
		commutes, then so does the hexagon:
			\begin{center}
			\begin{tikzpicture}[thick]
				\node (A) at (0.25,1) {$Q$};
				\node (B) at (2,0) {$F(d,d)$};
				\node (B1) at (4,0) {$G(d,d)$};
				\node (C) at (5.75,1) {$G(c,d)$};
				\node (D) at (2,2) {$F(c,c)$};
				\node (D1) at (4,2) {$G(c,c)$};
				
				\draw [->] (A)  to [bend left =-30] node [below] {$q_d$} (B);
				\draw [->] (B)  to [bend left =-5] node [below] {$\gamma_d$} (B1);
				\draw [->] (D)  to [bend left =5] node [above] {$\gamma_c$} (D1);
				\draw [->] (B1)  to [bend left =-30] node [below, sloped] {\footnotesize$G(f,d)$} (C);
				\draw [->] (A) to [bend left =30] node [above] {$q_c$} (D);
				\draw [->] (D1)  to [bend left =30] node [above, sloped] {\footnotesize$G(c,f)$} (C);

			\end{tikzpicture}
		\end{center}
\end{definition}

For functors $F, G, H : \cat{C}^\op \times \cat{C} \to \cat{D}$ and strong dinatural transformations $\gamma: F \dinato G$ and $\delta : G \dinato H$, the transformation 
\[
\{ \delta_c \circ \gamma_c : F(c,c) \to H(c,c) \}_{c\in \cat{C}}
\]
is also strongly dinatural so that strong dinatural transformations form a category.
To obtain the $2$-dimensional analogue, we first need to consider the notion of lax wedge~\cite{bozapalides1976theorie,bozapalides1977fins, climent20102, corner2017universal, hirata2022notes}:
 
\begin{definition}[lax wedge]
	Let $\bicat{C}, \bicat{D}$ be $2$-categories and $F : \bicat{C}^{\op} \times \bicat{C} \to \bicat{D}$ be a $2$-functor. For an object $Q$ of $\bicat{D}$, a \emph{lax wedge of $Q$ over $F$} consists of:
	\begin{itemize}
		\item a family of $1$-cells $\{ q_c : Q \to F(c,c) \}_{c \in \bicat{C}}$ indexed by the objects $c$ in $\bicat{C}$;
		\item for every $1$-cell $f: c \to d$ in $\bicat{C}$, a $2$-cell:
		\begin{center}
			\begin{tikzpicture}[thick]
				\node (A) at (0,1) {$Q$};
				\node (B) at (2,0) {$F(d,d)$};
				\node (C) at (4,1) {$F(c,d)$};
				\node (D) at (2,2) {$F(c,c)$};
				
				\draw [->] (A)  to [bend left =-30] node [below] {$q_d$} (B);
				\draw [->] (B)  to [bend left =-30] node [below, sloped] {\footnotesize$F(f,d)$} (C);
				\draw [->] (A) to [bend left =30] node [above] {$q_c$} (D);
				\draw [->] (D)  to [bend left =30] node [above, sloped] {\footnotesize$F(c,f)$} (C);
				\node at (2,1) {$\Downarrow q_f$};
				
			\end{tikzpicture}
		\end{center}
	\end{itemize}
	satisfying the following axioms:
	\begin{enumerate}
		\item unity: $q_{1_c}= \id_{q_c}$
		\item $1$-naturality: for $1$-cells $f : c\to d$ and $g : d \to e$,
		\begin{center}
			\begin{tikzpicture}[thick=,yscale=1.1,xscale=1.3]
				\node (A) at (0,1.5) {$Q$};
				\node (B) at (1.5,0.5) {$F(e,e)$};
				\node (C) at (3,1.5) {$F(c,e)$};
				\node (D) at (1.5,2.5) {$F(c,c)$};
				
				\draw [->] (A)  to [bend left =-30] node [below] {$q_e$} (B);
				\draw [->] (B)  to [bend left =-30] node [below,sloped] {\footnotesize$F(gf,d)$} (C);
				\draw [->] (A) to [bend left =30] node [above] {$q_c$} (D);
				\draw [->] (D)  to [bend left =30] node [above,sloped] {\footnotesize$F(c,gf)$} (C);
				\node at (1.5,1.5) {$\Downarrow q_{gf}$};
				\node at (4,1.5) {$=$};
				
				\begin{scope}[xshift = 5 cm]
					\node (B) at (3,-0.2) {$F(d,e)$};
					\node (B1) at (4. 5,1.5) {$F(c,e)$};
					\node (A) at (0,1.5) {$Q$};
					\node (C) at (1.25,0.2) {$F(e,e)$};
					\node (D) at (2,1.5) {$F(d,d)$};
					\node (D1) at (3,3.2) {$F(c,d)$};
					\node (C1) at (1.25,2.8) {$F(c,c)$};
					
					\draw [->] (A)  to [bend left =-10] node [left] {$q_e$} (C);
					\draw [->] (C)  to [bend left =-10] node [below,sloped] {\footnotesize$F(g,e)$} (B);
					\draw [->] (B)  to [bend left =-10] node [below,sloped] {\footnotesize$F(f,e)$} (B1);
					\draw [->] (A)  to  node [fill=white] {$q_d$} (D);
					\draw [->] (D)  to [bend left =-10] node [below,sloped] {\footnotesize$F(f,d)$} (D1);
					\draw [->] (A) to [bend left =10] node [left] {$q_c$} (C1);
					\draw [->] (C1) to [bend left =10] node [above,sloped] {\footnotesize$F(c,f)$} (D1);
					\draw [->] (D)  to [bend left =10] node [above,sloped] {\footnotesize$F(d,g)$} (B);
					\draw [->] (D1)  to [bend left =10] node [above,sloped] {\footnotesize$F(c,g)$} (B1);
					\node at (1.5,0.8) {$\Downarrow q_g$};
					\node at (1.5,2.2) {$\Downarrow q_f$};
					
				\end{scope}
			\end{tikzpicture}
		\end{center}
		\item $2$-naturality: for every $2$-cell $\alpha : f\Rightarrow f'$,
		\begin{center}
			\begin{tikzpicture}[thick, yscale=1.2, xscale=1.2 ]
				\node (A) at (0,1) {$Q$};
				\node (B) at (2,0) {$F(d,d)$};
				\node (C) at (4,1) {$F(c,d)$};
				\node (D) at (2,2) {$F(c,c)$};
				
				\draw [->] (A)  to [bend left =-30] node [below] {$q_d$} (B);
				\draw [->] (B)  to [bend left =-30] node [below, sloped] {\footnotesize$F(f',d)$} (C);
				\draw [->] (A) to [bend left =30] node [above] {$q_c$} (D);
				\draw [->] (D)  to [bend left =35] node [above, sloped] {\footnotesize$F(c,f)$} (C);
				\draw [->] (D)  to [bend left =-30] node [below, sloped] {\footnotesize$F(c,f')$} (C);
				\node at (2,0.7) {$\Downarrow q_{f'}$};
				\node at (3.2,1.35) {\footnotesize$\Swarrow $};
				\node[rotate=45] at (3,1.55) {\footnotesize$F(c, \alpha)$};
				\node at (5,1) {$=$};
				
				\begin{scope}[xshift = 5.5 cm]
					\node (A) at (0,1) {$Q$};
					\node (B) at (2,0) {$F(d,d)$};
					\node (C) at (4,1) {$F(c,d)$};
					\node (D) at (2,2) {$F(c,c)$};
					
					\draw [->] (A)  to [bend left =-30] node [below] {$q_d$} (B);
					\draw [->] (B)  to [bend left =-35] node [below, sloped] {\footnotesize$F(f',d)$} (C);
					\draw [->] (B)  to [bend left =30] node [above, sloped] {\footnotesize$F(f,d)$} (C);
					\draw [->] (A) to [bend left =30] node [above] {$q_c$} (D);
					\draw [->] (D)  to [bend left =30] node [above, sloped] {\footnotesize$F(c,f)$} (C);
					\node[rotate=-45] at (3,0.5) {\footnotesize$ F(\alpha,d)$};
					\node at (2.8,0.3) {\footnotesize$\Searrow $};
					\node at (2,1.3) {$\Downarrow q_f$};
				\end{scope}
			\end{tikzpicture}
		\end{center}
	\end{enumerate}
\end{definition}

\begin{definition}[morphism of lax wedges]
	For two lax conwedges $(Q,q)$ and $(P,p)$ over $F: \bicat{C}^{\op} \times \bicat{C} \to \bicat{D}$, a lax morphism from $(P,p)$ to $(Q,q)$ consists of 
	\begin{itemize}
		\item a $1$-cell $u: P \to Q$ in $\bicat{D}$,
		\item a family of $2$-cells $\{\Gamma_c : p_c \to q_c u\}_{c\in \bicat{C}}$ indexed by the objects $c$ in $\bicat{C}$                                                                         
	\end{itemize}
	such that for all $f : c\to d$ in $\bicat{C}$, the following two diagrams are equal:
	
	\begin{center}
		\begin{tikzpicture}[thick, scale=0.75]
			\node (P) at (-1.5,1) {$P$};
			\node (A) at (0,1) {$Q$};
			\node (B) at (2,-0.5) {$F(d,d)$};
			\node (C) at (5,1) {$F(c,d)$};
			\node (D) at (2,2.5) {$F(c,c)$};

			\draw [->] (P)  to [bend left =33] node [above left] {$p_c$} (D);
			\draw [->] (P)  to node [below] {$u$} (A);
			\draw [->] (A)  to [bend left =-30] node [left] {$q_d$} (B);
			\draw [->] (B)  to [bend left =-30] node [below, sloped] {\footnotesize$F(f,d)$} (C);
			\draw [->] (A) to [bend left =30] node [below right] {$q_c$} (D);
			\draw [->] (D)  to [bend left =30] node [above, sloped] {\footnotesize$F(c,f)$} (C);
			\node at (2,1) {$\Downarrow q_f$};
			\node at (-0.4,1.6) {$\Searrow \Gamma_c$};
			
			\node at (6.5,1) {$=$};
			
			\begin{scope}[xshift=9cm]
				\node (P) at (-1.5,1) {$P$};
				\node (A) at (0,-0.25) {$Q$};
				\node (B) at (2,-0.5) {$F(d,d)$};
				\node (C) at (5,1) {$F(c,d)$};
				\node (D) at (2,2.5) {$F(c,c)$};
				
				\draw [->] (P)  to [bend left =25] node [above] {$p_d$} (B);
				\draw [->] (P)  to [bend left =25] node [above] {$p_c$} (D);
				\draw [->] (P)  to [bend left =-10] node [left] {$u$} (A);
				\draw [->] (A)  to [bend left =-10] node [below] {$q_d$} (B);
				\draw [->] (B)  to [bend left =-25] node [below, sloped] {\footnotesize$F(f,d)$} (C);
				\draw [->] (D)  to [bend left =25] node [above, sloped] {\footnotesize$F(c,f)$} (C);
				\node at (2,1) {$\Downarrow p_f$};
				\node at (0.25,0.25) {$\Swarrow \Gamma_d$};
			\end{scope}
		\end{tikzpicture}
	\end{center}
	
\end{definition}

\begin{definition}[strong lax dinatural transformation]
	For $2$-categories $\bicat{C}$, $\bicat{D}$ and $2$-functors $F,G: \bicat{C}^{\op} \times \bicat{C} \to \bicat{D}$, a \emph{strong lax dinatural transformation} $\gamma: F 
\stdinato G$ consists of:
	\begin{itemize}
		\item a family of $1$-cells $\{ \gamma_c : F(c,c)\to G(c,c) \}_{c \in \bicat{C}}$ indexed by the objects $c$ in $\bicat{C}$;
		\item for every lax wedge $(Q,q)$ over $F$ and for every $1$-cell $f: c \to d$ in $\bicat{C}$, a $2$-cell:
		\begin{center}
			\begin{tikzpicture}[thick]
				\node (A) at (0.25,1) {$Q$};
				\node (B) at (2,0) {$F(d,d)$};
				\node (B1) at (4,0) {$G(d,d)$};
				\node (C) at (5.75,1) {$G(c,d)$};
				\node (D) at (2,2) {$F(c,c)$};
				\node (D1) at (4,2) {$G(c,c)$};
				
				\draw [->] (A)  to [bend left =-30] node [below] {$q_d$} (B);
				\draw [->] (B)  to [bend left =-5] node [below] {$\gamma_d$} (B1);
				\draw [->] (D)  to [bend left =5] node [above] {$\gamma_c$} (D1);
				\draw [->] (B1)  to [bend left =-30] node [below, sloped] {\footnotesize$G(f,d)$} (C);
				\draw [->] (A) to [bend left =30] node [above] {$q_c$} (D);
				\draw [->] (D1)  to [bend left =30] node [above, sloped] {\footnotesize$G(c,f)$} (C);
				\node at (3,1) {$\Downarrow \gamma_{q,f}$};
				
			\end{tikzpicture}
		\end{center}
	\end{itemize}
	satisfying the following axioms:
	
	\begin{enumerate}
		\item unity: for every object $c \in \bicat{C}$, $\gamma_{q,1_c}=\id_{\gamma_c \circ q_c}$,
		\item $1$-naturality: for every $1$-cells $f: c\to d$ and $g: d \to e$,
		\begin{center}
			\begin{tikzpicture}[thick, xscale=1.1, yscale=1.2]
				\node at (-3,0) {$\gamma_{q,gf}\quad =$};
				\node (A0) at (-1,0) {$Q$};

				\node (B) at (2,0) {$F(d,d)$};
				\node (B1) at (4,0) {$G(d,d)$};
				\node (B2) at (4,-2) {$G(e,e)$};
				\node (C) at (5.75,1) {$G(c,d)$};
				\node (C1) at (5.75,-1) {$G(d,e)$};
				\node (C2) at (2,-2) {$F(e,e)$};
				\node (E) at (7.5,0) {$G(c,e)$};
				\node (D) at (2,2) {$F(c,c)$};
				\node (D1) at (4,2) {$G(c,c)$};

				\draw [->] (A0)  to [bend left =-5] node [below] {$q_d$} (B);
				\draw [->] (B)  to [bend left =-5] node [below] {$\gamma_d$} (B1);
				\draw [->] (C2)  to [bend left =-5] node [below] {$\gamma_e$} (B2);
				\draw [->] (B2)  to [bend left =-30] node [below, sloped] {\footnotesize$G(g,e)$} (C1);
				\draw [->] (D)  to [bend left =5] node [above] {$\gamma_c$} (D1);
				\draw [->] (B1)  to [bend left =20] node [above, sloped] {\footnotesize$G(f,d)$} (C);
				\draw [->] (A0)  to [bend left =20] node [above left] {$q_c$} (D);
				\draw [->] (C)  to [bend left =20] node [above, sloped] {\footnotesize$G(c,g)$} (E);
				\draw [->] (B1)  to [bend left =-20] node [below, sloped] {\footnotesize$G(d,g)$} (C1);
				\draw [->] (A0)  to [bend left =-20] node [below left] {$q_e$} (C2);
				\draw [->] (C1)  to [bend left =-20] node [below, sloped] {\footnotesize$G(f,e)$} (E);

				\draw [->] (D1)  to [bend left =30] node [above, sloped] {\footnotesize$G(c,f)$} (C);
				\node at (3,1.2) {$\Downarrow \gamma_{q,f}$};		
				\node at (3,-1.2) {$\Downarrow \gamma_{q,g}$};
				\node at (5.75,0) {$=$};
				
			\end{tikzpicture}
		\end{center}
		\item $2$-naturality: for every $2$-cell $\alpha: f\Rightarrow f'$ in $\bicat{C}$:
		\begin{center}
			\begin{tikzpicture}[thick, yscale=1.3,xscale=1.1]
				\node (A) at (0.25,1) {$Q$};
				\node (B) at (2,0) {$F(d,d)$};
				\node (B1) at (4,0) {$G(d,d)$};
				\node (C) at (5.75,1) {$G(c,d)$};
				\node (D) at (2,2) {$F(c,c)$};
				\node (D1) at (4,2) {$G(c,c)$};
				
				\draw [->] (A)  to [bend left =-30] node [below left] {$q_d$} (B);
				\draw [->] (B)  to [bend left =-5] node [below] {$\gamma_d$} (B1);
				\draw [->] (D)  to [bend left =5] node [above] {$\gamma_c$} (D1);
				\draw [->] (B1)  to [bend left =-30] node [below, sloped] {\footnotesize$G(f',d)$} (C);
				\draw [->] (A) to [bend left =30] node [above, sloped] {$q_c$} (D);
				
				\draw [->] (D1)  to [bend left =35] node [above, sloped] {\footnotesize$G(c,f)$} (C);
				\draw [->] (D1)  to [bend left =-35] node [below, sloped] {\footnotesize$G(c,f')$} (C);
				\node at (3,1) {$\Downarrow \gamma_{q,f'}$};
				\node at (4.95,1.45) {\footnotesize$\Swarrow $};
				\node[rotate=45] at (4.75,1.6) {\footnotesize$G(c, \alpha)$};
				\node at (6.75,1) {$=$};
				\begin{scope}[xshift=7.5cm]
					\node (A) at (0.25,1) {$Q$};
					\node (B) at (2,0) {$F(d,d)$};
					\node (B1) at (4,0) {$G(d,d)$};
					\node (C) at (5.75,1) {$G(c,d)$};
					\node (D) at (2,2) {$F(c,c)$};
					\node (D1) at (4,2) {$G(c,c)$};
					
					\draw [->] (A)  to [bend left =-30] node [left] {$q_d$} (B);
					\draw [->] (B)  to [bend left =-5] node [below] {$\gamma_d$} (B1);
					\draw [->] (D)  to [bend left =5] node [above] {$\gamma_c$} (D1);
					\draw [->] (B1)  to [bend left =-35] node [below, sloped] {\footnotesize$G(f',d)$} (C);
					\draw [->] (A) to [bend left =30] node [ left] {$q_c$} (D);
					
					\draw [->] (D1)  to [bend left =30] node [above, sloped] {\footnotesize$G(c,f)$} (C);
					\draw [->] (B1)  to [bend left =35] node [above, sloped] {\footnotesize$G(f,d)$} (C);
					\node at (3,1) {$\Downarrow \gamma_{q,f}$};

					\node[rotate=-45] at (4.95,0.6) {\footnotesize$ G(\alpha,d)$};
					\node at (4.75,0.45) {\footnotesize$\Searrow $};
					
				\end{scope}
			\end{tikzpicture}
		\end{center}
		\item for every morphism of lax wedges $(u, \Gamma): (P,p) \to (Q,q)$ over $F$ and for every $f : c \to d$, 
		\begin{center}
			\begin{tikzpicture}[thick, yscale=1.1, xscale=0.9]
				\node (A0) at (-1,1) {$P$};
				\node (A) at (0.25,1) {$Q$};
				\node (B) at (2,0) {$F(d,d)$};
				\node (B1) at (4,0) {$G(d,d)$};
				\node (C) at (5.75,1) {$G(c,d)$};
				\node (D) at (2,2) {$F(c,c)$};
				\node (D1) at (4,2) {$G(c,c)$};
				
				\draw [->] (A)  to [bend left =-30] node [below] {$q_d$} (B);
				
				\draw [->] (B)  to [bend left =-5] node [below] {$\gamma_d$} (B1);
				\draw [->] (D)  to [bend left =5] node [above] {$\gamma_c$} (D1);
				\draw [->] (B1)  to [bend left =-30] node [below, sloped] {\footnotesize$G(f,d)$} (C);
				\draw [->] (A) to [bend left =10] node [above] {$q_c$} (D);
				\draw [->] (A0) to [bend left =35] node [above] {$p_c$} (D);
				\draw [->] (A0) to node [below] {$u$} (A);
				\draw [->] (D1)  to [bend left =30] node [above, sloped] {\footnotesize$G(c,f)$} (C);
				\node at (3,1) {$\Downarrow \gamma_{q,f}$};
				\node at (0,1.5) {$\Searrow \Gamma_c$};
				\node at (7,1) {$=$};
				
				\begin{scope}[xshift=9cm]
					\node at (0.2,0.5) {$\Swarrow \Gamma_d$};
					\node (A0) at (-1,1) {$P$};
					\node (A) at (0.25,0) {$Q$};
					\node (B) at (2,0) {$F(d,d)$};
					\node (B1) at (4,0) {$G(d,d)$};
					\node (C) at (5.75,1) {$G(c,d)$};
					\node (D) at (2,2) {$F(c,c)$};
					\node (D1) at (4,2) {$G(c,c)$};
					\draw [->] (A0)  to [bend left =-25] node [below] {$u$} (A);
					\draw [->] (A)  to [bend left =-5] node [below] {$q_d$} (B);
					\draw [->] (A0)  to [bend left =20] node [above] {$p_d$} (B);
					\draw [->] (B)  to [bend left =-5] node [below] {$\gamma_d$} (B1);
					\draw [->] (D)  to [bend left =5] node [above] {$\gamma_c$} (D1);
					\draw [->] (B1)  to [bend left =-30] node [below, sloped] {\footnotesize$G(f,d)$} (C);
					\draw [->] (A0) to [bend left =20] node [above] {$p_c$} (D);
					\draw [->] (D1)  to [bend left =30] node [above, sloped] {\footnotesize$G(c,f)$} (C);
					\node at (3,1) {$\Downarrow \gamma_{p,f}$};
					
				\end{scope}
			\end{tikzpicture}
		\end{center}
	\end{enumerate}
	
\end{definition}

\begin{definition}
	A \emph{modification between strong lax dinatural transformation} $\Phi : \gamma \Rrightarrow
	\delta : F \stdinato G$ consists of a family of $2$-cells $\{\Phi_c : \gamma_c \Rightarrow \delta_c \}_{c \in \bicat{C}}$ such that for every lax wedge $(Q,q)$ over $F$ and $1$-cell $f : c\to d$ in $\bicat{C}$ the following equality holds:
	
	\begin{center}
		\begin{tikzpicture}[thick]
			\node (A) at (0.25,1) {$Q$};
			\node (B) at (2,0) {$F(d,d)$};
			\node (B1) at (4,0) {$G(d,d)$};
			\node (C) at (5.75,1) {$G(c,d)$};
			\node (D) at (2,2) {$F(c,c)$};
			\node (D1) at (4,2) {$G(c,c)$};
			
			\draw [->] (A)  to [bend left =-30] node [below] {$q_d$} (B);
			\draw [->] (B)  to [bend left =-35] node [below] {$\delta_d$} (B1);
			\draw [->] (B)  to [bend left =35] node [above] {$\gamma_d$} (B1);
			\draw [->] (D)  to [bend left =5] node [above] {$\gamma_c$} (D1);
			\draw [->] (B1)  to [bend left =-30] node [below, sloped] {\footnotesize$G(f,d)$} (C);
			\draw [->] (A) to [bend left =30] node [above] {$q_c$} (D);
			\draw [->] (D1)  to [bend left =30] node [above, sloped] {\footnotesize$G(c,f)$} (C);
			\node at (3,1.3) {$\Downarrow \gamma_{q,f}$};
			\node at (3,0) {$\Downarrow \Phi_d$};
			\node at (6.75,1) {$=$};
			
			\begin{scope}[xshift=7.25cm]
				\node (A) at (0.25,1) {$Q$};
				\node (B) at (2,0) {$F(d,d)$};
				\node (B1) at (4,0) {$G(d,d)$};
				\node (C) at (5.75,1) {$G(c,d)$};
				\node (D) at (2,2) {$F(c,c)$};
				\node (D1) at (4,2) {$G(c,c)$};
				
				\draw [->] (A)  to [bend left =-30] node [below] {$q_d$} (B);
				\draw [->] (B)  to [bend left =-5] node [below] {$\gamma_d$} (B1);
				\draw [->] (D)  to [bend left =35] node [above] {$\gamma_c$} (D1);
				\draw [->] (D)  to [bend left =-35] node [below] {$\delta_c$} (D1);
				\draw [->] (B1)  to [bend left =-30] node [below, sloped] {\footnotesize$G(f,d)$} (C);
				\draw [->] (A) to [bend left =30] node [above] {$q_c$} (D);
				\draw [->] (D1)  to [bend left =30] node [above, sloped] {\footnotesize$G(c,f)$} (C);
				\node at (3,0.7) {$\Downarrow \delta_{q,f}$};
				\node at (3,2) {$\Downarrow \Phi_c$};
			\end{scope}
		\end{tikzpicture}
	\end{center}
	
\end{definition}

\begin{definition}
	For $2$-categories $\bicat{C}$ and $\bicat{D}$, we denote by $\StDin(\bicat{C},\bicat{D})$ the following data:
	\begin{itemize}
		\item $0$-cells: mixed variance $2$-functors $F,G :  \bicat{C}^{\op} \times \bicat{C} \to \bicat{D}$;
		\item $1$-cells: strong lax dinatural transformations $\gamma : F \stdinato G$;
		\item $2$-cells: modifications $\Phi: \gamma \Rrightarrow \delta$ between them.
	\end{itemize}
\end{definition}

\begin{proposition}
	For $2$-categories $\bicat{C}$ and $\bicat{D}$, $\StDin(\bicat{C},\bicat{D})$ defined above forms a $2$-category.
\end{proposition}

\subsection{Proofs of Section \ref{sec:construction}}

\fixFinal*

	\begin{proof}
	By Lemma \ref{lem:PseudoInitAdjEquiv}, $R$ is part of an adjoint equivalence $(R : \oc \Phi \to \Phi, L : \Phi \to \oc \Phi, \eta: \id \xRightarrow{\cong} RL, \varepsilon: L R \xRightarrow{\cong} \id)$
	where $L : \Phi \to \oc \Phi$ is pseudo-final. Therefore, there exists a $1$-cell $t : \oc \One \to \Phi$ and a $2$-cell $\xi$ in $\bicat{C}$ as below:
	\vspace{-0.2cm}
	\begin{center}

	\end{center}
	\vspace{-0.2cm}
	and we define $\tau$ as the following $2$-cell in $\bicat{C}$
		\vspace{-0.2cm}
	\begin{center}
		%
	\end{center}
	\vspace{-0.2cm}
	which corresponds to a $2$-cell with boundaries  $t \Rightarrow Rt$ in $\bicat{D}$ as desired.

	Assume now that we have $2$-cells $\nu: v \Rightarrow R v$ and $\omega: w \Rightarrow R w$ in $\bicat{D}$.
	By the universal property of $L$, there exists a unique $2$-cell $\psi:v \Rightarrow w$ such that 
	\begin{center}
		%
	\end{center}
	in $\bicat{C}$. Cancelling the $2$-cells $\varepsilon$ on both sides using the adjunction identities
		\begin{center}
		%
	\end{center} and rewriting the equality above in $\bicat{D}$, we obtain that there exists a unique invertible $2$-cell $\psi : v \Rightarrow w$ in $\bicat{D}$ such that:
	\begin{center}
		%
	\end{center}    
\end{proof}

\fixNat*

\begin{proof}
	Immediate unfolding of the corresponding $2$-cells.
\end{proof}
\unifstrongpsdinat*

\begin{proof}
	For the strong pseudo-dinatural axioms, we will only prove the $2$-naturality axiom, as the other cases work similarly. Assume that we have the following equality in $\bicat{D}$:
	\begin{center}
		%
	\end{center}
	\vspace{-0.3cm}
	Let $\phi : r \circ u_f  \Rightarrow u_g$ be the unique $2$-cell in $\bicat{C}$ such that:
\vspace{-0.2cm}
\begin{center}
	%
\end{center}
\vspace{-0.2cm}
so that $\unif_{\rho} = J\phi \cdot t : J(r) \circ J(u_f) \circ t \Rightarrow J(u_g) \circ t$. We want to show that $\unif_{\gamma} = \unif_{\rho} \circ (J\theta \cdot (J(u_f) \circ t))$ which is immediate from the equality below:
\vspace{-0.4cm}
\begin{center}
	%
\end{center}
\vspace{-0.2cm}
	
	To prove that the operator $((-)^*, \fix, \unif)$ we constructed is a pseudo-fixpoint operator on $\bicat{D}$ uniform with respect to $J$, it only remains to show the coherence axiom between $\fix$ and $\unif$ \emph{i.e.} that for a $2$-cell 
	\vspace{-0.2cm}
	\begin{center}
		%
	\end{center}
	\vspace{-0.2cm}
			The left-hand side is equal to:
			\begin{center}
				%
			\end{center} 
		where $\phi : s \circ u_f  \Rightarrow u_g$ is the unique invertible $2$-cell in $\bicat{C}$ such that:
		\vspace{-0.2cm}
		\begin{center}
			%
		\end{center}
		\vspace{-0.2cm}
			so that the left-hand side is equal to the diagram below
			\begin{center}
				%
			\end{center}
			which corresponds to the right-hand side of the desired equality.
\end{proof}

\contractiblefix*

\begin{proof}

	We first show that $\delta$ is a morphism of pseudo-fixpoint operators by proving that it commutes with the $\fix$ $2$-cells:
	\begin{center}
	%
\end{center}
\vspace{-0.3cm}
	
	The left hand diagram is equal to:
	
	\begin{center}
		%
	\end{center}
 where the equality above follows from the coherence between $\fix^\dag$ and $\unif^\dag$. By definition of $\delta_0$, the right-hand diagram is equal to:
 		\vspace{-0.2cm}
	\begin{center}
		%
			\vspace{-0.2cm}
	\end{center}
	and we obtain the desired equality.
	We now need to show that $\delta$ is a morphism of uniform pseudo-fixpoint operators, \emph{i.e.} for every square in $\bicat{D}$:
			\vspace{-0.2cm}
	 \begin{center}
		%
	\end{center}
			\vspace{-0.2cm}
	By definition of $\unif^*_{\gamma}$ and the $2$-naturality axiom for $\unif^\dag$, the left hand diagram is equal to:
	\begin{center}
		%
	\end{center}
	where $\phi$ is the unique $2$-cell $ s \circ u_f  \Rightarrow u_g$ such that:
	\vspace{-0.2cm}
\begin{center}
	%
\end{center}
\vspace{-0.2cm}
	The right hand diagram is equal to 
	\begin{center}
		%
	\end{center}
	where the equality above follows from the $1$-naturality axiom for $\unif^\dag$.

	For uniqueness, assume that there exists another morphism of uniform pseudo-fixpoint operators $\beta : (-)^*  \to (-)^\dag$. We proceed to show that for every $f :A \to A$, $\delta_f = \beta_f$. Let $\phi_0$ be the unique invertible $2$-cell $1_{\Phi} \Rightarrow u_R$ in $\bicat{C}$ such that:
			\vspace{-0.2cm}
	\begin{center}
		%
	\end{center}
		\vspace{-0.2cm}
which we obtain from Lemma \ref{lem:fixInitial}. We show that 
 \vspace{-0.1cm}
\begin{center}
	%
\end{center} 
\vspace{-0.3cm}
by proving the following equality:

\begin{center}
	%
\end{center}
and using the universal property of $\delta_0$.
By definition of $\phi_0$, the left-hand side is equal to:

\begin{center}
	%
\end{center}

\begin{minipage}{\textwidth}
Since $\beta$ is a morphism of pseudo fixpoint operators, the right-hand side is equal to:
\vspace{-0.2cm}
\begin{center}
	%
\end{center}
\vspace{-0.2cm}
so both sides are equal as desired by definition of $\fix_R^*$. We therefore obtain that 
\end{minipage}
 \vspace{-0.1cm}
\begin{center}
	%
\end{center} 
\vspace{-0.3cm}
Since $\beta$ is a morphism of uniform pseudo-fixpoint operators, $\delta_f$ is equal to:
		\vspace{-0.2cm}
\begin{center}
	%
\end{center} 
		\vspace{-0.2cm}
Using the universal property of $\unif^*_{\mu_f}$, we obtain that $ \unif^*_{\mu_f} \circ (J\phi_0 \cdot t)=\id$ giving us the desired conclusion.
\end{proof}

\dinatbifree*
\begin{minipage}{\textwidth}
	\begin{proof}
	The $2$-cell $\nu$ in $\bicat{D}$ has the following boundary in $\bicat{C}$:
	\begin{center}
		%
	\end{center}
	Since $\oc L \circ L$ is a pseudo-final $\oc \oc$-coalgebra, there is a unique isomorphism $\lambda : w \Rightarrow v$ such that:
	\begin{center}
		%
	\end{center}
 Cancelling the $2$-cells $\varepsilon$ and $\oc \varepsilon$ on both sides using the adjunction identities and rewriting the equality above in $\bicat{D}$, we obtain the desired equality in $\bicat{D}$.
\end{proof}
\end{minipage}

\unifpsdinat*

\begin{proof}
	
	\noindent\textbf{Coherence between $\dinat$ and $\unif$:} for two squares in $\bicat{D}$ of the form

	\begin{center}
		%
	\end{center}
\begin{minipage}{\textwidth}
	where $\gamma \star \rho$ corresponds to the following $2$-cell:

	\begin{center}
		%
	\end{center} 
\end{minipage}
	The right-hand diagram is equal to:

		\begin{center}
			%
		\end{center}
	and the left-hand diagram is equal to:

		\begin{center}
		%
	\end{center}
\begin{minipage}{\textwidth}
	Let $\phi: (s\times r) \circ u_{\sigma (f \times g)}  \Rightarrow u_{\sigma (h \times k)}$ be the unique invertible $2$-cell in $\bicat{C}$ such that 

	\begin{center}
	%
\end{center}
\end{minipage}
		\vspace{-0.2cm}
	The left-hand diagram of the desired equality is then equal to
			\vspace{-0.2cm}
		\begin{center}
		%
	\end{center}
	From the universal property of $\unif$, we can now verify that 
	\[
	\unif_{\Pi^1_{\sigma(h\times k)} } \circ (\pi_1 \cdot J\phi \cdot (RR)^*) = \unif_{\rho\star \gamma}   \circ (Js \cdot \unif_{\Pi^1_{\sigma(f\times g)} }) \qquad \text{and}
	\]
	\[
	(\pi_2 \cdot  J\phi^{-1}\cdot (RR)^*) \circ \unif^{-1}_{\Pi^2_{\sigma(h\times k)} } \circ \unif_{ 
		\gamma\star \rho} = Jr \cdot \unif^{-1}_{\Pi^2_{\sigma(f\times g)} }
	\] to obtain the desired equality.\\

	\noindent\textbf{Coherence between $\dinat$ and $\fix$:} 
	for $1$-cells $f : A \to B$ and $g: B\to A$ in $\bicat{D}$, we want to show:

	\begin{center}

	\end{center}
\begin{minipage}{\textwidth}
	The right-hand diagram is equal to:
			\vspace{-0.2cm}
		\begin{center}
		%
	\end{center} 
		\vspace{-0.2cm}
	and the left-hand diagram is equal to:
\end{minipage}
			\vspace{-0.2cm}
	\begin{center}
		%
	\end{center}
			\vspace{-0.2cm}
	We use a similar reasoning as before and let Let $\phi: \sigma \circ u_{\sigma (g \times f)}  \Rightarrow u_{\sigma (f \times g)}$ be the unique invertible $2$-cell in $\bicat{C}$ such that 
			\vspace{-0.2cm}
	\begin{center}
		%
	\end{center}
		\vspace{-0.2cm}
		so that we can rewrite the previous diagram as:
	\begin{center}
		%
	\end{center}
		\vspace{-0.2cm}
	Using the universal property of $\unif$, we can show that $\unif_{\Pi^2_{\sigma(g\times f)} }^{-1} \circ \unif_{\Pi^1_{\sigma(f\times g)} }\circ (\pi_1 \cdot J\phi \cdot t)=\id$ so that we can simplify the diagram to 
			\vspace{-0.2cm}
	\begin{center}
		%
	\end{center}
			\vspace{-0.2cm}
	Using the definition of $\unif_{\Pi^1_{\sigma(g\times f)} }$ and the coherence between $\unif$ and $\fix$ we rewrite the diagram above as:
			\vspace{-0.2cm}
		\begin{center}
		%
	\end{center}
		\vspace{-0.2cm}
It remains to verify that $\unif_{\Pi^1_{\sigma(g\times f)} } \circ (\pi_2 \cdot J\phi^{-1} \cdot (RR)^*) \circ \unif^{-1}_{\Pi^2_{\sigma(f\times g)} }= \id$ and we obtain the desired equality.\\

\noindent\textbf{Pseudo-dinaturality axioms for $\dinat$:} We only show the unity axiom as the others work similarly. We want to show that for all $f:A\to A$, $\dinat_{f}^{1_A}= \id_{f^*}$. We start by proving it for $f=R$ and obtain the general statement by uniformity. We first note that the two diagrams below 

	\begin{center}
		%
	\end{center} 
		\vspace{-0.2cm}
	are both equal to:
			\vspace{-0.2cm}
	\begin{center}
		%
	\end{center}
		\vspace{-0.2cm}
Using  Lemma \ref{lem:fixFinal}, we conclude that $\dinat_R^{1_\Phi} =\id_{R^*}$. 

\begin{minipage}{\textwidth}
For a general $f :A \to A$ in $\bicat{D}$, to show that $\dinat_f^{1_A} =\id_{f^*}$, we make use of the coherence axiom between $\unif$ and $\dinat$ we proved previously. Consider the two squares:
	\vspace{-0.2cm}
	\begin{center}
		%
	\end{center}
		\vspace{-0.2cm}
		since we have shown that $\dinat_{R}^{1_\Phi} =\id_{R^*}$, we obtain that $\dinat^{1_A}_{f} = \id_{f^*}$ from the equality above.
\end{proof}

\begin{lemma}\label{lem:dinatR}
	For the pseudo-bifree algebra $R$, the $2$-cell $\dinat_R^R$ is equal to:
			\vspace{-0.2cm}
	\begin{center}
		%
	\end{center}    
It suffices to shat that $\psi =\din$ by proving that the equality below holds:
	\begin{center}
	%
\end{center}

\begin{minipage}{\textwidth}
By definition of $\psi$ and the coherence axiom between $\dinat$ and $\fix$, the left-hand diagram is equal to 
	\begin{center}
	%
\end{center}
\end{minipage}
		\vspace{-0.2cm}
	and we obtain the desired equality.
\end{proof}

\contractibledinat*

\begin{proof}
	Let $((-)^\dag, \dinat^\dag, \unif^\dag)$ be another pseudo-dinatural fixpoint operators uniform with respect to $J$ on $\bicat{D}$.
	We show that the morphism $\delta :((-)^*,\unif^*) \to ((-)^\dag, \unif^\dag)$ we defined in Section \ref{sec:construction} also commutes with the structural $2$-cells $\dinat$, \emph{i.e.} it satisfies the following coherence for every $f: A \to B$ and $g: B \to A$:
			\vspace{-0.2cm}
	\begin{center}
		%
	\end{center}
	\vspace{-0.3cm}
	As before, we first show it for $f=g=R$ 
		\begin{center}
		%
	\end{center}
	\vspace{-0.1cm}
	and obtain the general result using the coherence between $\unif$ and $\dinat$. 
	Using Lemma \ref{lem:dinatR}, the right-hand diagram is equal to:
		
		\begin{center}
			%
		\end{center}
	where $\phi_0:1_{\Phi} \Rightarrow J(u_R)$ is the unique invertible $2$-cell such that $R \cdot J\phi_0 = \mu_R \circ (J\phi_0 \cdot R)$ (see proof of Proposition \ref{prop:contractiblefix}).

\begin{minipage}{\textwidth}
	Let $\zeta : R^\dag \Rightarrow (RR)^\dag$ be the unique invertible $2$-cell (obtained from Lemma \ref{lem:dinatbifree}) such that
	\begin{center}
		%
	\end{center}
\end{minipage}
	It plays a similar role as $\din : t \Rightarrow (RR)^*$ for the operator $(-)^\dag$ as we can show that:
		\begin{center}
		%
	\end{center}
	We can also prove that:
	 \vspace{-0.1cm}
	\begin{center}
		%
\end{center} 

We can now conclude:
	\begin{center}
	%
\end{center}

\begin{minipage}{\textwidth}
For general morphisms $f:A \to B$ and $g: B \to A $ in $\bicat{D}$, we consider the squares:
\begin{center}
	%
\end{center}
\end{minipage}
and obtain from the coherence between $\unif$ and $\dinat$:
	\begin{center}
	%
\end{center}
Therefore, we have:
	\begin{center}
	%
\end{center}
\end{proof}